\documentclass[a4paper,fleqn,usenatbib]{mnras}
\usepackage{setspace}

\usepackage{newtxtext,newtxmath}

\usepackage{xcolor}

\usepackage{bm}
\usepackage{booktabs}
\usepackage{threeparttable}

\usepackage[T1]{fontenc}
\usepackage{ae,aecompl}

\usepackage{graphicx}	
\usepackage{amsmath}	
\usepackage{amssymb}	

\newcommand{\teff}{$T_{\rm eff}$}
\newcommand{\edd}{$\lambda_{\rm Edd}$}
\newcommand{\lbol}{$L_{\rm bol}$}
\newcommand{\ledd}{$L_{\rm Edd}$}
\newcommand{\ld}{$L_{\rm disk}$}
\newcommand{\lnu}{$L_{\nu}$}
\newcommand{\msun}{$M_{\odot}$}
\newcommand{\kh}{$\kappa_{2-10}$}
\newcommand{\ko}{$\kappa_{5100}$}
\newcommand{\lopt}{$L_{5100}$}
\newcommand{\lx}{$L_{2-10}$}
\newcommand{\aox}{$\alpha_{\rm OX}$}
\newcommand{\aoptuv}{$\alpha_{\rm opt-UV}$}
\newcommand{\gx}{$\Gamma_{\rm X}$}
\newcommand{\ha}{$\rm H{\alpha}$}
\newcommand{\hb}{$\rm H{\beta}$}
\newcommand{\oiii}{[O\,{\footnotesize III}]}
\newcommand{\feii}{{\rm Fe\,{\footnotesize II}}}
\newcommand{\ebv}{$E(B-V)$}
\newcommand{\mbh}{$M_{\rm BH}$}
\newcommand{\angstrom}{\mbox{\normalfont\AA}}
\newcommand{\sw}{\it Swift}
\newcommand{\nh}{$N_{\rm H}$}
\newcommand{\mdot}{$\dot{M}$}

\title[Broadband SEDs for {\it Swift}-observed AGN]{Modeling accretion disk emission with generalized temperature profile and its effect on AGN spectral energy distribution}

\author[H.Q. Cheng et al.]{
Huaqing Cheng,$^{1,2}$\thanks{E-mail: hqcheng@nao.cas.cn}
Weimin Yuan,$^{1,2}$
He-Yang Liu,$^{1,2}$
A.A. Breeveld,$^{3}$
\newauthor
Chichuan Jin$^{1}$
and Bifang Liu,$^{1,2}$\thanks{Email: bfliu@nao.cas.cn}
\\
\\
$^{1}$Key Laboratory of Space Astronomy and Technology, National Astronomical Observatories, Chinese Academy of Sciences, \\
20A Datun Road, Chaoyang District, Beijing, 100101, People's Republic of China\\
$^{2}$School of Astronomy and Space Sciences, University of Chinese Academy of Sciences, 19A Yuquan Road, Beijing, 100049, \\People's Republic of China\\
$^{3}$Mullard Space Science Laboratory, University College London, Holmbury St. Mary, Dorking, Surrey RH5 6NT, UK
}

\date{Accepted 2019 May 30. Received 2019 May 29; in original form 2018 December 2}

\pubyear{2019}

\begin{document}
\label{firstpage}
\pagerange{\pageref{firstpage}--\pageref{lastpage}}
\maketitle

\begin{abstract}
The broadband spectral energy distribution (SED) of Active Galactic Nuclei (AGN) is investigated for a well-selected sample composed of $23$ Seyfert 1 galaxies observed simultaneously in the optical/UV and X-ray bands with the Neil Gehrels {\it Swift} Observatory. The optical to UV continuum spectra are modeled, for the first time, with emission from an accretion disk with a generalized radial temperature profile, in order to account for the intrinsic spectra which are found to be generally redder than the model prediction of the standard Shakura-Sunyaev disk (SSD) ($F_\nu\propto\nu^{+1/3}$). The power-law indices of the radial temperature profile ($T_{\rm eff}(R)\propto R^{-p}$, $R$ is the radius of the accretion disk) are inferred to be $p=0.5$ -- $0.75$ (a median of $0.63$), deviating from the canonical $p=0.75$ for the SSD model as widely adopted in previous studies. A marginal correlation of a flatter radial temperature profile (a smaller $p$ value) with increasing the Eddington ratio is suggested. Such a model produces generally a lower peak of accretion disk emission and thus a smaller bolometric luminosity in some of the AGN, particularly those with high Eddington ratios, than that based on the SSD model by a factor of several. The broadband SED, the bolometric correction factors and their dependence on some of the AGN parameters are re-visited. We suggest that such non-standard SSD disks may operate in AGN and are at least partly responsible for the reddened optical/UV spectra as observed. One possible explanation for these flattened temperature profiles is the mass loss process in form of disk winds/outflows.
\end{abstract}




\section{Introduction} \label{sec:intro}

The study of the broadband energy distributions (SEDs) of active galactic nuclei (AGN) is important for understanding the central engine and the physical processes in supermassive black holes (SMBHs) \citep[e.g.,][]{1994ApJS...95....1E, 2006ApJS..166..470R, 2012MNRAS.425..907J, 2017MNRAS.471..706J, 2013ApJS..206....4K}. The SED of radio-quiet AGN is mainly dominated by two wavebands, i.e., the optical/UV and the X-ray band. The former is believed to originate from a geometrically thin and optically thick accretion disk, and the latter from a hot, optically thin corona. In the study of the AGN SED, the modeling of the optical/UV spectra is crucial since, as a common practice, the fitted model is used to determine the bolometric luminosity by extrapolating it to the extreme-UV (EUV) band, where the peak of typical AGN emission is located but can not be observed due to strong absorption by interstellar medium in the Galaxy.

Conventionally, the standard accretion disk (or Shakura-Sunyaev disk, SSD) \citep{1973A&A....24..337S, 1981ARA&A..19..137P} model is widely utilized to fit the optical/UV spectra. This model predicts a specific dependence of the disk effective temperature on radius $T_{\rm eff}(R)\propto R^{-0.75}$ and consequently a hump of a power-law spectrum $F_\nu\propto\nu^{\alpha} \ (\alpha=+1/3)$ with a high energy cutoff throughout much of the optical-to-EUV waveband. This emission is generally thought to account for the observed `big blue bump' (BBB) feature in AGN \citep[e.g.,][]{1978Natur.272..706S,2005ApJ...619...41S}. On the contrary, however, continua with softer power-laws in optical/UV band  are often observed in AGN ($\alpha\approx-0.7$ -- $-0.3$) \citep[e.g.,][]{2001AJ....122..549V, 2004ApJ...615..135S, 2007ApJ...668..682D, 2012ApJ...752..162S, 2014ApJ...794...75S, 2015ApJ...806..109J, 2016A&A...585A..87S, 2017MNRAS.467.4674L}, showing much redder optical/UV spectra than the theoretical model prediction.

The optical/UV spectra with shallower slopes ($\alpha<+1/3$), as observed in many AGN, may result from a number of factors \citep[see][for a detailed review]{1999PASP..111....1K}. Firstly, host galaxy starlight can contribute to the optical spectra especially at the red end \citep[e.g.,][]{2006ApJ...644..133B, 2009ApJ...697..160B}. Secondly, the dust extinction effect can also redden the spectra by scattering the UV and optical photons \citep[e.g.,][]{2016ApJ...824...38X,2017MNRAS.467..226G}. The latter has been widely advocated to explain the observed redder optical/UV spectra in AGN, assuming the emission is from an accretion disk of the SSD-type \citep[e.g.,][]{2009MNRAS.399.1553V, 2012A&A...539A..48M}. However, the amount of the dust extinction intrinsic to the source is hard to measure independently. One way is to make use of the broad-line Balmer decrement measured from optical spectroscopy, which is suggested to be a good indicator for dust extinction in the broad-line region \citep[e.g.,][]{2008MNRAS.383..581D,  2016MNRAS.462.3570S, 2017MNRAS.467..226G}. There have been efforts made in recent years to carefully assess the contributions from these two effects in order to recover the intrinsic optical/UV spectra in AGN. As examples, \citet{2009MNRAS.399.1553V} employed 2-D optical image decomposition to exclude host galaxy contamination for a sample of $29$ AGN with black hole masses determined from reverberation mapping. On the other hand, \citet{2010ApJS..187...64G} tried to correct for the dust extinction effect using the Balmer decrement for a sample of $92$ soft-X-ray-selected Seyfert 1 galaxies.

Alternatively, the reddening of the optical/UV spectra can also be explained as an intrinsic emission feature of an accretion disk \citep[e.g.,][]{2004ApJ...616..147G}. For instance, observations of NGC 5548 show a significantly reddened optical/UV spectra even after removing the host galaxy starlight and correcting for dust extinction \citep{2007arXiv0711.1025G}. Theoretically, the canonical radial temperature profile $T_{\rm eff}(R)\propto R^{-0.75}$ in the SSD solution is only valid under the assumption that mass is accreted from the outer disk all the way to the inner part without mass loss or gain of the accretion flow, and the accretion energy is solely dissipated into the disk and radiated away efficiently throughout the course of mass accretion. However, in reality this may not be true since there are likely other processes involved during the accretion process such as the disk-wind/outflow \citep[e.g.,][]{2013MNRAS.430.1102T,2015ARA&A..53..115K,2016MNRAS.460.1716D}, disk evaporation and condensation \citep[e.g.,][]{2009ApJ...707..233L, 2013ApJ...777..102Q, 2015ApJ...806..223L}, energy advection into the BH \citep[the `Advection-Dominated Accretion Flow (ADAF), e.g.,][]{1977ApJ...214..840I, 1994ApJ...428L..13N, 1995ApJ...444..231N, 1995ApJ...452..710N, 2014ARA&A..52..529Y} and photon trapping effect \citep[the `slim' disk, e.g.,][]{1988ApJ...332..646A, 2001PASJ...53..915W}. In particular, the temperature distribution for a `slim' disk model is much flatter \citep[$T_{\rm eff}(R)\propto R^{-0.5}$, e.g.,][]{1999ApJ...522..839W, 1999PASJ...51..725W} than that of the SSD (in this paper a temperature profile with the power-law index less than $0.75$ is referred to as `flat' or `flattened', and those with smaller power-law indices as `flatter'). Other factors, such as the general relativity effects \citep[e.g.,][]{1993PASJ...45...97Y} and irradiation from the central object \citep[e.g.,][]{1993PASJ...45..443S, 2008ChJAS...8..302C}, may also alter the radial temperature profile. In order to take these possible effects into account, a generalized disk model has been proposed in \citet{1994ApJ...426..308M} by parameterizing the disk radial temperature profile as a power-law ($T_{\rm eff}(R)\propto R^{-p}$) with the slope $p$ being a free parameter, different from the SSD solution which has $p=0.75$. It should be noted that even a small deviation of $p$ from $0.75$ can make a noticeable difference in the slope of the power-law regime of the disk spectrum \citep[$\alpha=3-2/p$, e.g.,][]{1972A&A....21....1P, 2002apa..book.....F, 2008bhad.book.....K}, which, falls into the optical/UV band in the case of AGN. \citet{2008RMxAC..32....1G} pointed out that an optical/UV spectral slope of $\alpha=-0.5$ implies a radial temperature profile of $T_{\rm eff}(R)\propto R^{-0.57}$. The idea of changing the radial temperature profiles has been adopted to generate a reddened optical/UV spectra in various studies, either by radially changing the mass flow rate inside the accretion disk through disk-wind \citep[e.g.,][]{2012MNRAS.426..656S, 2014MNRAS.438.3024L}, or introducing an additional irradiating (heating) term mostly to the outer disk region \citep[e.g.,][]{2002MNRAS.329..456S, 2004MNRAS.355.1080L}. 

The appropriate modeling of the observed optical/UV emission in AGN is essential for constructing the broadband SED. This is because the EUV emission, which is thought to be the peak of the AGN continuum (in the $\log\nu F_{\nu}-\log\nu$ manifestation), is completely obscured due to Galactic absorption and is usually estimated by extrapolating the optical/UV spectral model obtained from data fitting. This has become a common practice in recent studies of AGN SED \citep[e.g.,][]{2007MNRAS.381.1235V, 2009MNRAS.392.1124V, 2009MNRAS.399.1553V,2012MNRAS.425..907J,2012A&A...539A..48M}. However, in most of these studies the SSD model was employed, and any spectral deviation is presumably attributed to some external processes, such as dust extinction. Clearly, if the actual disk radial temperature profile is indeed flattened compared to the SSD solution, the bolometric luminosities \lbol \ may have been largely overestimated, especially in sources with strong optical/UV emission. For instance, \citet{2012MNRAS.426..656S} argued that in such a case the estimated disk luminosity can differ by a factor of as large as $2$, and so is the estimated bolometric luminosity \lbol \ and the Eddington ratio \edd \ (defined as \lbol/\ledd\ where $L_{\rm Edd}=1.3\times10^{38} (M/M_{\odot}) \ {\rm erg \ s^{-1}}$ is the Eddington luminosity). As such, there is a possibility that some of the bolometric correction factors $\kappa_{\nu}$ (defined as \lbol/$\nu$\lnu, $\nu L_{\nu}$ is the monochromatic luminosity at frequency $\nu$) obtained in previous work may be subject to systematic uncertainties. Some of these factors have been widely used in the literature to estimate the bolometric luminosities of AGN from luminosity measurement in a single waveband.

In this work, we study the broadband SED of AGN using a well-selected sample composed of $23$ Seyfert 1 galaxies with simultaneous optical/UV and X-ray data\footnote{We note that the emission in the mid- and far- infrared (IR) waveband is not included in this study, as it is known to be re-processed from the UV emission by torus \citep[e.g.,][]{1993ARA&A..31..473A, 2015ARA&A..53..365N}.} obtained by the Neil Gehrels {\it Swift} Observatory \citep{2004ApJ...611.1005G}. The sample objects are selected in such a way that they are subject to no or at most mild dust extinction (as estimated from the Balmer decrement); for the latter, the extinction effect is also corrected. Furthermore, the host galaxy starlight is eliminated by applying 2-D image decomposition, as also did in \citet{2009MNRAS.399.1553V}. We model the optical to UV spectra by adopting, for the first time, an accretion disk model with a generalized radial temperature profile. The X-ray spectra of the sample objects from simultaneous observations are also modeled. We derive the bolometric correction factors for both the optical and X-ray bands and investigate their dependence on some of the AGN parameters.

The paper is organized as follows. A brief description of the sample selection and multi-waveband data reduction are presented in Section \ref{sec:sample} and Section \ref{sec:datas}, respectively. In Section \ref{sec:optuv} we describe the spectral modeling in optical/UV band. The X-ray spectral analysis, broadband SEDs and bolometric corrections are investigated in Section \ref{sec:broadbandsed}. The main results are discussed in Section \ref{sec:discuss}, followed by a summary in Section \ref{sec:sum}. A flat universe model with a Hubble constant of $H_0=75 \ {\rm km \ s^{-1} \ Mpc^{-1}}$, $\Omega_{\rm M}=0.27$ and $\Omega_{\rm \Lambda}=0.73$ is adopted throughout the paper.
\vspace{-0.2cm}

\section{sample}
\label{sec:sample}

For the study of AGN broadband SED, it is essential to use simultaneous observational data in both the optical/UV and X-ray bands, in which AGN radiate most of their energy. At present, there are two space telescopes capable of achieving this goal, i.e., {\it Swift} and XMM-{\it Newton}. In this work, we draw our sample from the published AGN catalogue  on Vizier observed with {\it Swift}. We choose the {\it Swift} database for two reasons. Firstly, the data provided by UV and optical telescope (UVOT) \citep{2005SSRv..120...95R} onboard {\it Swift} are generally in a better quality compared to those of optical mirror (OM) \citep{2001A&A...365L..36M} equipped on XMM-{\it Newton}: (1) the data images are less influenced by the scattered light, which may lead to an unpleasant background noise and ghost image (2) the effective areas of the detectors for two UV filters ({\it uvm2}, {\it uvw2}) are larger than those in OM, which makes the UV throughput a factor of $10$ higher than that of the OM instrument. Secondly, the X-ray spectra obtained by its X-ray telescope (XRT) \citep{2005SSRv..120..165B} are sufficient for our SED study as long as the exposure is long enough (a few kiloseconds). The sample is compiled of AGN with simultaneous observations with data from both the UVOT and XRT instruments. We consider only type $1$ AGN classified as radio-quiet in the literature. The modeling of the optical/UV spectra with an accretion disk model requires the knowledge of the black hole masses and the amount of dust extinction along the line of sight (see Section \ref{sec:dustandmass}). We therefore limit our sample to those having optical spectroscopic data from the Sloan Digital Sky Survey (SDSS), based on which both the broad-line Balmer decrements and black hole masses can be estimated in a reliable and homogeneous way. This leads to $103$ objects in total. Among these sources, $29$ objects are abandoned since there are no distinguishable \hb \ broad-line components in the SDSS spectra. $18$ objects are abandoned since the UVOT observations are operated in less than $2$ filters (the disk model employed in our work have two free parameters, see Section \ref{sec:pfree}). $2$ sources (SDSS J165430.72+395419.7, UM 614) are abandoned as they lie on the edge of UVOT image, which can introduce large uncertainties to the photometric measurements. $2$ sources (2MASX J11475508+0902284, 2MASX J15505317+0521119) are abandoned since their XRT spectra have less than $50$ source counts, which can not be fitted to determine the X-ray spectral slope (see Section \ref{sec:xrayfit}).

Next, we discard $20$ sources suffering from substantial dust extinction, which is indicated by their broad-line Balmer decrements. We assume a zero point \ha/\hb$=3.06$ for AGN having no intrinsic dust extinction \citep[][see Section \ref{sec:dustandmass}]{2008MNRAS.383..581D}, and regard those sources with \ha/\hb$>3.4$ as substantial reddening \citep[also adopted in][]{2010ApJS..187...64G}. Since 2-D imaging AGN-host galaxy decomposition is required in this work, we also discard the following objects: (1) $6$ faint AGN embedded within a dominant bulge of the host galaxy (2) $3$ sources too close to a nearby star. In these $29$ sources only the data in $3$ optical filters (for the former $20$ sources) or $3$ UV filters (for the latter $9$ sources) can be used for spectral modeling. However, this may lead to uncertainties in the determination of the optical/UV spectral slope and model parameters. We perform simple tests to the uncertainty caused by using photometric data in only 3 optical or UV filters and find that the spectral slope and SED parameters can differ a lot. For instance, in PG 1138+222, the optical/UV spectral slopes derived are $-0.32$ (using {\it v}, {\it b} and {\it u} filters), $+0.70$ (using {\it uvw1}, {\it uvm2} and {\it uvw2} filters) and $+0.02$ (using all the $6$ filters), and the difference in the fitted model parameters leads to a difference in the Eddington ratio by a factor of $\sim10$ ($0.221$, $1.952$ and $0.352$). This is because in actual cases the spectral shapes in optical and UV bands may not be the same (which can be recognized in Figure \ref{fig:seds}). Therefore, in order to get a more accurate estimation of the optical/UV spectral shape as well as a more convincing result of the spectral modeling, we do not include the above sources in our final sample. The final sample is composed of $23$ sources, whose basic data are listed in Table \ref{tab:information}.

\begin{table*}
\centering
\caption{Basic information of our sample AGN.}
\label{tab:information}
\begin{threeparttable}
\small
\renewcommand\arraystretch{1.0}
\begin{tabular}{cccccccc} 
\hline
\hline
No. & Object & RA & DEC & Redshift & $E(B-V)_{\rm G}^{\dagger}$ & Swift ObsID & SDSS SpecObjlD \\
&       &	     &      &         &             &             & (Plate-MJD-Fiber) \\
\hline
1 & Mrk 1018 & 02 06 15.9 &  -- 00 17 29 & 0.0424 &  0.028 & 00035166001 & 0404--51812--0141 \\
2 & MCG 04-22-042 & 09 23 43.0 & +22 54 33 & 0.0323 & 0.043 & 00035263001 & 2290--53727--0578 \\
3 & Mrk 705 & 09 26 03.3 & +12 44 04 & 0.0292 & 0.041 & 00090998001 & 2578--54093--0195 \\
4 & RX J1007.1+2203 & 10 07 10.2 & +22 03 02 & 0.0820 & 0.032 & 00036537002 & 2364--53737--0458 \\
5 & CBS 126 & 10 13 03.2 & +35 51 24 & 0.0791 & 0.011 & 00035306001 & 1951--53389--0614 \\
6 & Mrk 141 & 10 19 12.6 & +63 58 03 & 0.0417 & 0.010 & 00035765001 & 0488--51914--0161 \\
7 & Mrk 142 & 10 25 31.3 & +51 40 35 & 0.0449 & 0.016 & 00036539002 & 1008--52707--0558 \\
8 & Ton 1388 & 11 19 08.7 & +21 19 18 & 0.1765 & 0.022 & 00035767001 & 6430--56299--0516 \\
9 & SBS 1136+594 & 11 39 08.9 & +59 11 55 & 0.0601 & 0.014 & 00035265001 & 7099--56666--0869 \\
10 & PG 1138+222 & 11 41 16.1 & +21 56 21 & 0.0632 & 0.027 & 00036541001 & 2504--54179--0639 \\
11 & KUG 1141+371 & 11 44 29.9 & +36 53 09 & 0.0381 & 0.019 & 00091632001 & 1997--53442--0126 \\
12 & Mrk 1310 & 12 01 14.3 & -- 03 40 41 & 0.0196 & 0.031 & 00091002003 & 0331--52368--0121\\
13 & RX J1209.8+3217 & 12 09 45.2 & +32 17 02 & 0.1444 & 0.017 & 00035769006 & 2004--53737--0466 \\
14 & Mrk 50 & 12 23 24.1 & +02 40 45 & 0.0234 & 0.016 & 00080077001 & 0519--52283--0487 \\
15 & Mrk 771 & 12 32 03.6 & +20 09 29 & 0.0630 & 0.027 & 00080082001 & 2613--54481--0342 \\
16 & PG 1307+085 & 13 09 47.0 & +08 19 48 & 0.1550 & 0.033 & 00037571001 & 1795--54507--0457 \\
17 & Ton 730 & 13 43 56.7 & +25 38 48 & 0.0866 & 0.013 & 00037572002 & 2246--53767--0066 \\
18 & RX J1355.2+5612 & 13 55 16.6 & +56 12 45 & 0.1219 & 0.007 & 00036547001 & 1323--52797--0443\\
19 & Mrk 1392 & 15 05 56.5 & +03 42 26 & 0.0361 & 0.047 & 00081174001 & 0589--52055--0111 \\
20 & Mrk 290 & 15 35 52.3 & +57 54 09 & 0.0296 & 0.015 & 00080152003 & 0615--52347--0108 \\
21 & Mrk 493 & 15 59 09.6 & +35 01 47 & 0.0313 & 0.024 & 00035080001& 1417--53141--0078 \\
22 & KUG 1618+410 & 16 19 51.3 & +40 58 48 & 0.0379 & 0.007 & 00036548001 & 1171--52753--0166 \\
23 & RX J1702.5+3247 & 17 02 31.1 & +32 47 20 & 0.1633 & 0.023 & 00035771002 & 0973--52426--0114 \\
\hline
\end{tabular}
{$\bm Notes.$} $\dagger$. Galactic reddening from \citet{1998ApJ...500..525S}.
\end{threeparttable}
\end{table*}

\section{data reduction}
\label{sec:datas}
The reduction procedures of the optical/UV photometric data, the X-ray data and the optical SDSS spectroscopic data are described in this section. We download the pipeline-processed `Level 2' UVOT and XRT \textsc{fits} files from ASI Science Data Center\footnote{\url{http://www.asdc.asi.it/mmia/index.php?mission=swiftmastr}}(ASDC). For sources with multiple observations we utilize the data with the maximum exposure time. The SDSS optical spectra are downloaded from SDSS Science Archive Server\footnote{\url{https://dr13.sdss.org/home}}(SAS).

\begin{table*}
\centering
\caption{UVOT photometry.}
\label{tab:galfit}
\begin{threeparttable}
\small
\renewcommand\arraystretch{1.0}
\begin{tabular}{ccccccccccc} 
\hline
\hline
No. & Object & m$_{v,AGN}$ & m$_{v,ga}$ & m$_{b,AGN}$ & m$_{b,ga}$ & m$_{u,AGN}$ & m$_{u,ga}$ & 
m$_{w1,AGN}$ & m$_{m2,AGN}$ & m$_{w2,AGN}$\\
(1)&(2)&(3)&(4)&(5)&(6)&(7)&(8)&(9)&(10)&(11)\\
\hline
1 & Mrk 1018 & $15.12$ & $14.24$ & $15.48$ & $15.09$ & $14.00$ & $14.70$ & $13.54$ & $13.42$ & $13.44$\\
2 & MCG 04-22-042 & $14.95$ & $14.22$ & $15.31$ & $14.93$ & $13.80$ & $12.61$ & $13.46$ & $13.44$ & $13.45$ \\
3 & Mrk 705 & $15.11$ & $14.25$ & $15.59$ & $15.34$ & $14.28$ & $15.04$ & $14.04$ & $13.95$ & $14.02$ \\
4 & RX J1007.1+2203 & $16.86$ & - & $17.26$ & - & $16.12$ & - & $16.02$ & $15.91$ & $15.88$ \\
5 & CBS 126 & $15.49$ & - & $15.82$ & - & $14.61$ & - & $14.47$ & $14.40$ & $14.35$ \\
6 & Mrk 141 & $15.94$ & $14.97$ & $16.42$ & $15.79$ & $15.47$ & $15.74$ & $15.29$ & $15.34$ & $15.40$ \\
7 & Mrk 142 & $15.70$ & $16.06$ & $15.96$ & $17.17$ & $14.82$ & - & $14.59$ & $14.41$ & $14.39$ \\
8 & Ton 1388 & $14.43$ & - & $14.55$ & - & $13.23$ & - & $13.00$ & $12.82$ & $12.92$\\
9 & SBS 1136+594 & $15.59$ & $16.53$ & $16.08$ & - & $14.62$ & $16.28$ & $14.46$ & $14.46$ & $14.45$\\
10 & PG 1138+222 &$15.49$ & - & $16.01$ & - & $14.69$ & - & $14.33$ & $14.25$ & $14.13$ \\
11 & KUG 1141+371 & $16.90$ & $15.53$ & $17.24$ & $16.34$ & $16.08$ & $16.80$ & $15.84$ & $15.99$ & $16.04$ \\
12 & Mrk 1310 & $16.74$ & $15.07$ & $17.41$ & $15.83$ & $15.54$ & $15.22$ & $15.52$ & $15.59$ & $15.64$\\
13 & RX J1209.8+3217 & $16.95$ & - & $17.33$ & - & $16.23$ & - & $16.11$ & $15.97$ & $15.98$\\
14 & Mrk 50 & $16.00$ & $15.21$ & $16.25$ & $15.85$ & $14.76$ & $15.66$ & $14.56$ & $14.60$ & $14.63$ \\
15 & Mrk 771 & $15.29$ & - & $15.62$ & - & $14.15$ & -  & $13.84$ & $13.76$ & $13.74$ \\
16 & PG 1307+085 & $15.51$ & - & $15.61$ & - & $14.30$ & - & $14.09$ & $13.93$ & $13.83$\\
17 & Ton 730 & $16.35$ & $16.55$ & $16.60$ & $17.38$ & $15.18$ & - & $14.94$ & $14.79$ & $14.71$ \\
18 & RX J1355.2+5612 & $16.37$ & - & $16.93$ & - & $15.86$ & - &$15.76$ & $15.64$ & $15.72$ \\
19 & Mrk 1392 & $15.21$ & $14.25$ & $15.60$ & $15.02$ & $14.31$ & $15.15$ & $13.87$ & $13.77$ & $13.81$\\
20 & Mrk 290 & $15.17$ & $15.33$ & $15.48$ & $15.49$ & $14.18$ & - & $14.05$ & $14.09$ & $14.16$ \\
21 & Mrk 493 & $15.38$ & $14.16$ & $15.63$ & $14.81$ & $15.42$ & $14.66$ & $14.19$ & $14.15$ & $14.22$ \\
22 & KUG 1618+410 & $17.29$ & $15.21$ & $17.81$ & $15.97$ & $16.71$ & $15.99$ & $16.26$ & $16.18$ & $16.17$ \\
23 & RX J1702.5+3247 & $15.45$ & - & $15.71$ & - & $14.54$ & - & $14.46$ & $14.36$ & $14.47$ \\
\hline
\end{tabular}
{$\bm Notes.$} Column (1): identification number assigned in this paper. Column (2): NED name of the sample objects. Column (3): AGN magnitude in {\it v} band. Column (4): total magnitude of the host galaxy in {\it v} band. Column (5): AGN magnitude in {\it b} band. Column (6): total magnitude of the host galaxy in {\it b} band. Column (7): AGN magnitude in {\it u} band. Column (8): total magnitude of the host galaxy in {\it u} band. Column (9): AGN magnitude in {\it uvw1} band. Column (10): AGN magnitude in {\it uvm2} band. Column (11): AGN magnitude in {\it uvw2} band. The magnitudes are the Vega magnitudes of {\it Swift} system \citep{2008MNRAS.383..627P}. All the magnitudes have been corrected for the Galactic dust extinction.
\end{threeparttable}
\end{table*}

\begin{figure}
	\includegraphics[width=\columnwidth]{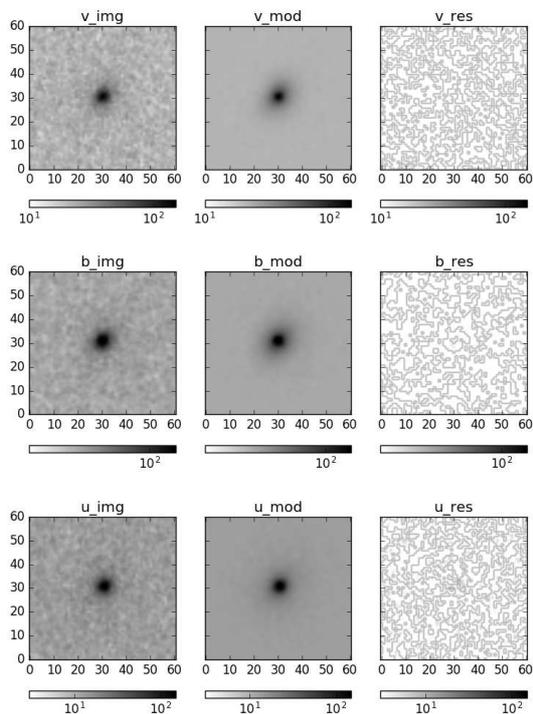}
\vspace{-1cm}
\caption{An example of 2-D image decomposition results for Mrk 1310. The rows from top to bottom represent the results in {\it v}, {\it b} \& {\it u} bands respectively. In each row the first column shows the data image, the second column shows the best-fit model (including the background component) and the third column shows the residual image derived by subtracting the model from the original data.}
\label{fig:galfitexample}
\end{figure}

\subsection{UVOT data}
\label{sec:uvotdata}

All the sources in our sample have $6$ UVOT filter photometric measurements in the optical ({\it v}, {\it b}, {\it u}) and UV ({\it uvw1}, {\it uvm2}, {\it uvw2}) bands. We follow the UVOT reduction threads\footnote{\url{http://www.swift.ac.uk/analysis/uvot/index.php}} for data reduction. The `Level $2$' UVOT \textsc{fits} files for each of the filters are summed together by using the procedure \texttt{uvotimsum}, and source magnitudes are extracted by using \texttt{uvotsource}. We use a circle source region with a radius of $5''$ centering on the source, and a background region with a radius of $20''$ selected from a source-free region close-by. At the UV wavelengths the starlight contribution within the aperture is negligible, and we simply consider the measurement as being dominated by the AGN radiation. For the {\it v}, {\it b} and {\it u} magnitudes, contamination from the host galaxy starlight may not be negligible, and we thus perform 2-D image decomposition on the obtained images to decompose the AGN from the host galaxy. Such a procedure was also adopted in a similar work by \citet{2009MNRAS.399.1553V} for UVOT data.

We use the \textsc{galfit} software developed by \citet{2002AJ....124..266P} to perform 2-D image decomposition. For each of the image, a point spread function (PSF) is constructed from nearby stars in the field of view \citep[USNO A$2.0$ catalogue,][]{1998AAS...19312003M} with count rates comparable to that of the target AGN. The background is calculated by selecting a source-free region near the source and is always included. The AGN is modeled with the point-like source (as an instrumental PSF), with an initial magnitude and position from the results of \texttt{uvotsource}. A potential galaxy component is added in, as either an exponential or a Sersic model, whichever results in a better fit. Such a model is fitted to the image data of the three optical filters, resulting in good fits in all but one object (PG 1138+222), which is judged by visually examining the residual images. Among these, $8$ objects can be well fitted with a point-like source model in all of the three filters (i.e., the host galaxy starlight contribution is negligible), while in the remaining $14$ objects an additional galactic component (exponential) is needed to improve the fits for at least two filters. An example of the 2-D imaging fitting results is shown in Figure \ref{fig:galfitexample}. For PG 1138+222, no good fit can be achieved due to the poor data quality of the UVOT images. Since no host galaxy is seen apparently in the images, we consider them to be dominated by AGN emission, and simply use the $5''$ aperture photometric measurements as the AGN magnitudes.

Finally, with the PSF magnitudes derived from \textsc{galfit}, we perform correction for coincidence-loss \citep[the phenomenon when two or more photons arrive at a similar location on the detector within the same CCD readout interval, see][]{2008MNRAS.383..627P, 2010MNRAS.406.1687B} in the optical filters\footnote{We find that for a few cases the PSF flux after the coincidence correction exceeds that obtained by \texttt{uvotsource} using the standard $5''$ aperture. Considering that there is some scope for error in the coincidence loss correction when the PSF image is extended, we simply choose the $5''$ aperture photometry.}. The best-estimated optical-to-UV magnitudes of the AGN for each of the filter bands are listed in Table \ref{tab:galfit}, along with the magnitudes of the host galaxies in the optical filters for those with non-negligible host galaxy starlight. The magnitudes are corrected for the Galactic extinction by employing the IDL code \texttt{fm\_unred.pro} with $E(B-V)_{\rm G}$ given in Table \ref{tab:information}, assuming the Milky Way (MW) extinction curve. No correction is made for the contamination to the continuum from AGN emission lines, which is considered to be negligible\footnote{Considering the relatively low redshifts and UVOT coverage (1928, 2246, 2600, 3465, 4392, 5468 \angstrom), Mg II (2798 \angstrom, rest frame) is the main contaminant in {\it uvw1} and {\it u} filters. AI III (1958 \angstrom)+ C III (1909 \angstrom) might be another contaminant in {\it uvw2} filter. However, for the majority of the sample objects, the emission line contamination was estimated to be a negligible effect in previous studies \citep{2009MNRAS.399.1553V, 2010ApJS..187...64G}}.

\begin{table*}
\centering
\caption{Log of XRT data reduction.}
\label{tab:xrtreduce}
\begin{threeparttable}
\small
\renewcommand\arraystretch{1.0}
\begin{tabular}{ccccccccccc} 
\hline
\hline
No. & Object & Mode & Exposure time & Source count rate & Background count rate \\
(1)&(2)&(3)&(4)&(5)&(6)\\
\hline
1 & Mrk 1018 & PC & $5208$ & $3.20$ & $1.46$ \\
2 & MCG 04-22-042 & PC & $9129$ & $2.96$ & $1.22$  \\
3 & Mrk 705 & WT & $1773$ & $5.94$ & $10.3$ \\
4 & RX J1007.1+2203 & PC & $11641$ & $0.43$ & $1.07$ \\
5 & CBS 126 & PC & $4734$ & $1.64$ & $1.61$  \\
6 & Mrk 141 & PC & $8156$ & $1.18$ & $1.56$ \\
7 & Mrk 142 & PC &  $2861$ & $2.23$ & $1.05$ \\
8 & Ton 1388 & PC & $5090$ & $2.71$ & $1.38$ \\
9 & SBS 1136+594 & PC & $9202$ & $3.91$ & $1.49$ \\
10 & PG 1138+222 & PC & $4092$ & $3.75$ & $1.44$ \\
11 &KUG 1141+371 & PC & $2680$ & $1.39$ & $1.08$  \\
12 & Mrk 1310 & PC & $4054$ & $1.43$ & $2.02$ \\
13 & RX J1209.8+3217 & PC & $17210$ & $0.21$ & $1.12$\\
14 & Mrk 50 & PC & $6399$ & $3.95$ & $1.61$ \\
15 & Mrk 771 & PC & $6687$ & $2.63$ & $1.81$ \\
16 & PG 1307+085 & PC & $8769$ & $1.78$ & $2.12$\\
17 & Ton 730 & PC & $11910$ & $1.39$ & $1.28$ \\
18 & RX J1355.2+5612 & PC & $4398$ & $1.01$ & $1.55$ \\
19 & Mrk 1392 & PC & $6331$ & $2.60$ & $1.93$ \\
20 & Mrk 290 & PC & $5423$ & $2.47$ & $1.88$ \\
21 & Mrk 493 & PC & $6978$ & $2.15$ & $1.30$ \\
22 & KUG 1618+410 & PC & $2332$ & $0.44$ & $1.42$ \\
23 & RX J1702.5+3247 & PC & $9144$ & $2.00$ & $1.33$ \\
\hline
\end{tabular}
{$\bm Notes.$} Column (1): identification number assigned in this paper. Column (2): NED name of the sample objects. Column (3): observational mode of XRT. Column (4): exposure time of XRT observation in unit of s$^{-1}$. Column (5): count rate of the source in unit of $10^{-1}$ counts s$^{-1}$. Column (6): count rate of the background in unit of $10^{-2}$ counts s$^{-1}$.
\end{threeparttable}
\end{table*}

\subsection{XRT data}
\label{sec:xrtdata}

Of $23$ XRT observations of the sample objects, $22$ were performed in the Photon Counting (PC) mode and one in the Windowed Timing (WT) mode (Mrk 705). We follow the XRT analysis threads\footnote{\url{http://www.swift.ac.uk/analysis/xrt/index.php}} starting with `Level $2$' clean event \textsc{fits} files and employ \texttt{xselect} procedure to extract the source spectra. For the PC mode observation, the source photons are selected by using a circular region centered on the target with a typical radius of $47''$, and the background photons from a close-by source-free region with $r=188''$.  For the WT mode we use two circular regions with $r=47''$ for the source and background selection, respectively. We perform pile-up correction for sources with count rates larger than $0.5$ counts s$^{-1}$ observed in PC mode by excluding a small region in the center with a radius determined from \texttt{ximage} procedure. For Mrk 705, the observed count rate ($\sim0.6$ counts s$^{-1}$) is much less than the threshold of pile-up ($100$ counts s$^{-1}$) for WT mode. X-ray events with grades from $0$ -- $12$ are extracted to produce the X-ray spectra for the full energy range of the XRT (about $0.2$ -- $10$ keV).  For the extracted X-ray spectra, we use \texttt{xrtmkarf} to build the Ancillary Response Files (ARFs). The Response Matrix Functions (RMFs) are determined individually by using \texttt{quzcif}. The source spectra are rebinned to have a minimum of $25$ photons in each energy bin by using \texttt{grppha} (version 3.0.1). The information on the XRT data reduction is given in Table \ref{tab:xrtreduce}.

\subsection{SDSS optical spectra}
\label{sec:sdss}

\begin{table*}
\centering
\caption{SDSS spectral fitting results.}
\label{tab:sdssfit}
\begin{threeparttable}
\small
\renewcommand\arraystretch{1.0}
\begin{tabular}{ccccccccccc} 
\hline
\hline
No. & Object & $\rm FWHM_{H\alpha^b}$ & $\rm FWHM_{H\beta^b}$ & $\log L_{\rm H\alpha^b}$ & $\log L_{\rm H\beta^b}$
 & $\log L_{5100}$ & $L_{\rm H\alpha^b}/L_{\rm H\beta^b}$ & $\log (M_{\rm BH}/M_{\sun})$ \\
(1)&(2)&(3)&(4)&(5)&(6)&(7)&(8)&(9)\\
\hline
1 & Mrk 1018 & $4152$ & $6011$ & $42.17$ & $41.74$ & $43.86$ & $2.67$ & $8.40$ \\
2 & MCG 04-22-042 & $2062$ & $2989$ & $42.34$ & $41.97$ & $43.69$ & $2.33$  & $7.71$ \\
3 & Mrk 705 & $2062$ & $2039$ & $42.03$ & $41.57$ & $43.38$ & $2.88$  & $7.22$  \\
4 & RX J1007.1+2203 & $2092$ & $2407$ & $42.17$ & $41.67$ & $43.66$ & $3.17$ & $7.50$ \\
5 & CBS 126 & $2639$ & $2914$ & $43.00$ & $42.53$ & $44.18$ & $2.94$ & $7.93$  \\
6 & Mrk 141 & $3624$ & $6672$ & $41.91$ & $41.39$ & $43.51$ & $3.31$ & $8.31$ \\
7 & Mrk 142 & $1494$ & $1650$ & $42.00$ & $41.58$ & $43.60$ & $2.64$ & $7.15$   \\
8 & Ton 1388 & $2966$ & $4004$ & $44.11$ & $43.60$ & $45.27$ & $3.25$ & $8.75$  \\
9 & SBS 1136+594 & $2637$ & $3249$ & $42.72$ & $42.24$ & $43.86$ & $2.99$ & $7.86$ \\
10 & PG 1138+222 & $2246$ & $2246$ & $42.55$ & $42.02$ & $43.95$ & $3.34$ & $7.59$ \\
11 &KUG 1141+371 & $6795$ & $9049$ & $41.82$ & $41.36$ & $43.28$ & $2.83$ & $8.46$   \\
12 & Mrk 1310 & $2252$ & $2717$ & $41.31$ & $40.92$ & $42.75$ & $2.46$ & $7.16$ \\
13 & RX J1209.8+3217 & $2019$ & $2038$ & $42.54$ & $42.08$ & $44.09$ & $2.89$ & $7.57$ \\
14 & Mrk 50 & $4838$ & $5837$ & $41.78$ & $41.33$ & $43.07$ & $2.83$ & $7.98$ \\
15 & Mrk 771 & $2800$ & $3358$ & $42.66$ & $42.19$ & $43.91$ & $2.92$ & $7.92$ \\
16 & PG 1307+085 & $3650$ & $4612$ & $43.67$ & $43.18$ & $44.78$ & $3.13$ & $8.63$ \\
17 & Ton 730 & $3009$ & $3520$ & $42.49$ & $42.12$ & $43.95$ & $2.33$ & $7.98$   \\
18 & RX J1355.2+5612 & $1141$ & $1192$ & $42.67$ & $42.16$ & $44.16$ & $3.22$ & $7.14$ \\
19 & Mrk 1392 & $4183$ & $5371$ & $42.29$ & $41.83$ & $43.60$ & $2.91$ & $8.17$  \\
20 & Mrk 290 & $3612$ & $4533$ & $42.22$ & $41.82$ & $43.58$ & $2.53$ & $8.01$  \\
21 & Mrk 493 & $1095$ & $1095$ & $41.71$ & $41.29$ & $43.41$ & $2.63$ & $7.51$ \\
22 & KUG 1618+410 & $2045$ & $2418$ & $41.22$ & $40.75$ & $42.90$ & $2.92$ & $7.13$ \\
23 & RX J1702.5+3247 & $2330$ & $2442$ & $43.27$ & $42.82$ & $44.70$ & $2.82$ & $8.03$\\
\hline
\end{tabular}
{$\bm Notes.$} Column (1): identification number assigned in this paper. Column (2): NED name of the sample objects. Column (3): FWHM of the $\rm H\alpha$ broad component. Column (4): FWHM of the $\rm H\beta$ broad component. Column (5): logarithmic of the luminosity of the $\rm H\alpha$ broad component in units of erg s$^{-1}$. Column (6): logarithmic of the luminosity of the $\rm H\beta$ broad component in units of erg s$^{-1}$. Column (7): logarithmic of the 5100 \angstrom\ monochromatic luminosity in units of erg s$^{-1}$. Column (8): broad-line Balmer decrement derived from $L_{\rm H\alpha^b}$/$L_{\rm H\beta^b}$. Column (9): logarithmic of Black hole mass $M_{\rm BH}$ derived from $\rm FWHM({H\beta^b})$ and $L_{5100}$ \citep{2006ApJ...641..689V} in units of $M_{\sun}$.
\end{threeparttable}
\end{table*}

\begin{figure}
	\includegraphics[width=\columnwidth]{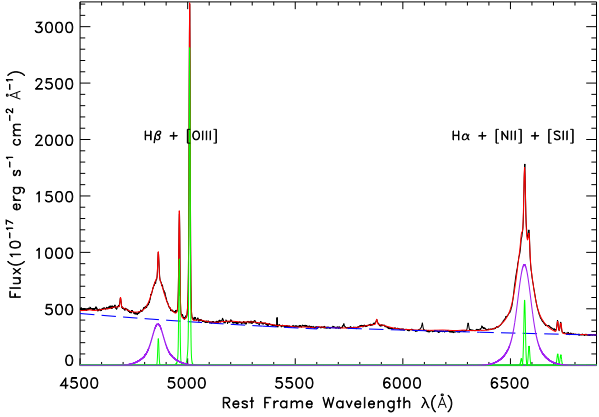}
\caption{Illustration of the spectral fitting for Mrk 290. The black solid line denotes the rest-frame SDSS spectrum. Purple and green lines represent the broad and narrow components of $\rm H\alpha$ and $\rm H\beta$ respectively. Red solid line shows the overall model.}
\label{fig:sdss}
\end{figure}

The SDSS optical emission line spectra are used to derive the Balmer decrements, as well as to estimate the masses of the black holes. The spectral analysis follows the procedure developed in our previous series work on SDSS AGN spectroscopic studies, as described in \citet{2006ApJS..166..128Z}, \citet{2008MNRAS.383..581D} and \citet{2018ApJS..235...40L}. For all the sources of the sample, the SDSS spectra are clearly dominated by emission from the AGN. We fit simultaneously the AGN continuum, the Balmer lines and the \feii \ emission lines. A broken power-law with a break frequency at $5600$ \angstrom \ is adopted to fit the AGN continuum in the \ha \ and \hb \ region. The optical \feii \ multiplets are modeled by two separate templates built by \citet{2008MNRAS.383..581D} based on the results of \citet{2004A&A...417..515V}. The Balmer lines are de-blended into a broad and a narrow component. The broad components are fitted with one or two gaussians assuming their redshifts to be the same, while the narrow lines are fitted with one gaussian except for the \oiii \ lines. The \oiii \ $\lambda4959,5007$ doublets are fitted with one or two gaussians for each assuming that they have the same profiles and redshifts, with the flux ratio fixed to the theoretical value. Figure \ref{fig:sdss} illustrates one example of the spectral fittings. Table \ref{tab:sdssfit} lists the results of SDSS spectral fitting for the sample objects, including the Full Width at Half Maximum (FWHM) and the luminosities of \ha \ and \hb \ broad components and the $5100$ \angstrom \ monochromatic luminosities \lopt.

\subsection{Estimation of intrinsic dust extinction and black hole mass}
\label{sec:dustandmass}

\begin{figure}
	\includegraphics[width=\columnwidth]{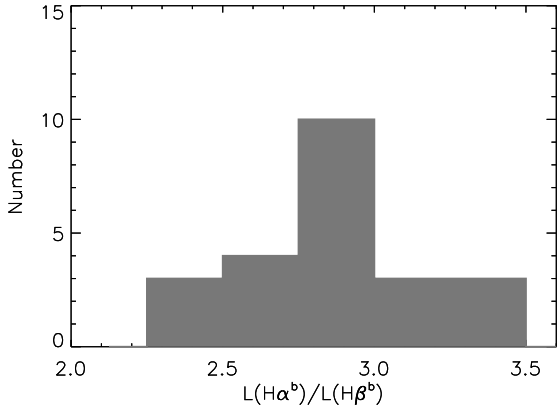}
\caption{The distribution of the broad-line Balmer decrements for the sample.}
\label{fig:dustdist}
\end{figure}

The intrinsic dust extinction in our sample AGN is estimated from the broad-line Balmer decrement, which is derived from the flux ratio of the broad components of \ha \ and \hb. The intrinsic value of  \ha/\hb\ of AGN with no dust reddening has been a topic of debate \cite[e.g.,][]{2005ApJ...620..629D, 2007ApJ...671..104L, 2008MNRAS.383..581D, 2016MNRAS.462.3570S, 2017MNRAS.467..226G}. In this work, we adopt an intrinsic \ha/\hb\ value of $3.06$ derive by \citet{2008MNRAS.383..581D} based on a sample of $446$ blue AGN, which is verified later by \citet{2016ApJ...832....8B} with a much larger AGN sample and has been employed in various studies \citep[e.g.,][]{2009AJ....137.4002W, 2010ApJS..187...64G, 2016ApJ...822...64L}. The color excess is derived as
\begin{equation}
	E(B-V)=2.12\log\frac{L_{\rm H\alpha^b}/L_{\rm H\beta^b}}{3.06}
	\label{equ:ebv}
\end{equation}
by using the Small Magellanic Cloud (SMC) extinction curve, which is suggested to be applicable in AGN \citep[e.g.,][see Section \ref{sec:slope} for the discussion of using other extinction curves]{2004AJ....128.1112H, 2007ASPC..373..586C, 2007ASPC..373..561L, 2015AJ....149..203K}. The derived broad-line Balmer decrements $L_{\rm H\alpha^b}/L_{\rm H\beta^b}$ are listed in Table \ref{tab:sdssfit}, and their distribution is plotted in Figure \ref{fig:dustdist}. The Balmer decrements are around $2.2$ -- $3.4$, indicating little or no dust extinction in most of the sample objects (\ha/\hb$\le3.06$), except for a few cases where some moderate extinction might be present. For the $6$ objects with \ha/\hb$>3.06$, the intrinsic continuum fluxes at the $6$ UVOT filters are derived by correcting for the excess reddening effect from the $E(B-V)$ derived above, using the IDL code \texttt{smc\_unred.pro}\footnote{\url{http://users.ynao.ac.cn/~xbdong/}}.

\begin{figure}
	\includegraphics[width=\columnwidth]{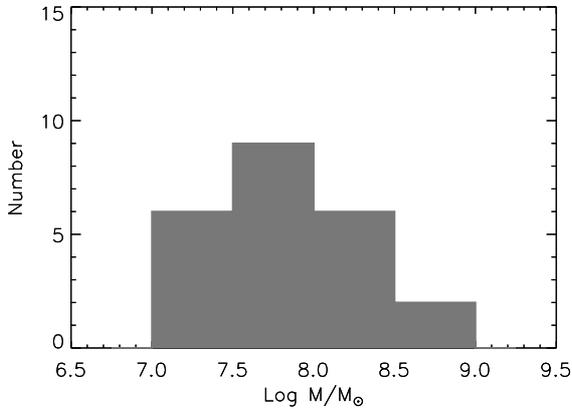}
\caption{The distribution of the black hole masses for the sample.}
\label{fig:massdist}
\end{figure}

The BH masses of AGN can be estimated using the spectral measurements based on the virial method \citep[e.g.,][]{2000ApJ...533..631K, 2009ApJ...707.1334W}. In this work, the BH masses are estimated using the FWHMs of the \hb \ broad component and monochromatic luminosity at $5100$ \angstrom\ \citep[Equation $5$ in][]{2006ApJ...641..689V}. It should be noted that the black hole mass estimated in this way is indirect and subject to systematics as large as $\sim0.3$ dex \citep[e.g.,][]{2000ApJ...543L...5G, 2006ApJ...641L..21G, 2013ApJ...773...90G} (the effect on our results is discussed in Section \ref{sec:pfree}). The values of the black hole masses are also listed in Table \ref{tab:sdssfit}, and the distribution is shown in Figure \ref{fig:massdist}. The black hole masses lie in the range of $10^7$ -- $10^9$\msun \, with a median around $10^8$\msun. 

\section{modeling the optical/UV continuum}
\label{sec:optuv}
\subsection{Optical/UV spectral slopes}

 Assuming the optical/UV continuum of an AGN (at a redshift $z$) is radiated from a geometrically thin, optically thick disk, the rest-frame luminosity at the frequency $\nu$ is calculated as
\begin{align}
\nu L_{\nu} &= \nu_0F_{\nu_0}\times\frac{2\pi d_{\rm L}^2}{\cos{i}}, \label{eq:fl}\\
\nu &= \nu_0(1+z),
\end{align}
where $F_{\nu_0}$ is the flux density (in units of $\rm erg \ cm^{-2} \ s^{-1} \ Hz^{-1}$) observed at a frequency $\nu_0$ in the observer's frame. $d_{\rm L}$ is the luminosity distance. $i$ is the viewing angle to the norm of the accretion disk ($i=0^\circ$ for face-on).
In this work we assume $i=30^\circ$, a typical value for type $1$ AGN as commonly adopted and also consistent with polarization measurement \citep{2014MNRAS.441..551M}. It should be noted that the optical/UV continuum spectral shapes derived below are independent of the assumption of the disk inclination. The effect of varying the inclination on our results is also discussed below.

\begin{figure}
\includegraphics[width=\columnwidth]{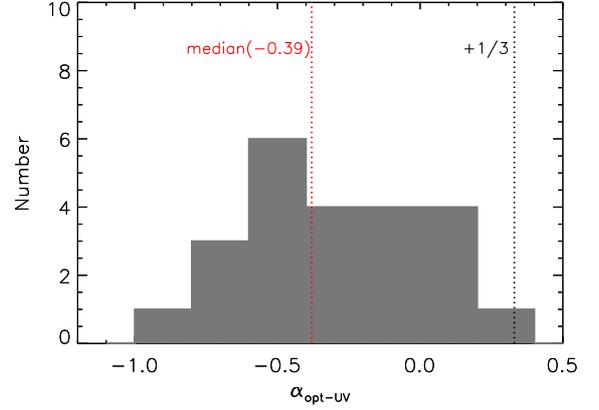}
\caption{The distribution of the optical/UV spectral index \aoptuv. The red dotted line represents $\alpha_{\rm opt-UV}=-0.39$ (the median). The black dotted line represents $\alpha_{\rm opt-UV}=+1/3$ (the SSD prediction).}
\label{fig:uv-opt-slope}
\end{figure}

First, a simple power-law model is used to fit the optical/UV spectra. The fitted slopes \aoptuv\ are listed in Table \ref{tab:sed_analysis}. Figure \ref{fig:uv-opt-slope} shows the distribution, which lie in the range of $-1.0$ -- $+0.3$ with a median of $-0.39$. It shows that most of the sources have a shallower \aoptuv\ than that of the SSD ($+1/3$), consistent with most of the previous results. Considering that the host galaxy contamination and possible dust reddening have been largely eliminated in this work, their effects on the observed spectral slopes are considered to be small, and hence cannot account for the reddening of the optical/UV spectra (see Section \ref{sec:slope} for discussion on the systematics). Here we consider that the derived continua are intrinsic emission from the AGN. As mentioned above, a shallower optical/UV spectral slope is at odds with that predicted from the SSD model.

\subsection{Accretion disk model with non-standard radial temperature profile}
\label{sec:pfree}

In this work, an alternative approach is adopted as an attempt to reconcile the observation with theoretical models of accretion disks. We consider that the observed optical/UV spectra originate intrinsically from a realistic accretion disk whose radial profile of the effective temperature deviates from the SSD solution. For this purpose, a generalized radial profile of the effective temperature is assumed, 
\begin{equation}
	T_{\rm eff}(R) = T_{\rm eff}(R_{\rm in})\times(\frac{R}{R_{\rm in}})^{-p}
	\label{eq:t_r}
\end{equation}
where $T_{\rm eff}(R_{\rm in})$ is the effective temperature at the inner disk radius $R_{\rm in}$ and the index $p$ a free parameter. Such a model is sometimes dubbed as p-free disk model in the literature \citep[e.g.,][]{2004ApJ...601..428K, 2005ApJ...631.1062K, 2011ApJ...727...31A}. This model was first introduced in the studies of black hole X-ray binaries (XRBs) to verify the disk solution in outbursts \citep{1994ApJ...426..308M, 1995ApJ...446..350H}. Other values of $p$ deviating from the SSD value $0.75$ have physical meanings. For instance, under the assumption of local thermal equilibrium (LTE), $T_{\rm eff}(R)\propto M_{\rm BH}^{1/4}\dot{M}^{1/4}R^{-3/4}$ \citep[e.g.,][]{2002apa..book.....F, 2008bhad.book.....K}, a non-standard radial temperature profile may result from a radius-dependent mass flow rate distribution ($\dot{M}(R)\propto R^{3-4p}$, where $p\ne0.75$), in contrast to the constant ($R$-independent) $\dot{M}$ in the standard disk formalism ($p=0.75$). In reality, this may well be the case for an realistic accretion disk with mass evaporation/outflow or condensation.

The total disk luminosity at a frequency $\nu$ can be calculated as the integral of the local blackbody radiation from the inner disk radius $R_{\rm in}$ to the outmost disk $R_{\rm out}$ 
\begin{align}
L_{\nu}=2\int_{R_{\rm in}}^{R_{\rm out}}\pi B_{\nu}(R)2\pi RdR \label{eq:lumi}
\end{align}
in which
\begin{align}
&B_{\nu}(R)=\frac{2 \rm h \nu^{3} }{{\rm c^2}{f_{\rm col}^4}} \frac{1}{e^{{\rm h} \nu/{{\rm k_{B}}T(R)}}-1}, \\
&T(R)=f_{\rm col}T_{\rm eff}(R) \label{eq:tc},
\end{align}
where $T(R)$ is the color temperature and $f_{\rm col}$ represents the electron scattering correction to the pure local blackbody radiation \citep[also refer to hardening factor, e.g.,][]{1995ApJ...445..780S, 2013MNRAS.431.3510S, 2018arXiv180905134D}. For X-ray binaries a typical value of $\sim1.7$ is often used \citep{1995ApJ...445..780S}. Little is known for the hardening factor for AGN, however. The characteristic disk temperature in AGN is about two orders of magnitude lower than that in XRBs and the color correction is suggested to be $1.0$ -- $1.6$ in some studies \citep[e.g.,][]{2011MNRAS.415.2942C, 2016ApJ...821..104Y}. Here we adopt $f_{\rm col}=1$, as in some similar previous studies \citep[e.g.,][]{2007MNRAS.381.1235V, 2009MNRAS.392.1124V}. For a disk extending all the way to the innermost stable circular orbit (ISCO), the inner disk radius $R_{\rm in}$ is determined by the black hole spin ($a_*$), from $\sim R_{\rm g}$ ($={\rm G}M_{\rm BH}/c^2$ is the gravitational radius) for a co-rotating disk around an extremely-spinning BH ($a_*=0.99$) to $3R_{\rm s}$ ($R_{\rm s}=2R_{\rm g}$, the Schwarzschild radius) for a non-spinning BH ($a_*=0$), and $9R_{\rm g}$ for a retrograde-rotating disk around an extremely spinning BH ($a_*=-1$). As a common practice here a non-spinning BH is assumed following \citet{2007MNRAS.381.1235V} and thus $R_{\rm in}=6R_{\rm g}$. $R_{\rm out}$ is the outer radius of the disk, which is determined by the self-gravity of the accretion disk \citep{1989MNRAS.238..897L} and set to a typical value $3000R_{\rm s}$\footnote{Generally, $R_{\rm out}\ga1000R_{\rm s}$ for a black hole with a mass of $10^8 M_{\sun}$. Considering that the temperature in lower-mass black holes ($M_{\rm BH}\approx10^7M_{\sun}$) is relative higher and the emission from outer radius can have a contribution to the emission in optical band, we take a commonly adopted value $3000R_{\rm s}$.}. The black hole masses derived from the optical spectral parameters (see Section \ref{sec:dustandmass}) are used to calculate $R_{\rm g}$. 

In this model, the inner disk temperature $T_{\rm eff}(R_{\rm in})$ and index $p$ in Equation \ref{eq:t_r} are free parameters. The value of $p$ allowed in the fitting is set to a reasonable range of [$0.45, 0.85$], considering the limits in the classical accretion theory ($p=0.5$ in `slim' disk and $0.75$ in SSD) and possible uncertainty in the spectral fitting procedure caused by other parameters such as black hole mass (see below). The IDL routine \texttt{mpfit} \citep{2009ASPC..411..251M}, which employs the Levenberg-Marquardt least-square method, is used to fit the optical/UV spectra obtained above with the p-free disk model. For most of the objects, the model can fit the optical/UV data reasonably well. The best-fit values of $p$ and $T_{\rm eff}(R_{\rm in})$ are listed in Table \ref{tab:sed_analysis}. The measured AGN luminosities and the best-fit models are shown in Figure \ref{fig:seds}. 

The above fitted $p$ and $T_{\rm eff}(R_{\rm in})$ values are obtained by assuming a specific set of parameters of the BH and the accretion disk, which might deviate in reality. Moreover, the black hole masses estimated in Section \ref{sec:dustandmass} from the single-epoch optical spectra are subject to systematic uncertainties as large as $\sim0.3$ dex. To investigate the systematics of the fitted $p$ and $T_{\rm eff}(R_{\rm in})$ values caused by these uncertainties, the above fitting procedure is repeated for $5000$ times using various combinations of the values of these parameters which are drawn randomly from reasonable ranges in the parameter space of \mbh, $a_*$ and $i$. Specifically, \mbh\ is drawn from a Gaussian distribution of $\log(M_{\rm BH})$ with a standard deviation $\sigma=0.3$, and $i$ and $a_*$ from a uniform distribution in the [$0^\circ$ -- $60^\circ$] and [$-0.99$ -- $+0.99$] ranges, respectively. The $1\sigma$ error of $p$ and $T_{\rm eff}(R_{\rm in})$ is derived from the simulated distribution and given in Table \ref{tab:sed_analysis}. 

\begin{figure}
\includegraphics[width=\columnwidth]{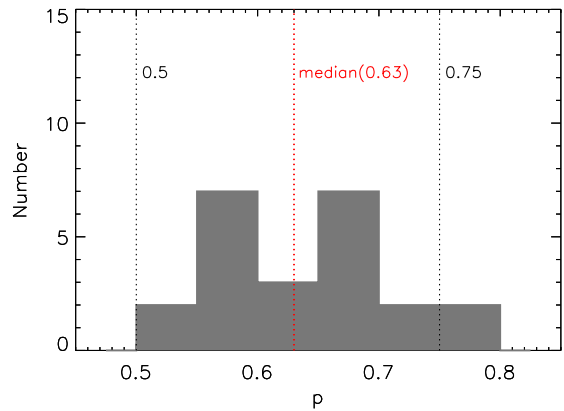}
\caption{The distribution of the radial temperature power-law index $p$, obtained from the optical/UV spectral fitting using the p-free disk model. The two black dotted lines represent $p=0.5$ (`slim' disk solution) and $p=0.75$ (SSD solution). The red dotted line represents $p=0.66$ (the median).}
\label{fig:p_hist}
\end{figure}

\begin{figure}
\includegraphics[width=\columnwidth]{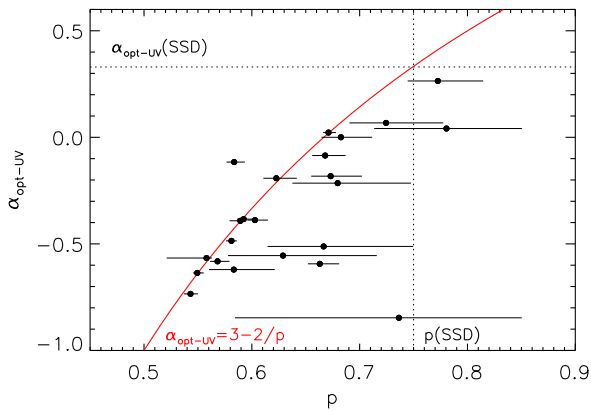}
\caption{Power-law index of the radial temperature profile $p$ versus optical/UV spectra slope \aoptuv. The vertical dotted line represents $p=0.75$ (SSD solution). The horizontal dotted line represents $\alpha_{\rm opt-UV}=+1/3$ (SSD prediction). The red solid line represents the theoretical relation of $p$ and $\alpha_{\rm opt-UV}$ assuming UVOT data sample the power-law regime of the disk spectrum. The error of $p$ are given in $1\sigma$.}
\label{fig:p-slope}
\end{figure}

Figure \ref{fig:p_hist} shows the distribution of the best-fit indices of $p$. The majority lie between $0.5$ and $0.75$ with a median of $0.63$, suggesting a flatter radial temperature profile than that of the SSD model ($0.75$) in most of the sample objects. Theoretically, the slope of the power-law regime of an emergent disk spectrum in optical/UV band can be determined by the radial temperature profile of the disk, and thus a relation between the two can be predicted, as $\alpha_{\rm opt-UV}=3-2/p$ \citep[e.g.,][]{2008RMxAC..32....1G}. The relationship between the obtained $p$ values and \aoptuv \ is shown in Figure \ref{fig:p-slope}, which is found to be consistent with the theoretical relation for most of the objects. In these sources, the UVOT data mainly sampled the power-law portion of the emergent disk spectra. The value of $p$ is mainly determined by \aoptuv\ and has a weak dependence on the black hole mass (thus has a small uncertainty). The Spearman's rank test confirms a strong correlation between these two parameters ($\rho_s=0.63$, $P=1.320\times 10^{-3}$). Figure \ref{fig:p-slope} also reveals a few outliers with rather redder spectra. By examining the spectral fits it is clear that for these sources the UVOT data sampled mostly the Wien rather than the power-law regime of the spectrum (this also leads to a strong dependence of $p$ on the black hole mass and a large uncertainty therein). Among these, KUG 1141+371 is an extreme case with the shallowest spectral slope ($\alpha_{\rm opt-UV}=-0.84$), while the index is fitted to be $p\approx0.74$, indicating a radial temperature profile very close to that of the SSD. The result shows that in some AGN the shallower optical/UV slopes may simply be caused by the fact that the peak or the high-energy turnover of the disk spectrum is observed, which is expected in AGN with high BH mass \mbh\ and relatively low Eddington ratio \edd\ (see Figure \ref{fig:seds} for the SED fitting for KUG 1141+371).

\section{broadband SED and bolometric correction factors}
\label{sec:broadbandsed}

It has been shown above that the optical/UV spectra in most of our AGN can well be reproduced by emission from accretion disks with a flattened radial temperature profile. As a result, such a disk produces relatively lower emission in the EUV band than that previously predicted for a SSD model with the $F_{\nu}\propto \nu^{+1/3}$ spectrum. This energy band is thought to be where the disk emission peaks, and thus contains a substantial fraction of the bolometric luminosity. In the rest of the paper, we investigate the broadband SED for our sample AGN assuming that a non-SSD radial temperature profile may operate in accretion disks.

Aside from the optical/UV band, another dominant emission bandpass is X-ray. In this section we first model the X-ray emission by making use of the simultaneous {\sw}/XRT observations. Then the broadband SEDs are constructed and the bolometric luminosities are derived. Finally the bolometric correction factors are studied for the X-ray and optical bands.

\begin{table*}
\centering
\caption{Spectral analysis of XRT spectra.}
\renewcommand\arraystretch{1.2}
\label{tab:xrtanalysis}
\begin{threeparttable}
\begin{tabular}{ccccccccccc} 
\hline
\hline
No. & Object & Model & $N_{\rm H}^{\rm Gal}$ & $\Gamma_{\rm X}$ & $T_{\rm bb}$
 & $\log F_{\rm 2-10}$ & $\chi^2/\nu$ & Count Rates \\
(1)&(2)&(3)&(4)&(5)&(6)&(7)&(8)&(9)\\
\hline
1 & Mrk 1018 & (b) & $2.43$ & $1.84_{-0.13}^{+0.13}$ & $0.12_{-0.04}^{+0.03}$ & $-10.91_{-0.05}^{+0.05}$ & $37/53$ & $0.302\pm0.008$ \\
2 & MCG 04-22-042 & (b) & $3.12$ & $1.78_{-0.10}^{+0.10}$ & $0.12_{-0.04}^{+0.03}$ & $-10.90_{-0.04}^{+0.03}$ & $78/83$ & $0.282\pm0.006$ \\
3 & Mrk 705 & (b) & $3.22$ & $1.72_{-0.27}^{+0.26}$ & $0.15_{-0.03}^{+0.02}$ &  $-11.11_{-0.09}^{+0.08}$ & $35/33$ & $0.473\pm0.020$ \\
4 & RX J1007.1+2203 & (b) & $2.60$ & $1.69_{-0.40}^{+0.39}$ & $0.12_{-0.02}^{+0.01}$ & $-12.24_{-0.14}^{+0.11}$ & $15/13$ & $0.037\pm0.002$ \\
5 & CBS 126 & (b) & $0.97$ & $1.71_{-0.21}^{+0.21}$ & $0.09_{-0.01}^{+0.01}$ &  $-11.65_{-0.09}^{+0.08}$ & $18/19$ & $0.136\pm0.005$ \\
6 & Mrk 141 & (b) & $1.18$ & $1.69_{-0.14}^{+0.13}$ & $0.08_{-0.03}^{+0.04}$ & $-11.59_{-0.06}^{+0.06}$ & $15/28$ & $0.105\pm0.004$ \\
7 & Mrk 142 & (b) & $1.18$ & $1.74_{-0.62}^{+0.62}$ & $0.13_{-0.02}^{+0.01}$ & $-11.38_{-0.23}^{+0.14}$ & $15/17$ & $0.201\pm0.008$ \\
8 & Ton 1388 & (a) & $1.38$ &  $2.24_{-0.07}^{+0.07}$ & - & $-11.50_{-0.05}^{+0.04}$ & $45/42$ & $0.255\pm0.007$  \\
9 & SBS 1136+594 & (b) & $0.94$ & $1.64_{-0.08}^{+0.08}$ & $0.12_{-0.01}^{+0.01}$ & $-11.10_{-0.03}^{+0.03}$ & $108/104$ & $0.370\pm0.006$  \\
10 & PG 1138+222 & (a) & $2.02$ & $1.92_{-0.07}^{+0.07}$ & - & $-11.10_{-0.04}^{+0.04}$ & $43/51$ & $0.364\pm0.009$ \\
11 & KUG 1141+371 & (a) & $1.65$ & $1.48_{-0.15}^{+0.15}$ & -  & $-11.39_{-0.09}^{+0.08}$ & $13/12$ & $0.132\pm0.007$ \\
12 & Mrk 1310 & (b) & $2.50$ & $1.33_{-0.33}^{+0.30}$ & $0.18_{-0.03}^{+0.03}$ & $-11.31_{-0.09}^{+0.08}$ & $10/17$ & $0.135\pm0.006$  \\
13 & RX J1209.8+3217 & (c) & $1.24$ & $2.57_{-0.06}^{+0.15}$ & - & $-12.81_{-0.11}^{+0.06}$ & $9/7$ & $0.017\pm0.001$ \\
14 & Mrk 50 & (a) & $1.58$ & $1.79_{-0.05}^{+0.05}$ & - & $-11.06_{-0.03}^{+0.03}$ & $73/82$ & $0.384\pm0.008$ \\
15 & Mrk 771 & (b) & $2.75$ & $2.04_{-0.13}^{+0.13}$ & $0.09_{-0.03}^{+0.02}$  & $-11.44_{-0.05}^{+0.05}$ & $64/53$ & $0.245\pm0.006$ \\
16 & PG 1307+085 & (b) & $2.21$ & $1.71_{-0.16}^{+0.16}$ & $0.12_{-0.03}^{+0.02}$ & $-11.45_{-0.05}^{+0.05}$ & $53/49$ & $0.146\pm0.006$ \\
17 & Ton 730 & (a) & $1.11$ & ${2.17}_{-0.08}^{+0.08}$ & - &$-11.80_{-0.05}^{+0.05}$ & $68/48$ & $0.119\pm0.003$ \\
18 & RX J1355.2+5612 & (b) & $1.02$ & $1.85_{-0.58}^{+0.54}$ & ${0.10}_{-0.01}^{+0.01}$ & $-11.99_{-0.21}^{+0.18}$ & $10/9$ & $0.078\pm0.004$  \\
19 & Mrk 1392 & (b) & $3.80$ & $1.82_{-0.11}^{+0.11}$ & $0.10_{-0.02}^{+0.02}$ & $-11.26_{-0.05}^{+0.04}$ & $54/52$ & $0.253\pm0.006$ \\
20 & Mrk 290 & (b) & $1.76$ & $1.40_{-0.11}^{+0.11}$ & $0.09_{-0.02}^{+0.02}$ & $-11.13_{-0.05}^{+0.04}$ & $58/43$ & $0.237\pm0.007$ \\
21 & Mrk 493 & (b) & $2.11$ & $2.12_{-0.18}^{+0.17}$ & $0.14_{-0.03}^{+0.03}$ & $-11.62_{-0.06}^{+0.06}$ & $29/45$ & $0.197\pm0.005$ \\
22 & KUG 1618+410 & (a) & $0.84$ & $1.64_{-0.46}^{+0.51}$& - & $-11.86_{-0.26}^{+0.21}$ & $1.5/1$ & $0.032\pm0.004$ \\
23 & RX J1702.5+3247 & (b) & $1.98$ & $2.50_{-0.18}^{+0.19}$ & $0.10_{-0.02}^{+0.01}$ & $-11.90_{-0.07}^{+0.06}$ & $40/46$ & $0.170\pm0.004$ \\
\hline
\end{tabular}
{$\bm Notes.$} Column (1): identification number assigned in this paper. Column (2): NED name of the sample objects. Column (3): the \textsc{xspec} models used: (a) \texttt{wabs$\times$powerlaw} (b) \texttt{wabs$\times$(bbody+powerlaw)} 
(c) \texttt{wabs$\times$absori$\times$powerlaw}. Column (4): the neutral hydrogen column density $N_{\rm H}^{\rm Gal}$ used in \texttt{wabs} in units of $10^{20}$ cm$^{-2}$. Column (5): the hard X-ray photon index $\Gamma_{\rm X}$ in \texttt{powerlaw}. Column (6): the blackbody temperature in units of keV in \texttt{bbody}. Column (7): logarithmic of the observed $2-10$ keV flux in units of erg cm$^{-2}$ s$^{-1}$. Column (8): the goodness of fit for the model. Column (9): the XRT count rates in the observed $0.3-10$ keV band in units of counts s$^{-1}$. The errors are given in 90\% confidence interval.
\end{threeparttable}
\end{table*}

\subsection{The 0.3 -- 10 keV X-ray band}
\label{sec:xrayfit}

We make use of the \textsc{xspec} software \citep[][version 12.9.1]{1996ASPC..101...17A} to model the XRT spectra of our sample objects in the observed $0.3$ -- $10$ keV energy band. An absorbed power-law model (\texttt{wabs$\times$powerlaw}) with a free absorption column density ($N_{\rm H}$) is fitted to the spectra at first, and the fitted \nh\ do not show evidence for excess absorption above the Galactic value. We thus fix the absorption column densities at the Galactic value for better constraining of the spectral parameters. The spectra of $6$ AGN can be well fitted with a simple absorbed power-law model. In other $16$ sources, a possible soft excess component below $2$ keV is indicated in the fitting residuals when the model is fitted to the $2$ -- $10$ keV data and is extrapolated below $2$ keV. We thus add a blackbody component in the model (\texttt{wabs$\times$(bbody+powerlaw)}), whose addition is justified by using the $F$-test and an acceptance $P$-value of $P<0.05$ is adopted. For these objects, an absorbed power-law plus a soft X-ray excess model is found to better describe the X-ray spectra. The temperature of the soft excess is found to be $T_{\rm bb}\approx0.1$ -- $0.2$ keV, consistent with the typcial values in AGN. In the remaining source RX J1209.8+3217 a distinct feature is noticed, with a possible absorption around $1$ -- $2$ keV. A power-law model plus an ionized absorption component (\texttt{wabs$\times$absori$\times$powerlaw}) with their photon indices tied with each other gives a satisfactory fit. The results of the X-ray spectral analysis are summarized in Table \ref{tab:xrtanalysis} (the errors are given in $90\%$ confidence interval). We note that the reflection component which is often observed in Seyfert galaxies should have a negligible effect to the determination of X-ray power-law continuum in this work, as the practical energy range of the XRT spectra for our sample objects are all below $8$ keV. 

\subsection{Broadband Spectral Energy Distributions}

Based on the spectral fitting results in both the optical/UV and X-ray bands, we construct the broadband SED by integrating the spectral models from the optical to X-ray band. For the optical to EUV band, the emergent spectrum from an accretion disk model with the best-fit $p$ and $T_{\rm eff}(R_{\rm in})$ parameters is calculated from Equation \ref{eq:t_r} -- \ref{eq:tc}. For the X-ray band, the power-law continuum, which is thought to originate from a hot corona, is extrapolated to both the lower and higher energy bands. At the lower-energy end (EUV to soft X-rays), an operational cutoff energy is applied, as often adopted in the literature \citep[e.g.,][]{2010ApJS..187...64G, 2013MNRAS.433..648F}, since the corona emission does not contribute significantly to the UV band. The e-folding energy is determined for each object individually based on the consideration of Compton scattering of the disk emission\footnote{The power-law model drops toward low energies as an exponential decay $\exp^{-E_{\rm cut, low}/{E}}$ in EUV band. The $E_{\rm cut, low}$ is determined for each individual source as (1) $4E_{\rm cut, low}\ge E_{\rm scat}$, where $E_{\rm scat}=2.4E_{\rm disk, peak}$ represents the minimum scattered energy after inverse Compton scattering; (2) the peak of the continuum is determined by the fitted accretion disk model above.}. It should be noted that the exact value of this cutoff energy within a reasonable range has almost no effect on the calculation of bolometric luminosity. At the higher energy end, observations have revealed an exponential cutoff in the power-law spectra of AGN in the hard X-ray band, with a cutoff energy $E_{\rm cut}=150$ -- $250$ keV for Seyfert galaxies \citep[e.g.,][]{2008A&A...485..417D, 2014ApJ...782L..25M}. We note that the choice of the exact $E_{\rm cut}$ value within this range has only little effect on the bolometric luminosities, causing a scatter of only $0.05$ dex. We thus assume a uniform high energy exponential cutoff $E_{\rm cut}=200$ keV for all the objects. 

\begin{figure*}
	\includegraphics[width=5.8cm]{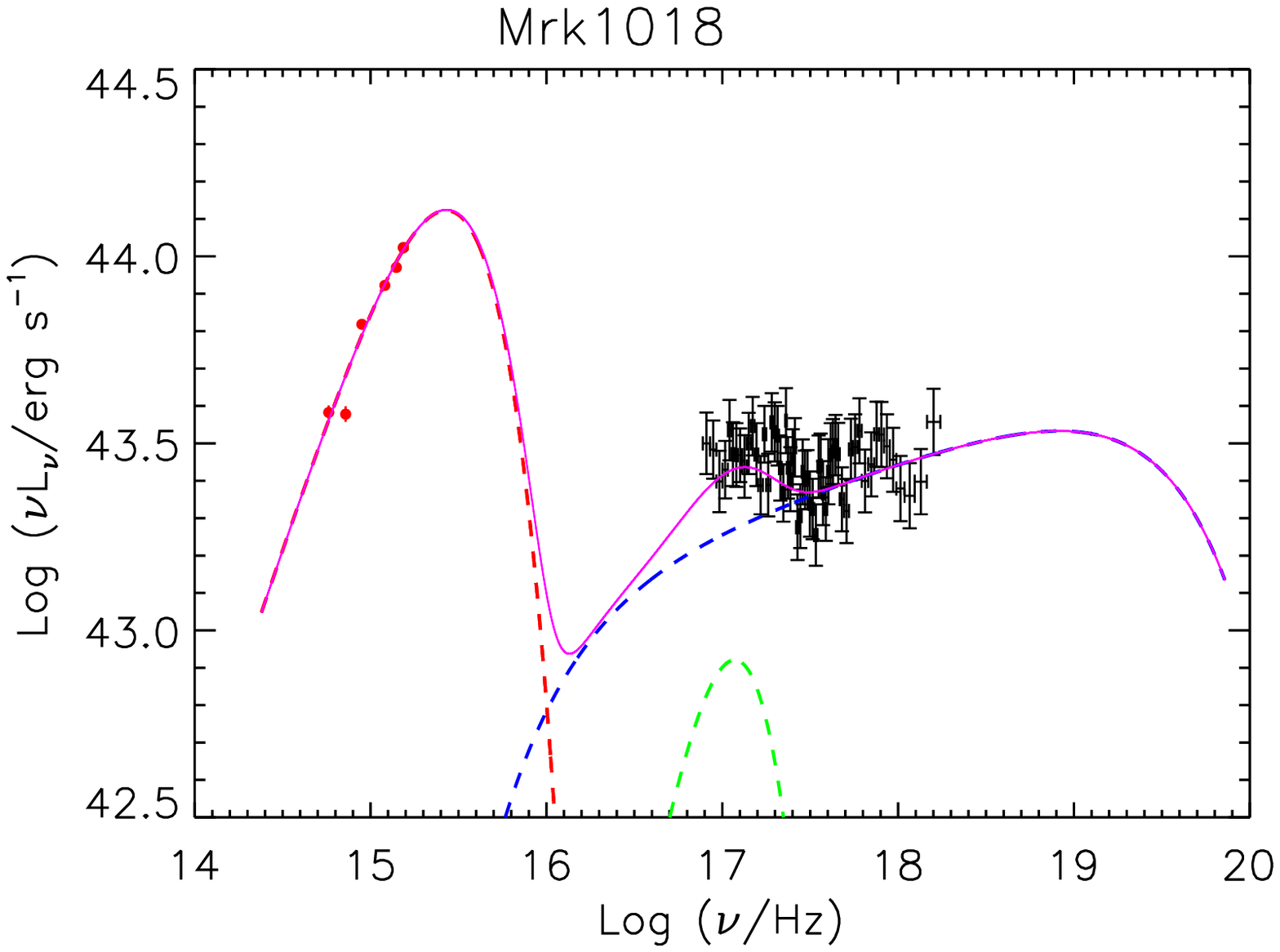}
	\includegraphics[width=5.8cm]{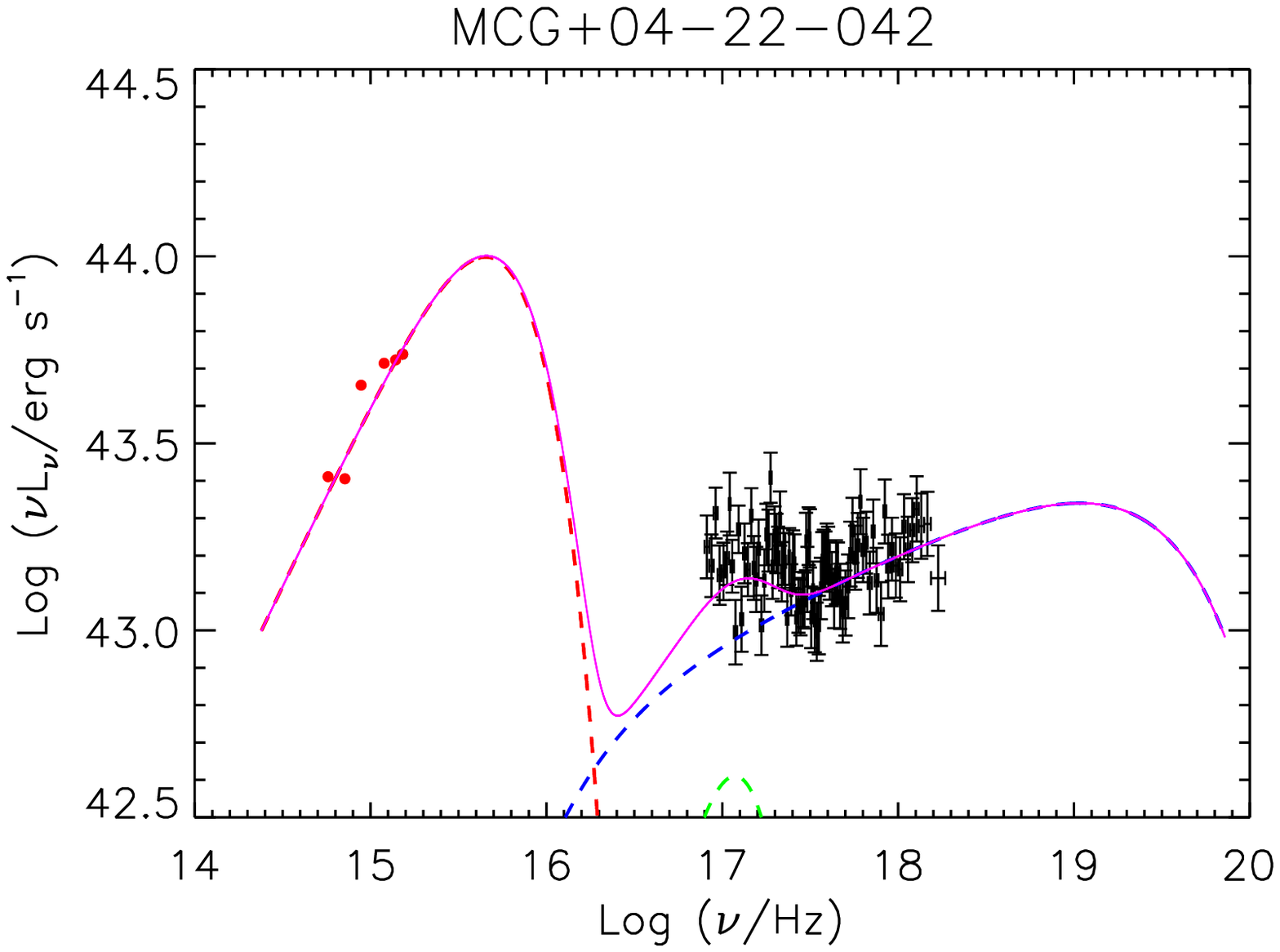}
	\includegraphics[width=5.8cm]{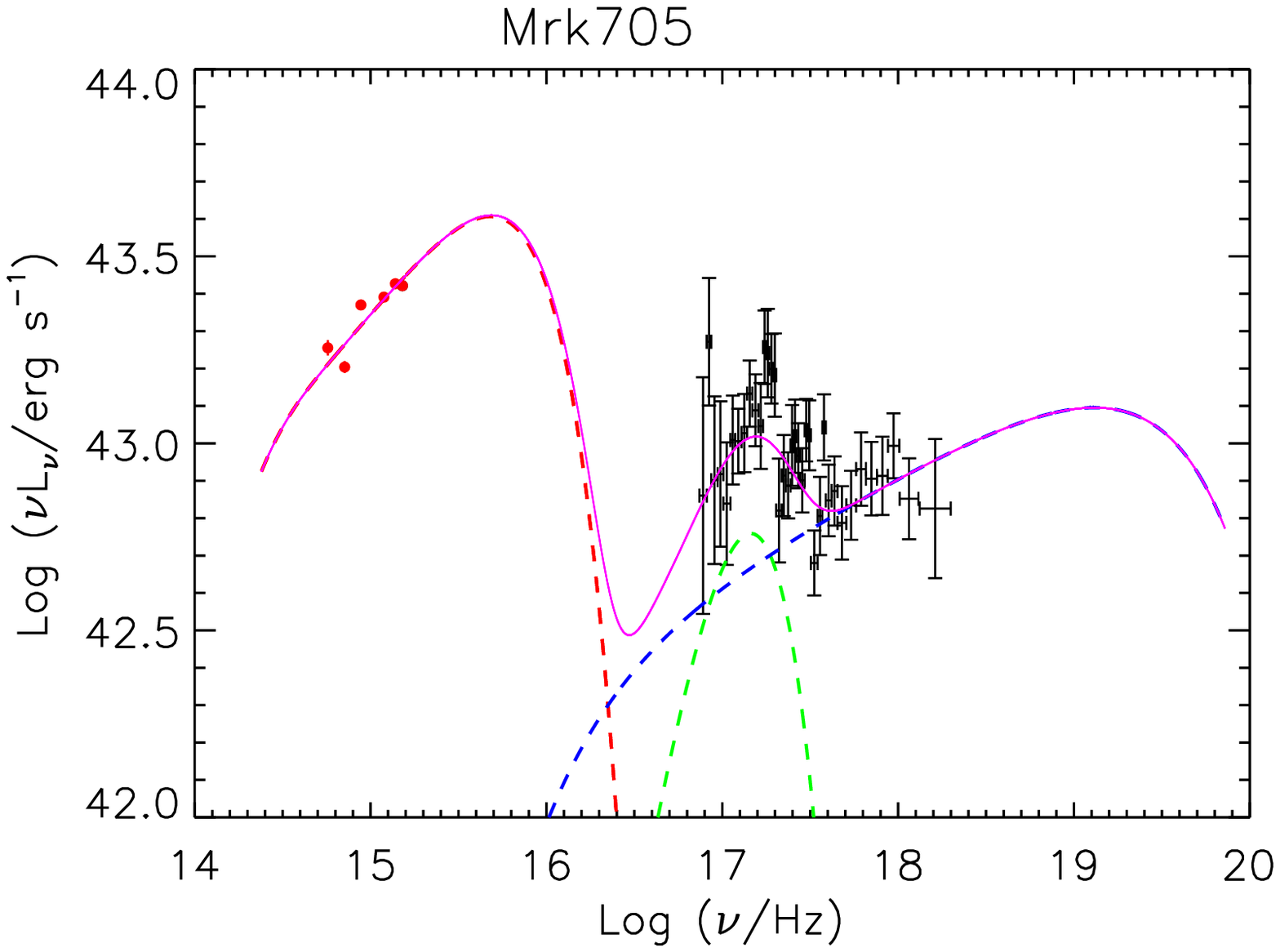}
	\includegraphics[width=5.8cm]{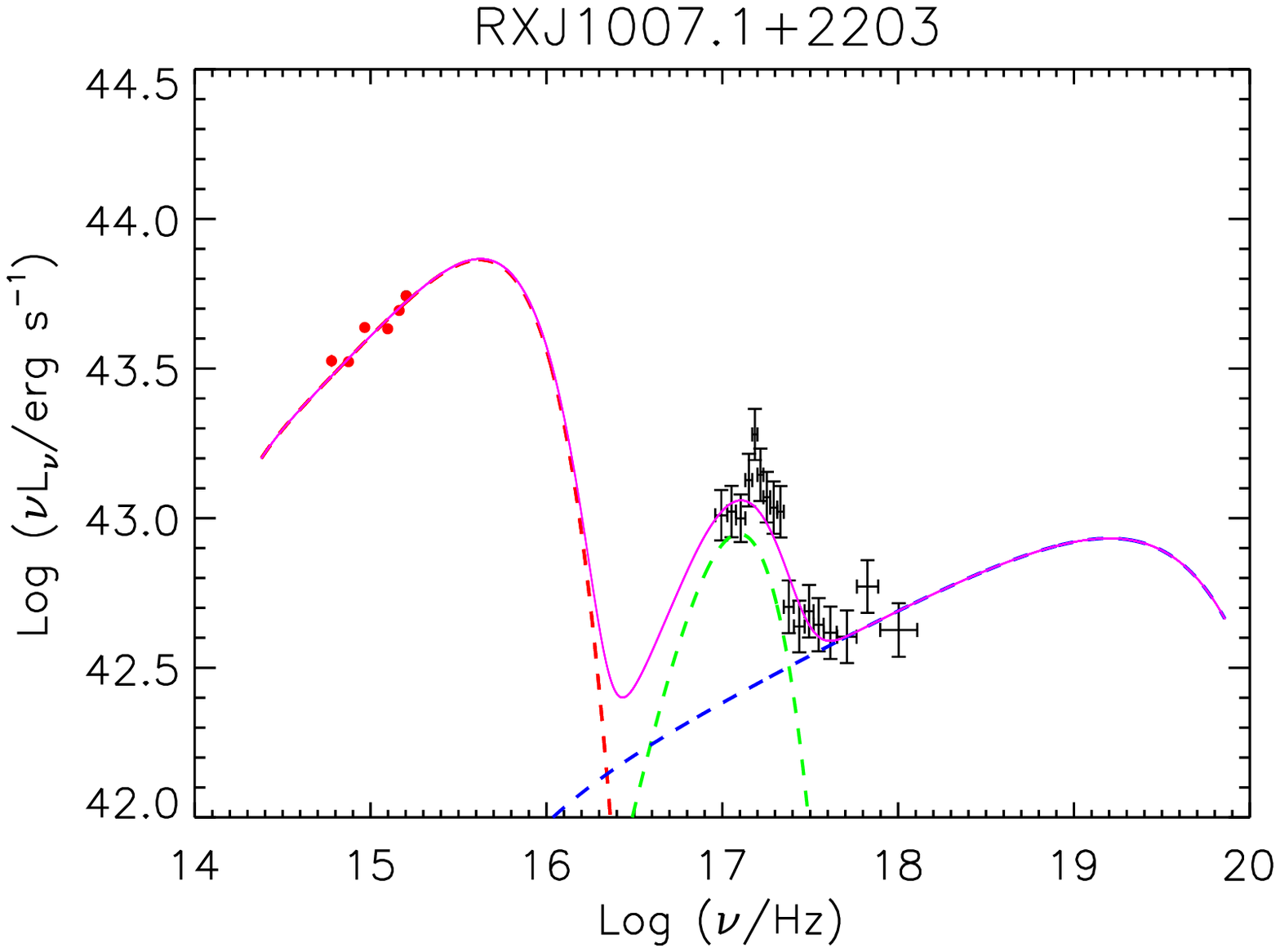}
	\includegraphics[width=5.8cm]{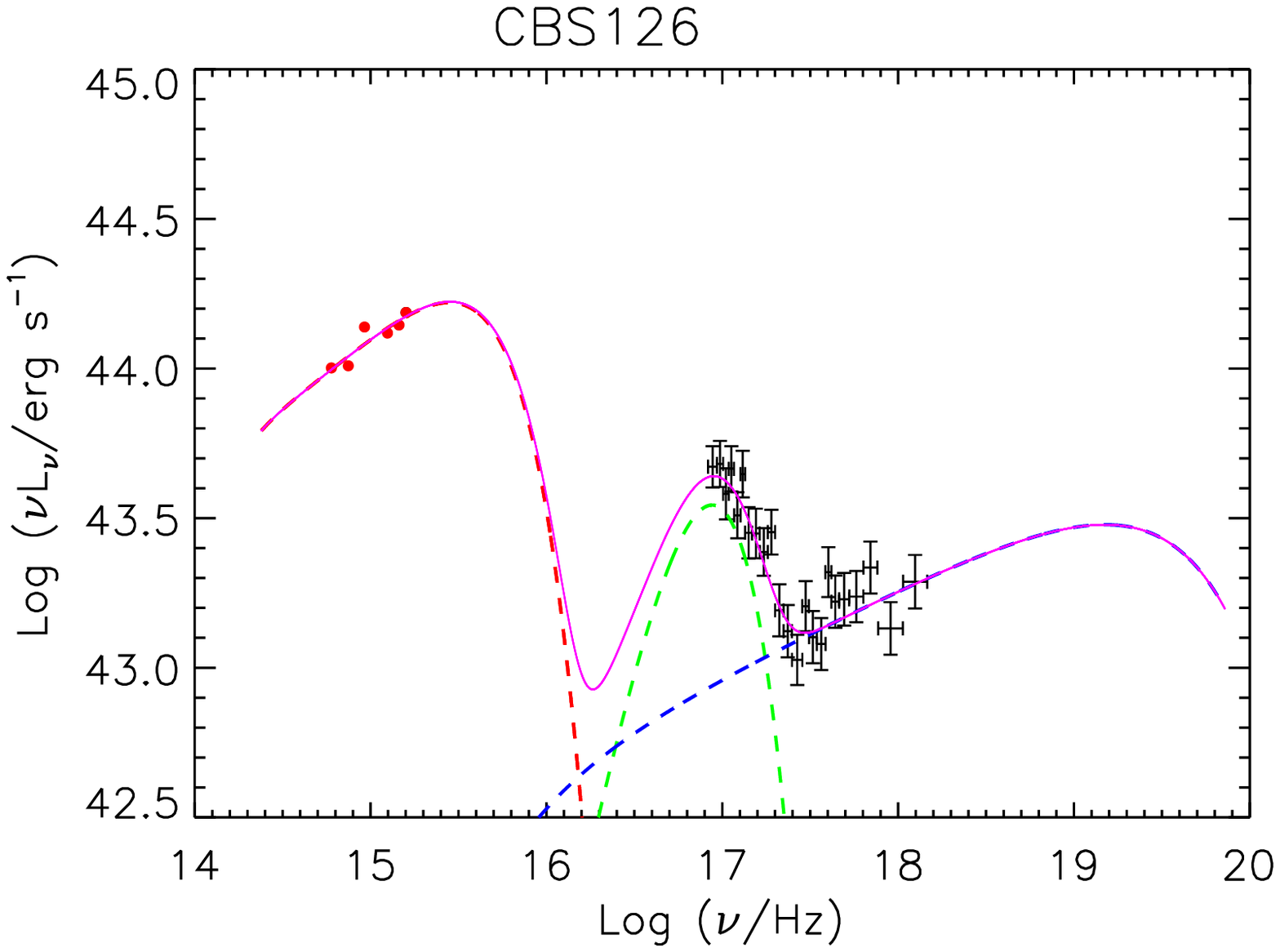}
	\includegraphics[width=5.8cm]{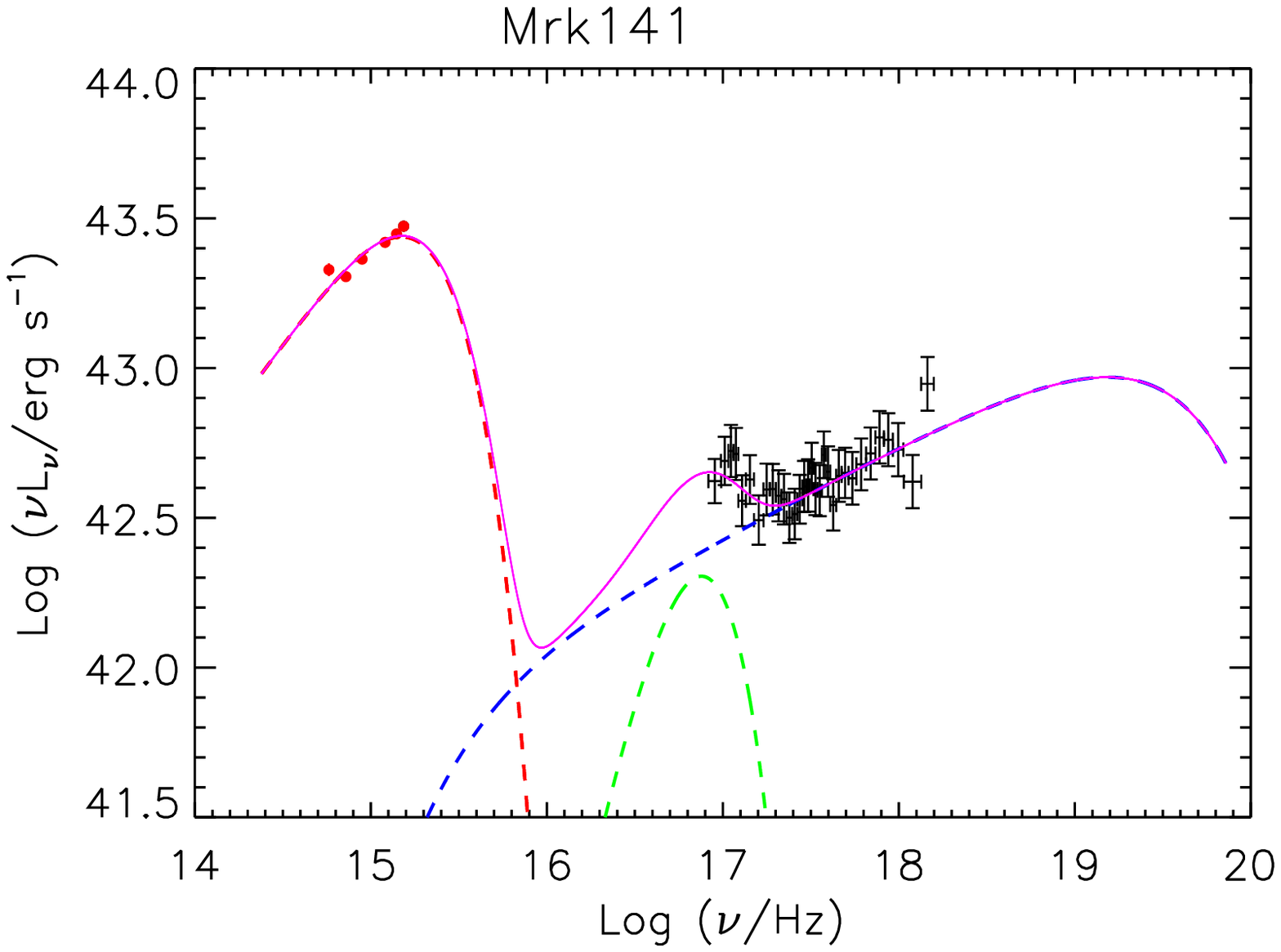}
	\includegraphics[width=5.8cm]{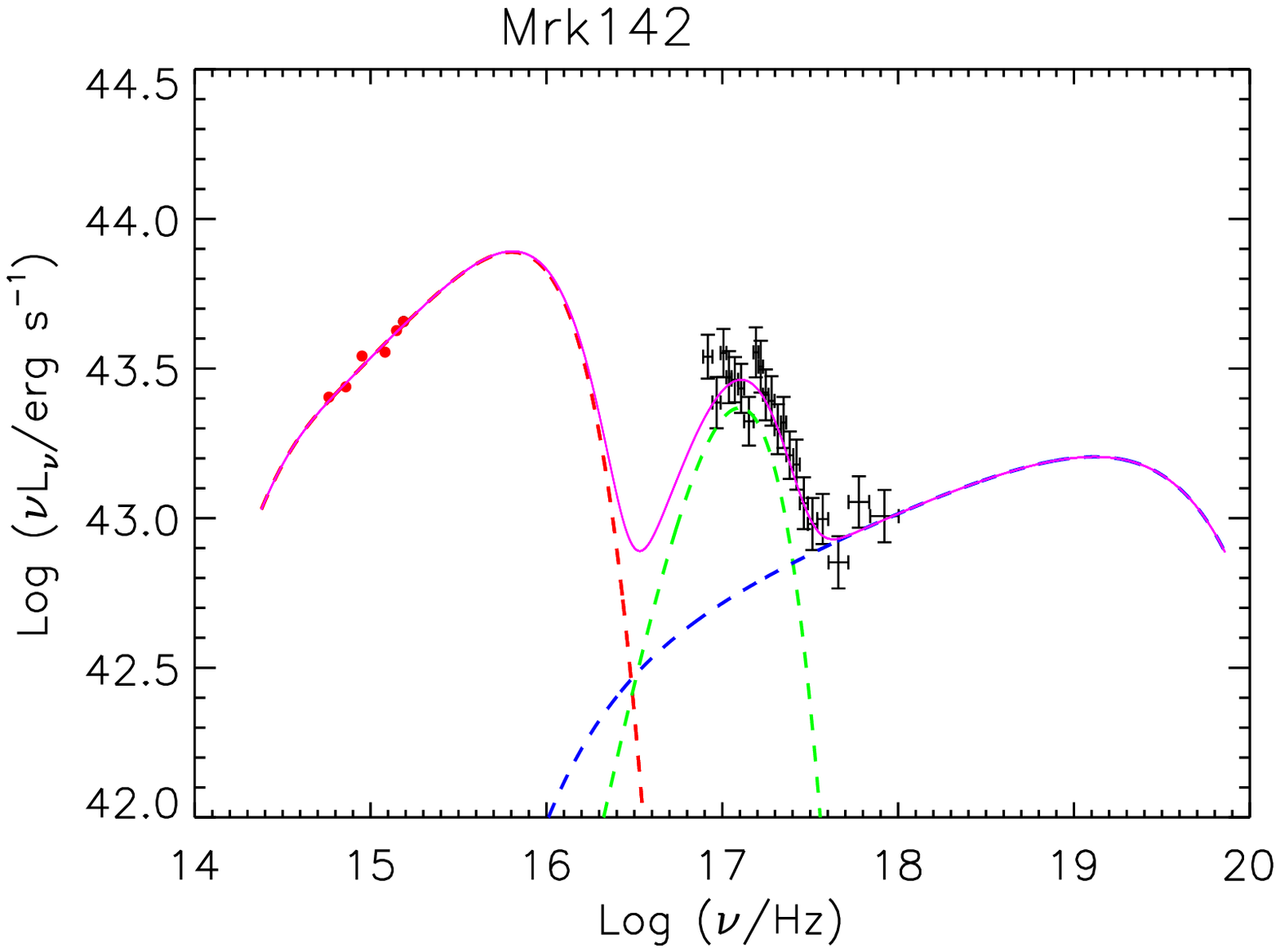}
	\includegraphics[width=5.8cm]{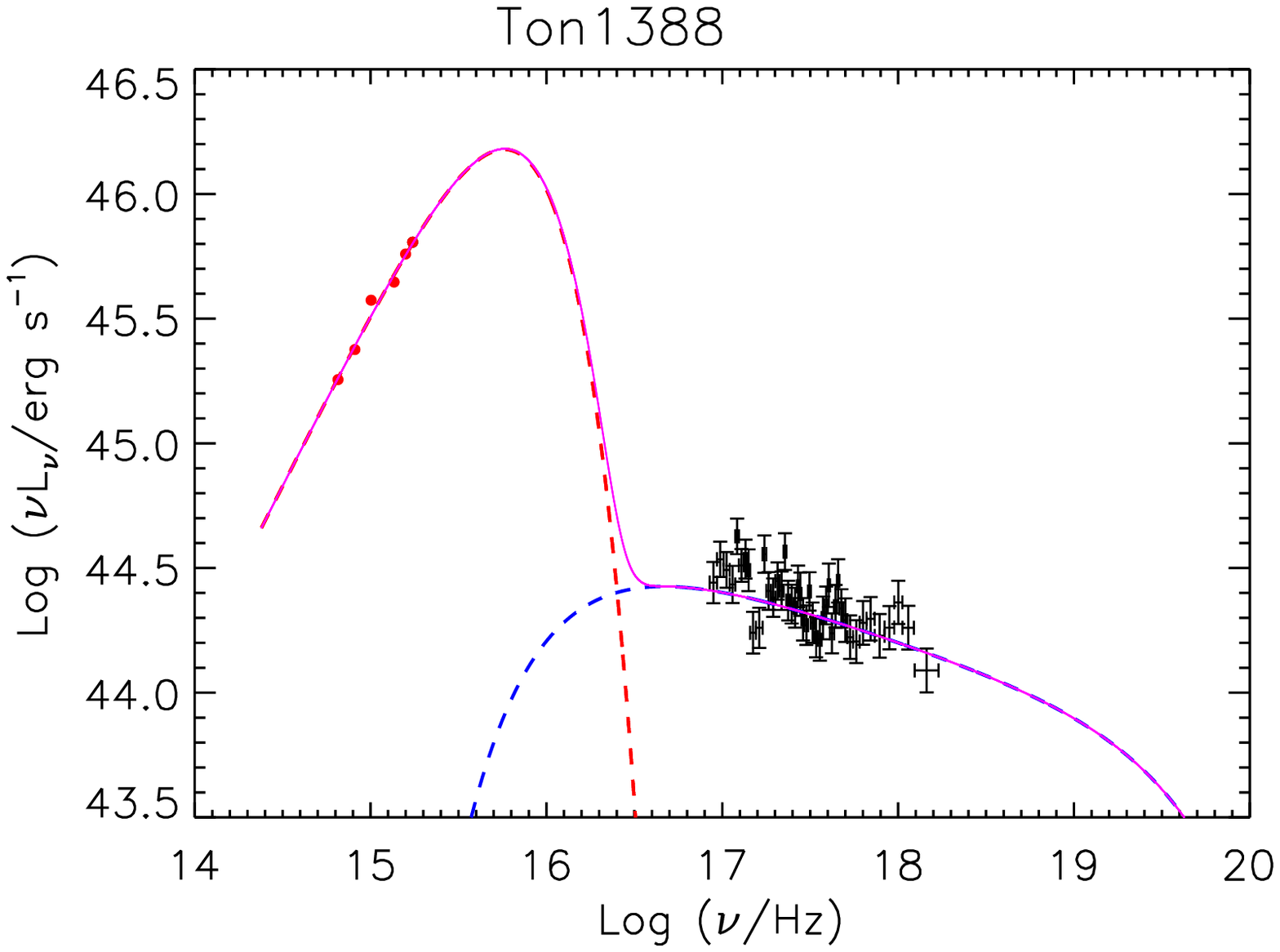}
	\includegraphics[width=5.8cm]{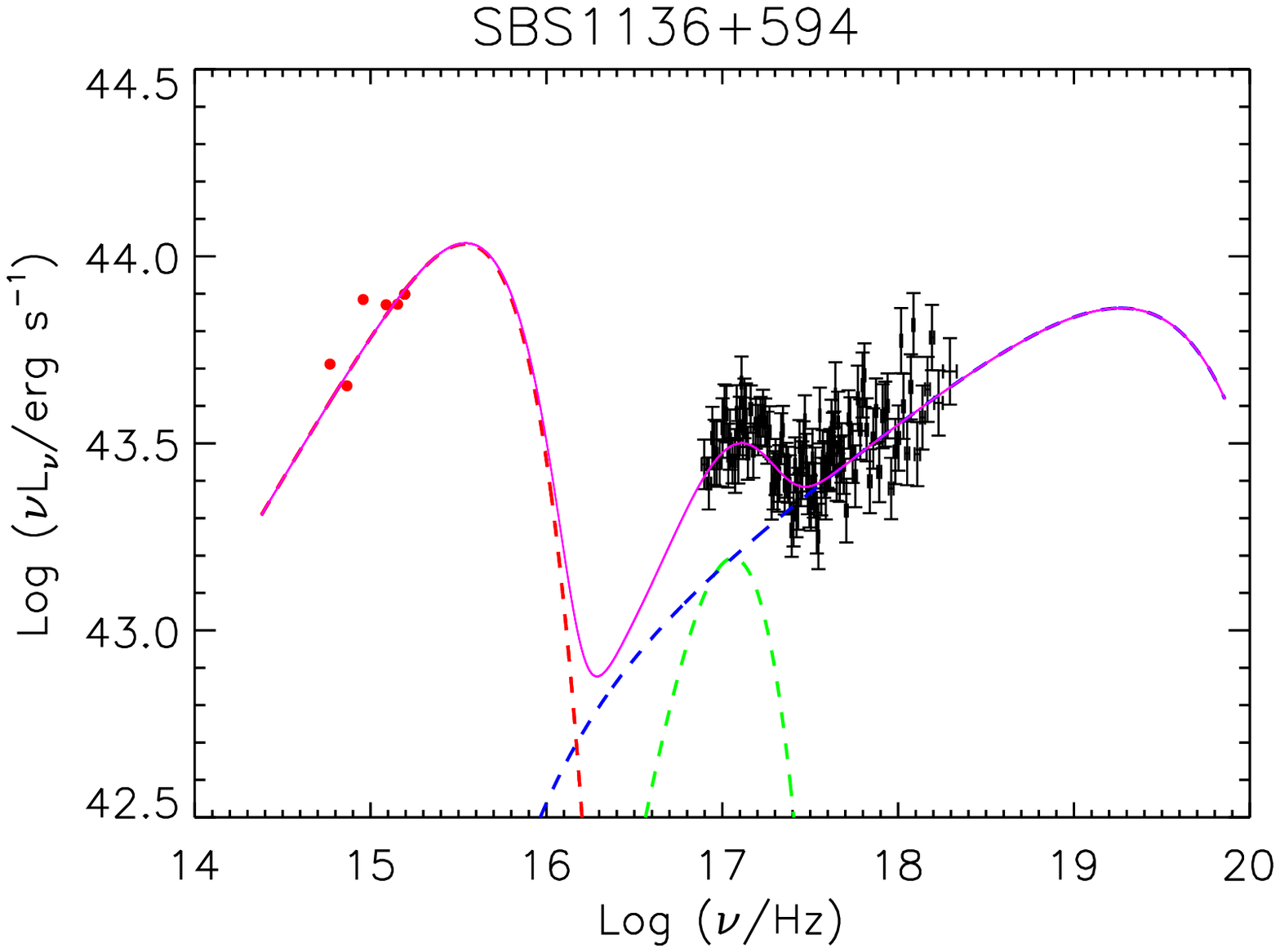}
	\includegraphics[width=5.8cm]{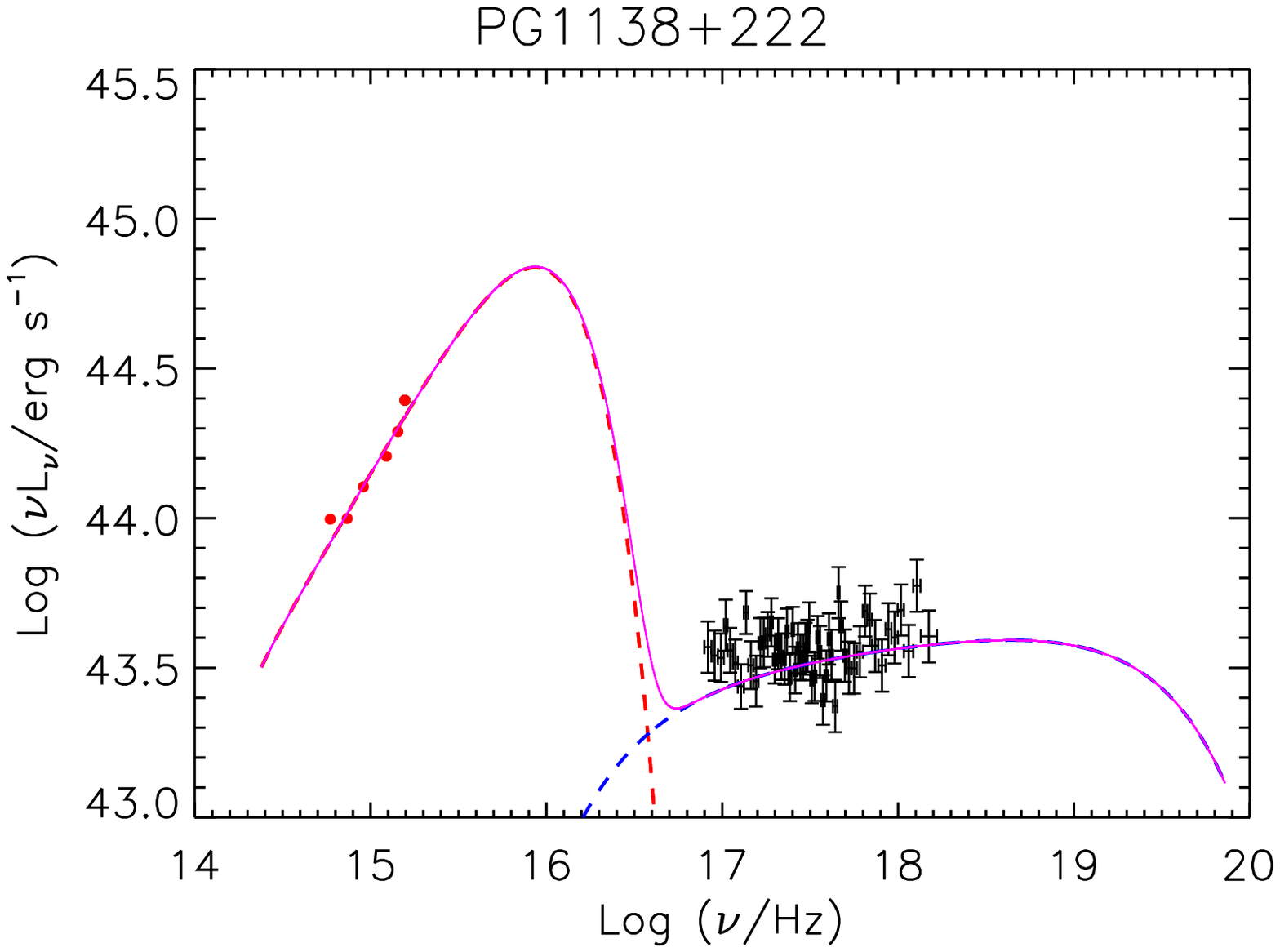}	
	\includegraphics[width=5.8cm]{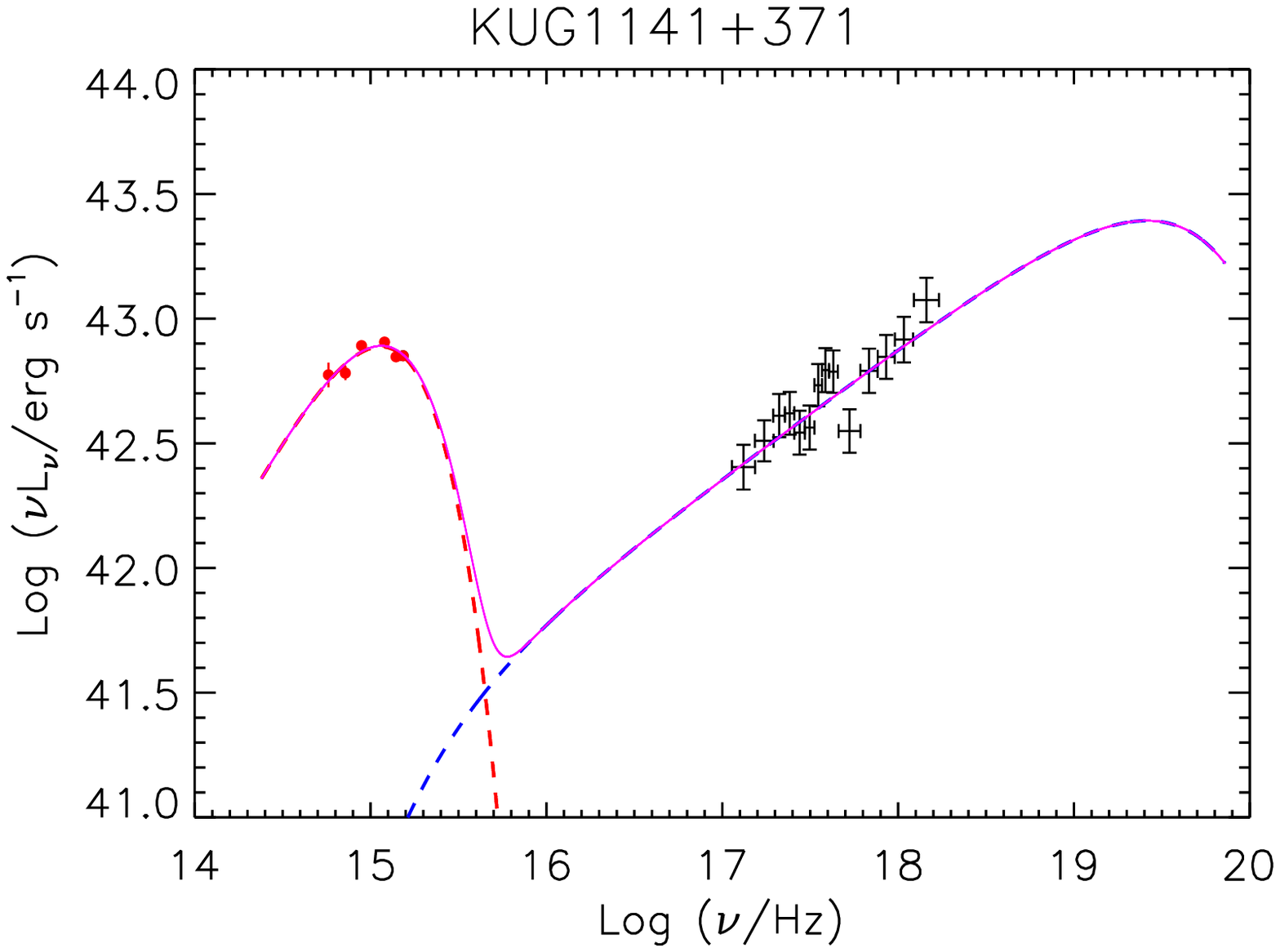}	
	\includegraphics[width=5.8cm]{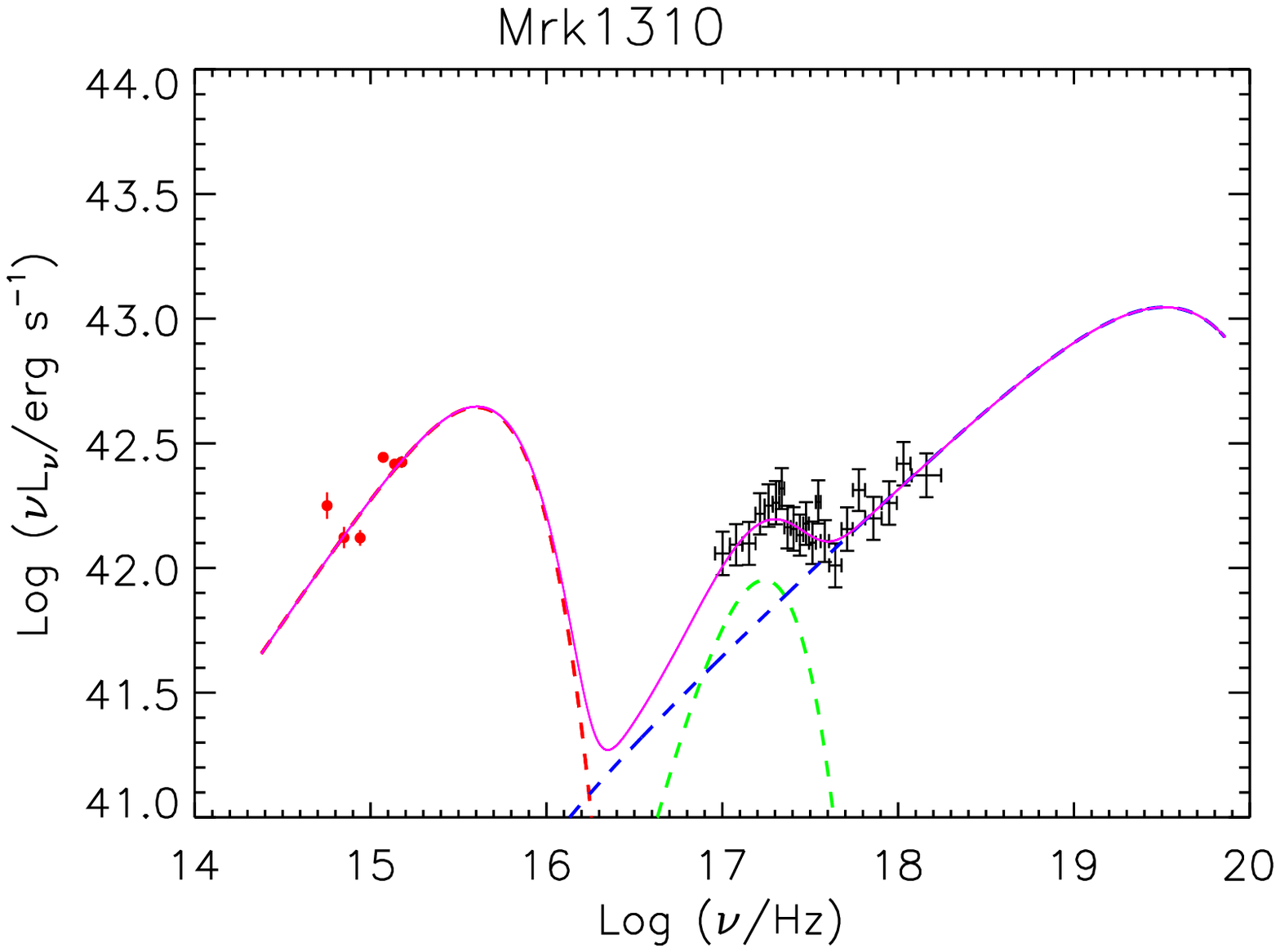}
	\includegraphics[width=5.8cm]{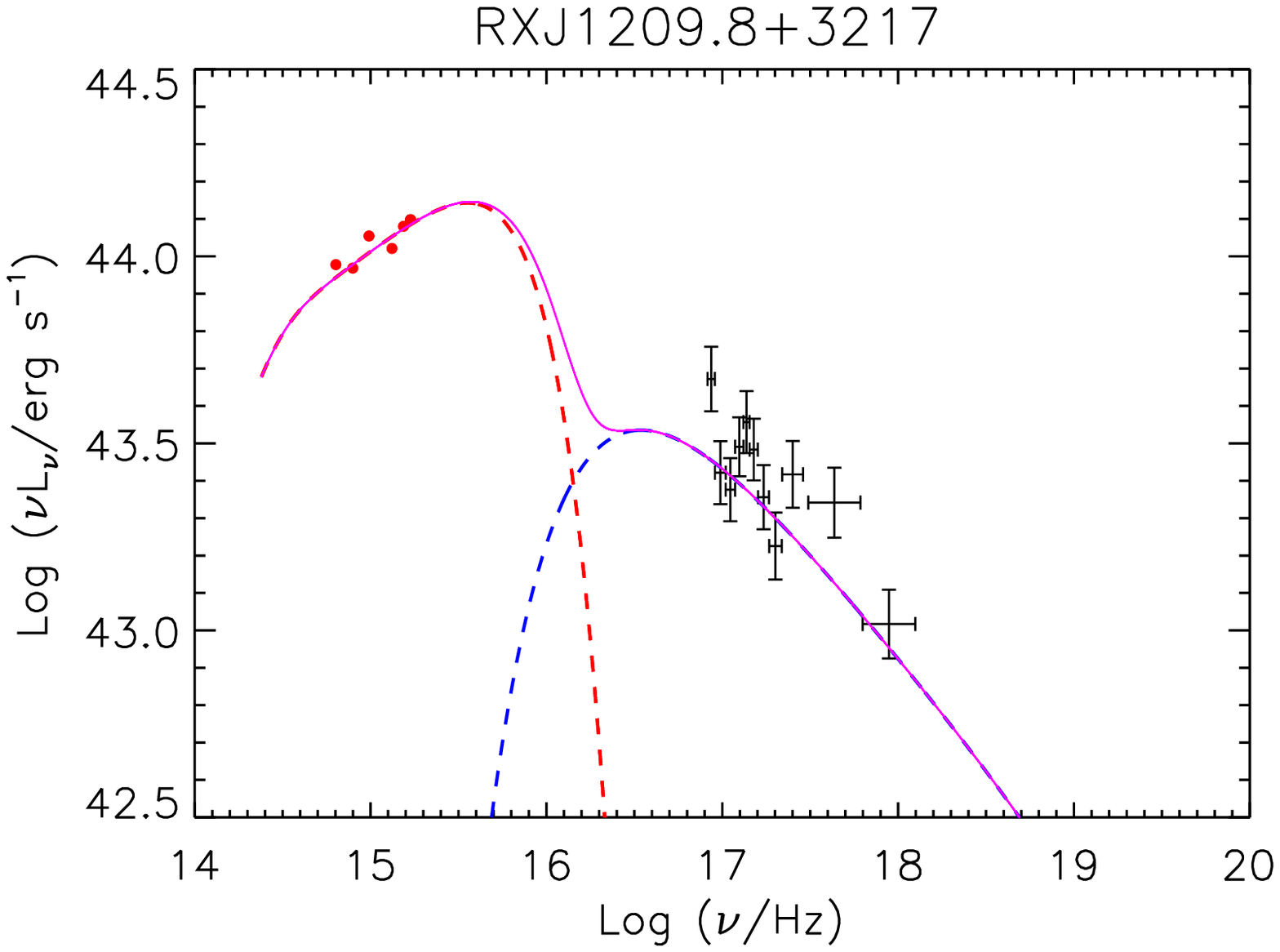}
	\includegraphics[width=5.8cm]{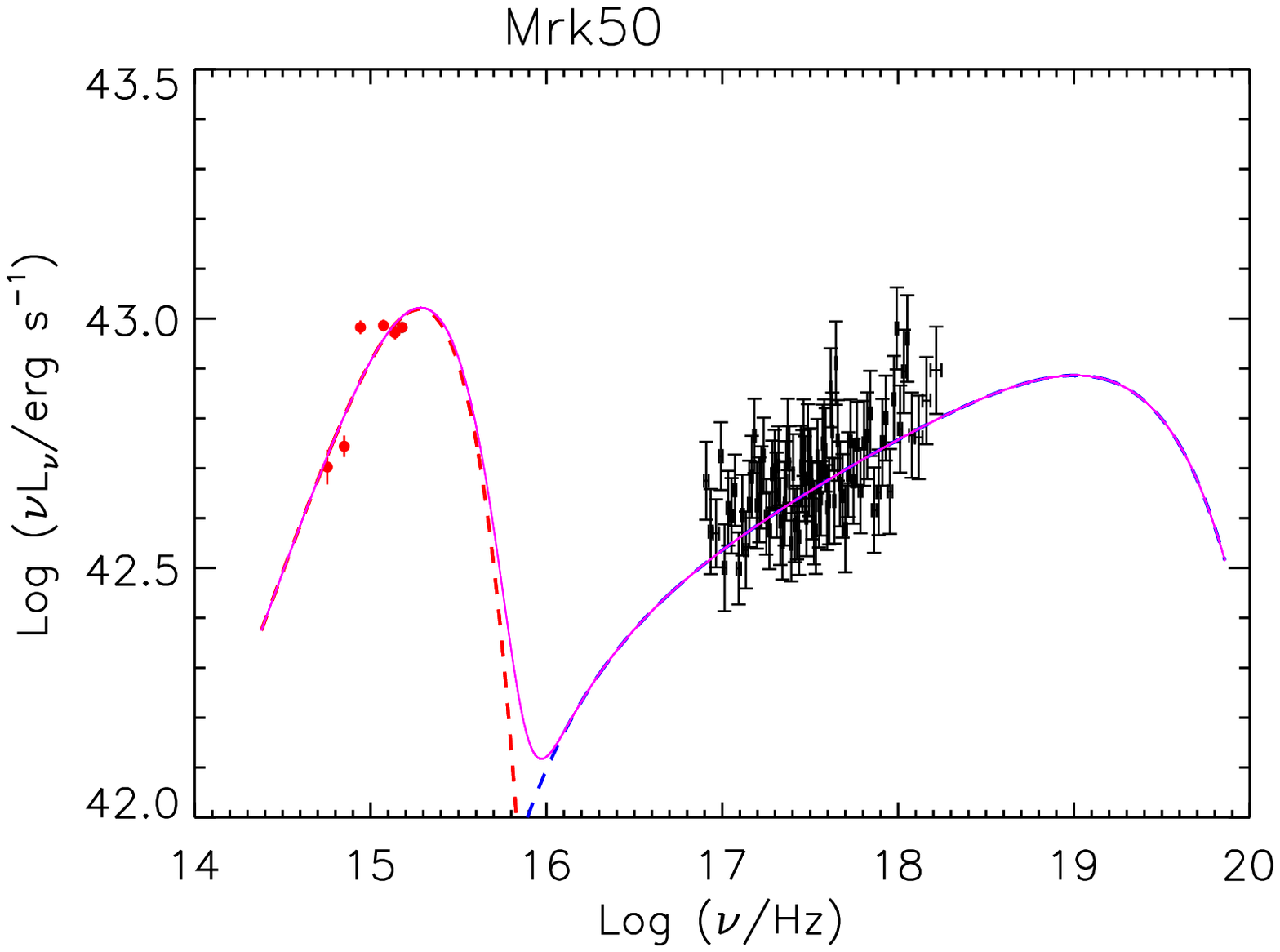}
	\includegraphics[width=5.8cm]{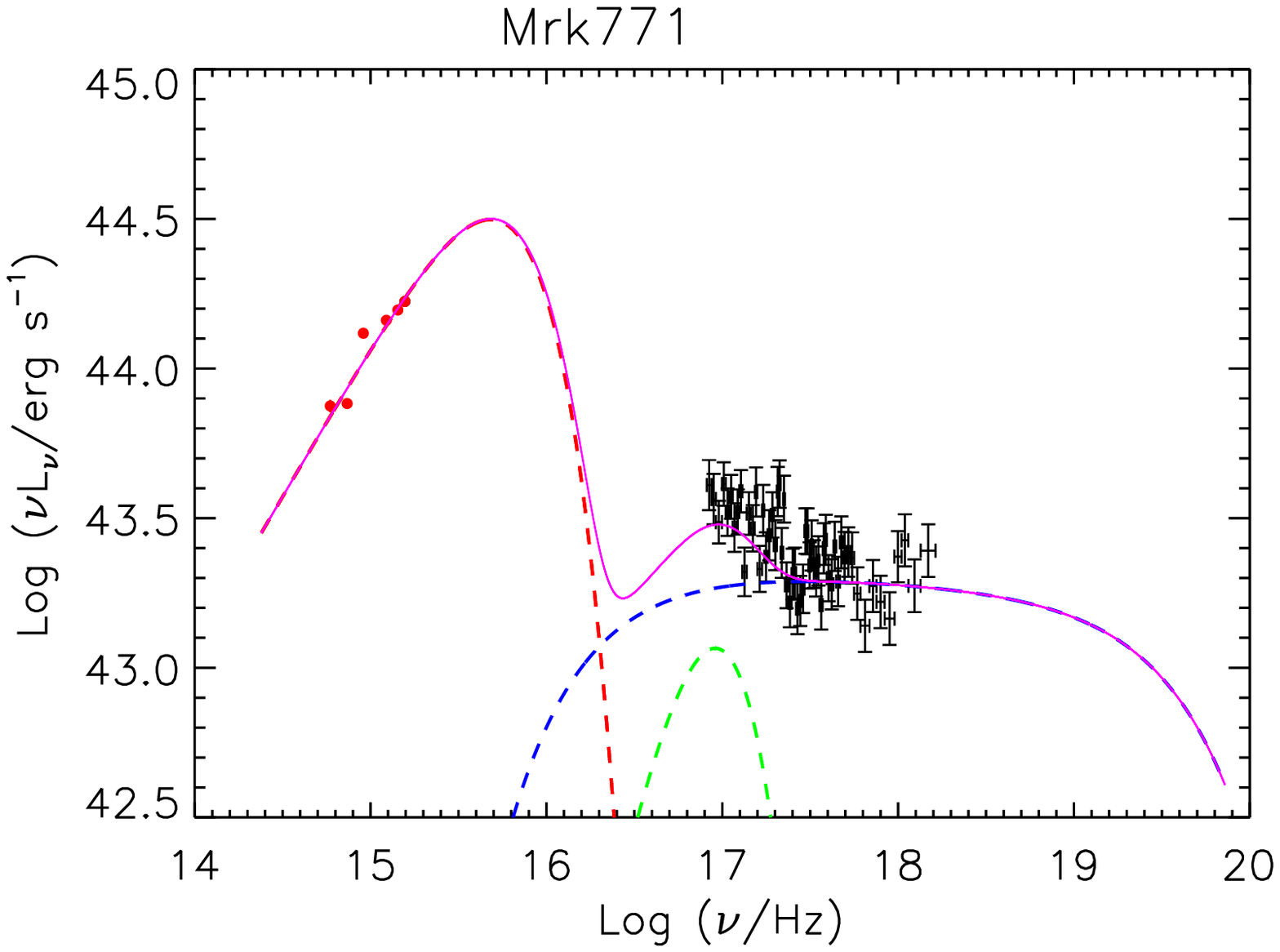}	
	\caption{The spectral energy distributions for our sample, including the measured luminosities and spectral models in optical/UV and X-ray bands. Red dots represent the measured luminosities in the 6 UVOT filters. Black crosses represent the X-ray luminosities obtained by XRT instrument. The spectral model components are denoted by dashed lines in different colors. The red dashed line represents the p-free disk model. The green dashed line represents the blackbody model (if required). The blue dashed line represents the power-law model, which is extrapolated by an exponential cutoff at both ends. The sum of all models is represented by a pink solid line. The data and models are in the source rest-frame.}
	\label{fig:seds}	
\end{figure*}

We overplot the derived broadband SED models for our sample in Figure \ref{fig:seds}, along with the measured luminosities and the best-fit spectral models in the optical/UV and X-ray bands. Some of the parameters of the SED are summarized in Table \ref{tab:sed_analysis} (the errors are given in $1\sigma$). The effective optical-to-X-ray spectral index \aox\ is calculated by measuring the luminosity densities at $2500$ \angstrom \ and $2$ keV \citep[$\alpha_{\rm OX}=-0.384\log L({\rm 2keV})/L(2500 \angstrom)$,][]{1979ApJ...234L...9T}. \lopt \ is the monochromatic luminosity at $5100$ \angstrom. \lx \ is the rest-frame $2$ -- $10$ keV bolometric luminosity. The bolometric luminosity \lbol\ is defined as the integral of the SED within the energy range of $0.001$ -- $300$ keV. The Eddington ratios \edd, defined as \lbol/\ledd\, are derived. The errors of $p$, $T_{\rm eff}(R_{\rm in}$), \lopt\ and \aox\footnote{We note that the uncertainty of \aox\ is dominated by that of $L_{2500}$ as it has a larger uncertainty than $L(\rm 2keV)$.}, are derived from the Monte Carlo simulations in the optical/UV band (see Section \ref{sec:pfree}). The error of \lx \ is derived by using \textsc{xspec} command \texttt{cflux} . The error of \lbol\ is derived from a simulated distribution (A group of $5000$ simulated X-ray luminosities is obtained by drawing them randomly from a Gaussian distribution, with a standard deviation $\sigma$ derived by using \texttt{cflux}. The luminosities are added to $5000$ Monte Carlo-simulated disk luminosities to generate a group of $5000$ simulated bolometric luminosities \lbol). The errors of other parameters which are dependent on \lbol, such as \edd, \ko\ and \kh\ (see Section \ref{sec:bc}), are also derived.

\begin{figure}
\includegraphics[width=\columnwidth]{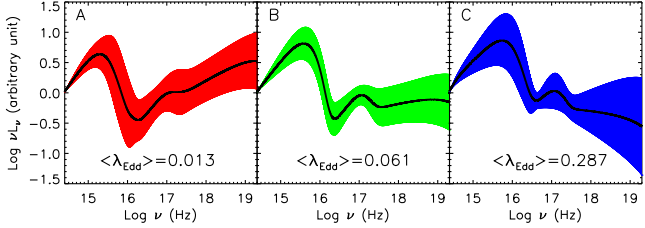}
\vspace{-1.5cm}
\caption{The averaged SEDs in different Eddington ratio bins, i.e., \edd$\la0.03$ (low-, panel A), $0.03\la$\edd$\la0.09$ (intermediate-, panel B) and \edd$\ga0.09$ (high-, panel C). The black solid line represents the averaged SED, and the colored regions denote the $68\%$ scatter. The average values of \edd\ in each bin are also shown in the panel. }
\label{fig:allseds-new}
\end{figure}

The averaged SED is often used to study the dependence of the SED shape on some key parameters, as it can minimize the intrinsic dispersions among individual objects. The sample objects are divided into $3$ groups according to their Eddington ratio \edd, i.e., low-, $\lambda_{\rm Edd}\la0.03$ ($8$ objects), intermediate-, $0.03\la\lambda_{\rm Edd}\la0.09$ ($8$ objects) and high-, $\lambda_{\rm Edd}\ga0.09$ ($7$ objects). The SEDs are all normalized at $1$ eV and the average of their logarithms are calculated in each \edd\ bin. The $3$ averaged SEDs are plotted in Figure \ref{fig:allseds-new}, together with the $68\%$ scatter. It shows clearly that, as \edd\ increases, the optical/UV (disk) emission gradually dominates the overall energy budget, with the peak luminosity shifts to higher frequencies in EUV band; meanwhile, the X-ray continuum (corona emission) becomes softer in higher-\edd\ objects. These results are generally consistent with previous findings \citep[e.g.,][]{2006ApJ...646L..29S, 2009MNRAS.392.1124V}.

\begin{figure*}
	\includegraphics[width=5.8cm]{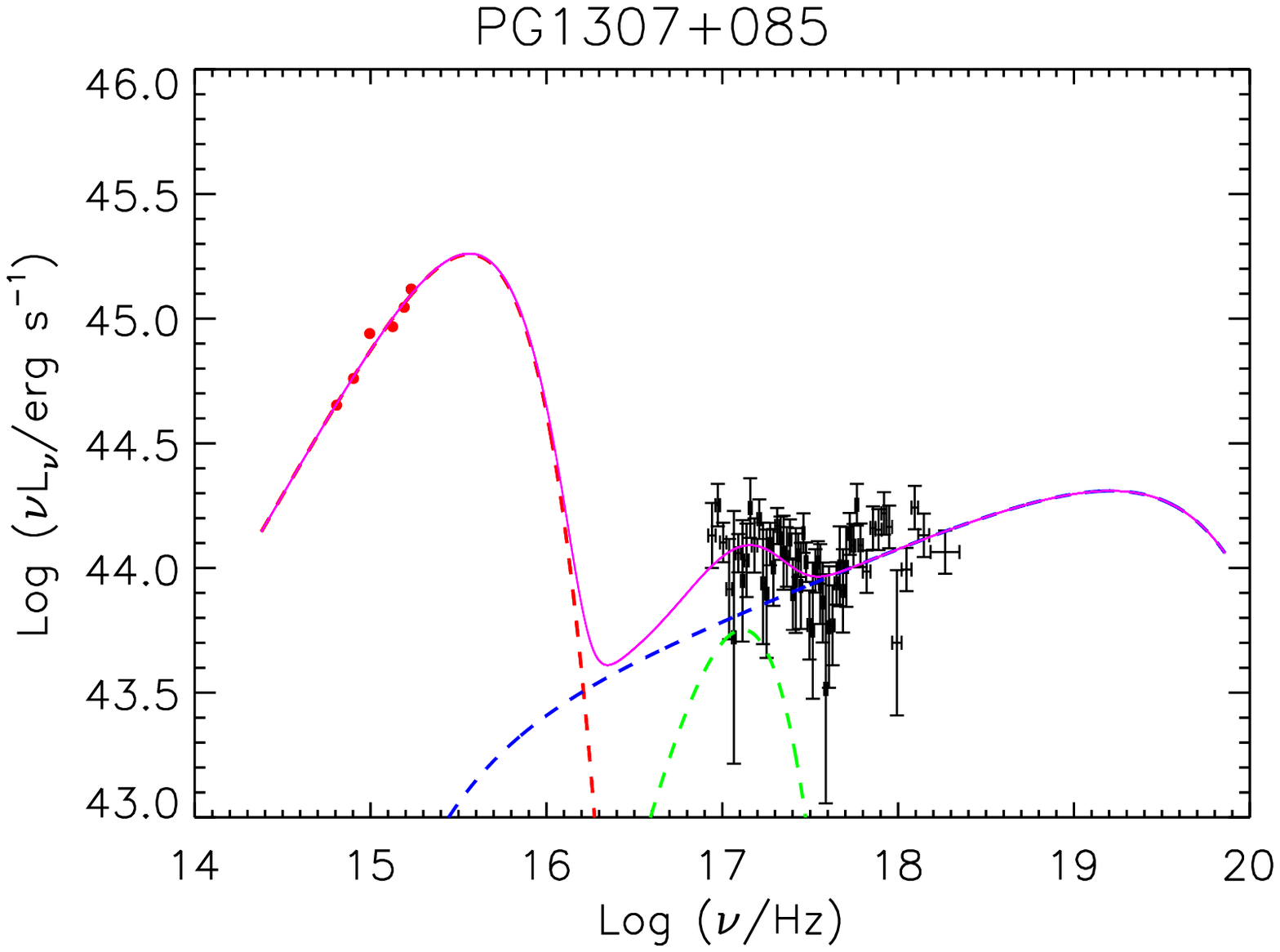}
	\includegraphics[width=5.8cm]{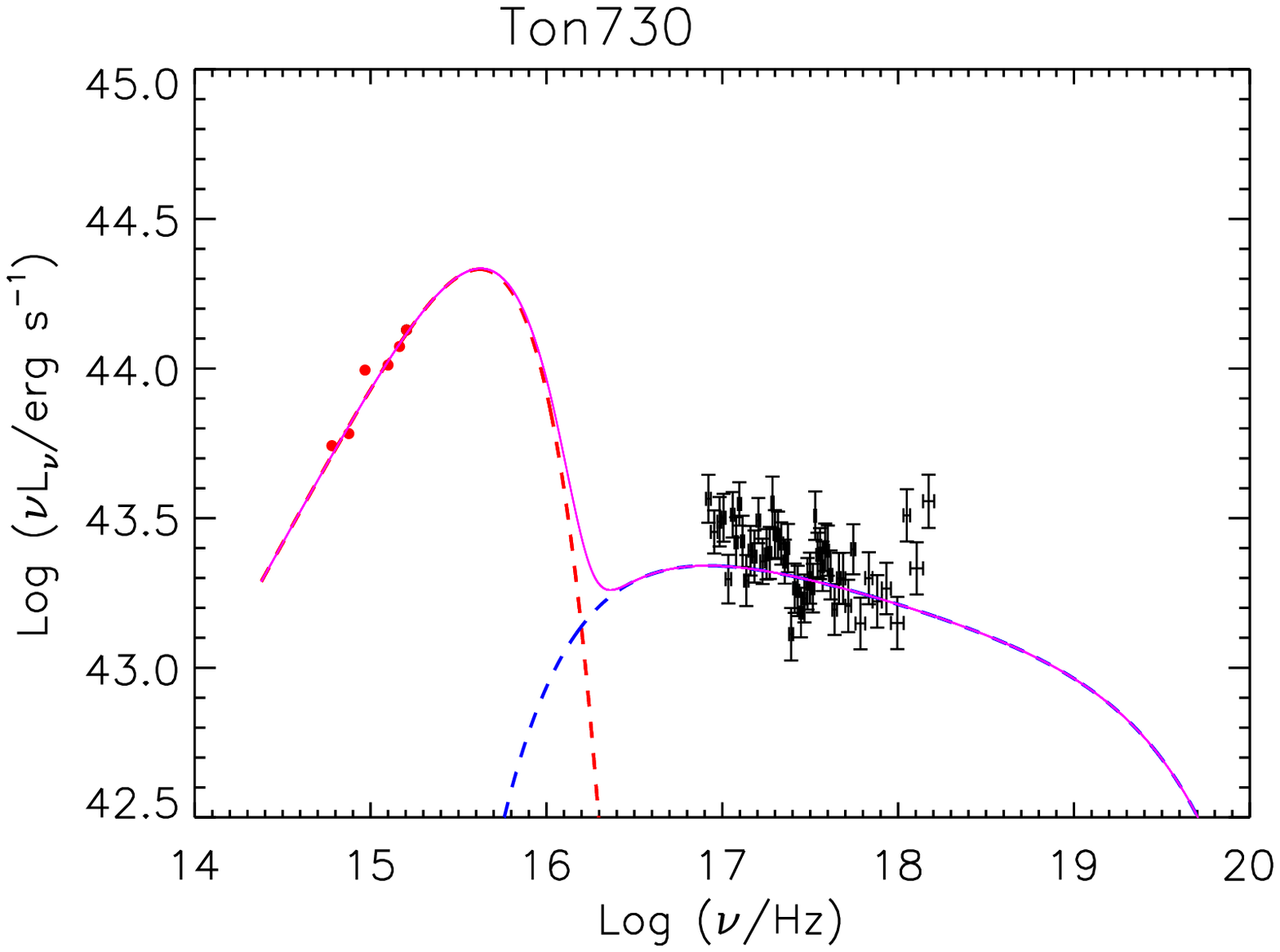}
	\includegraphics[width=5.8cm]{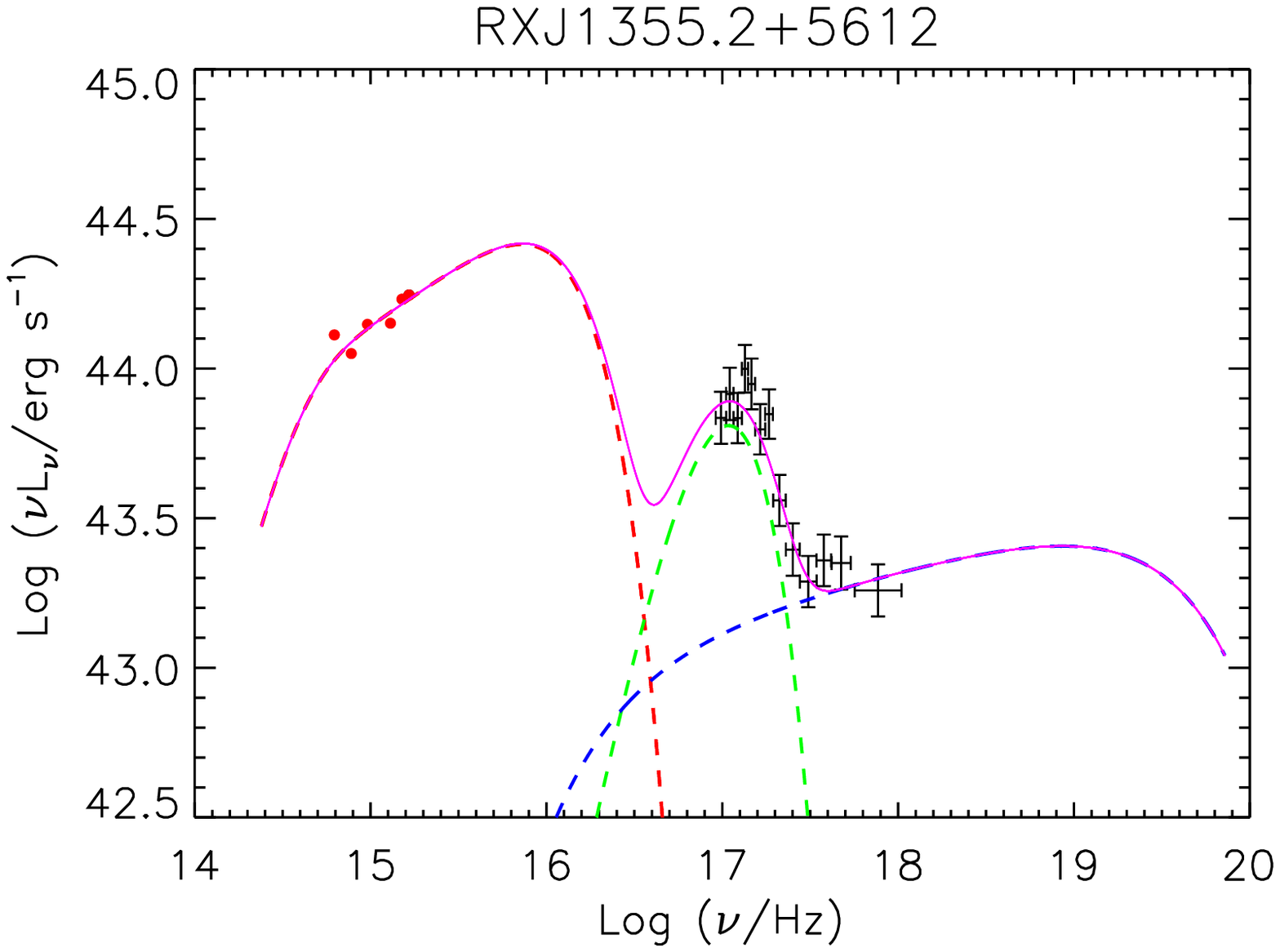}
	\includegraphics[width=5.8cm]{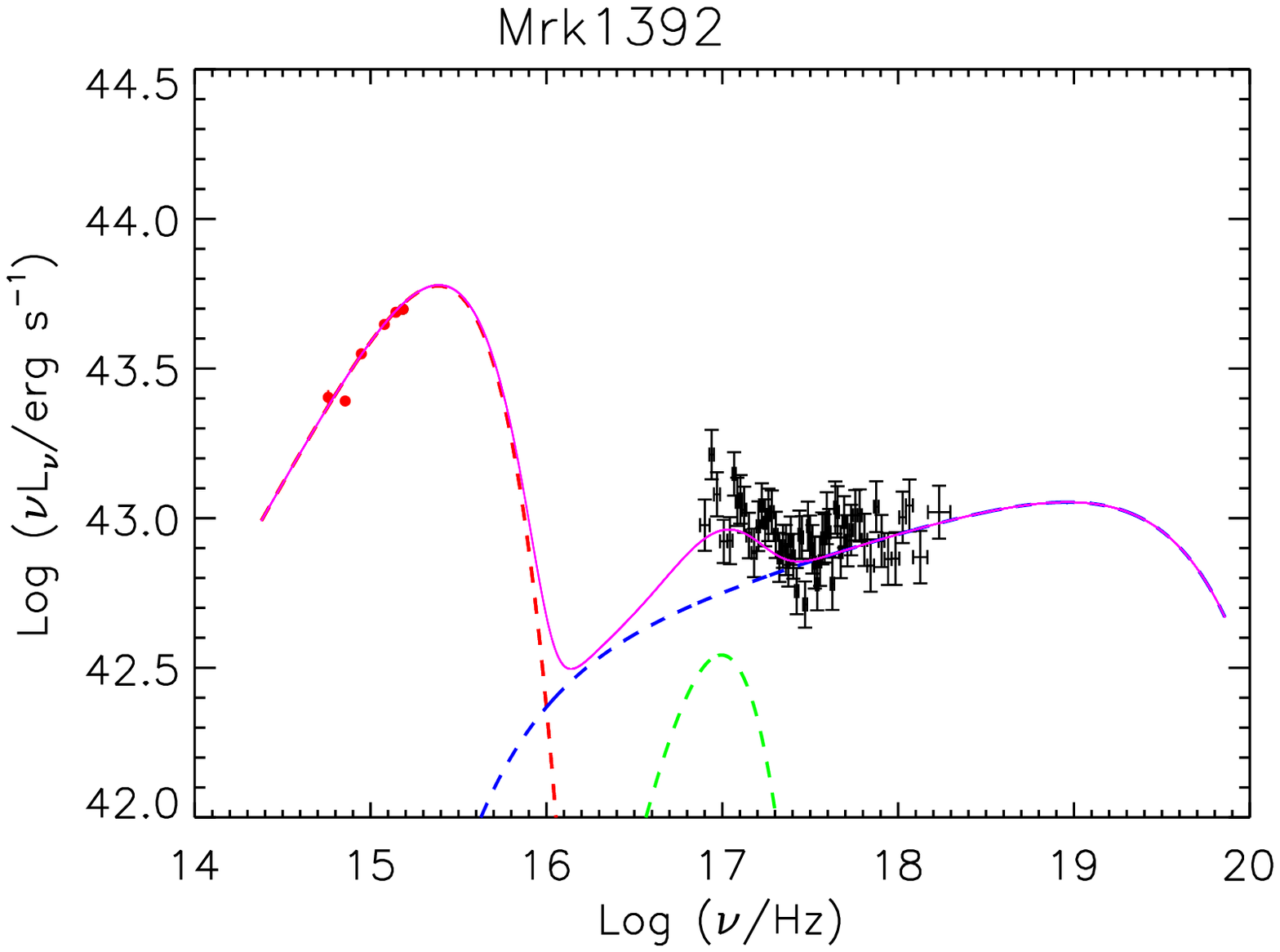}
	\includegraphics[width=5.8cm]{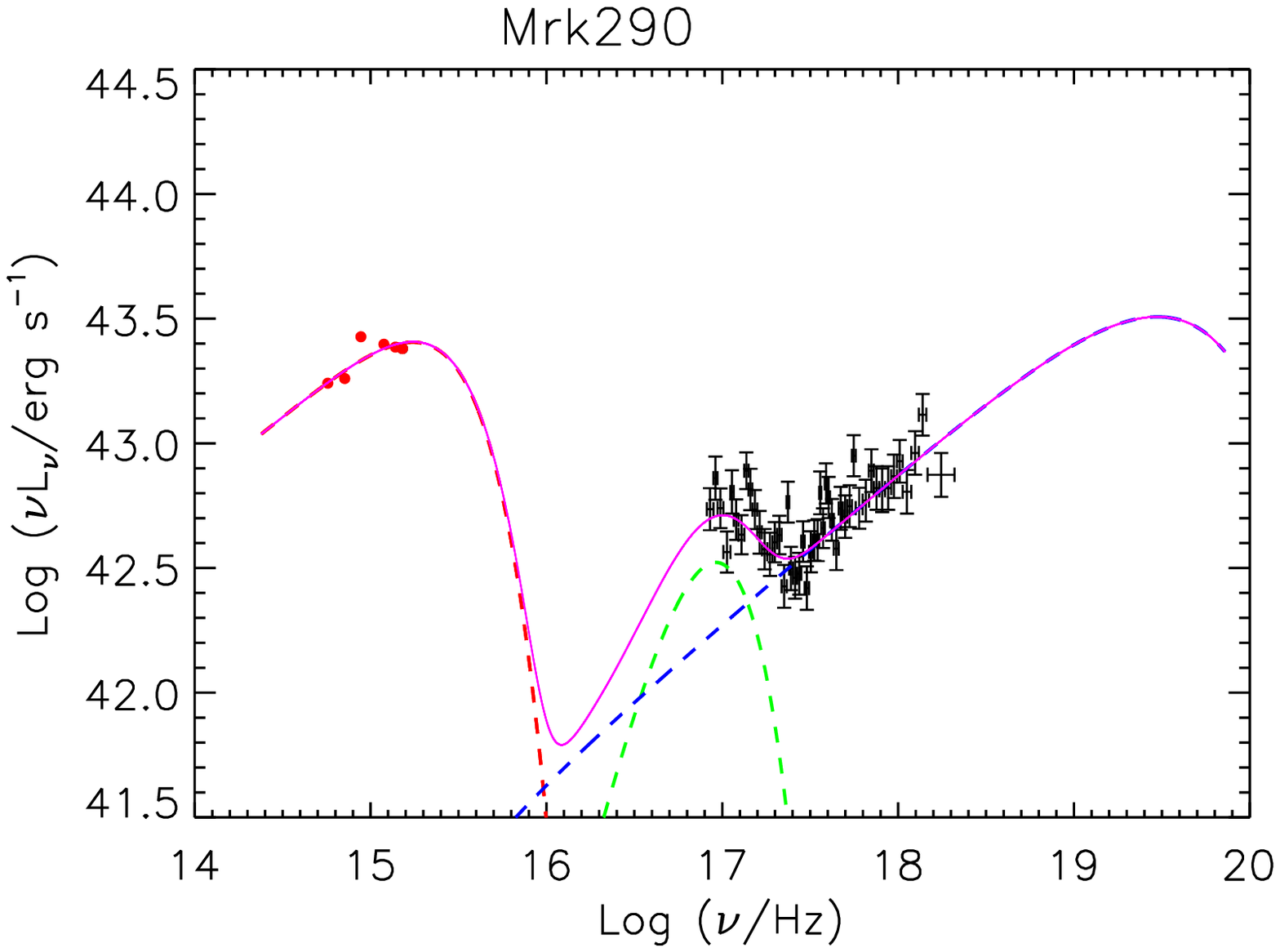}
	\includegraphics[width=5.8cm]{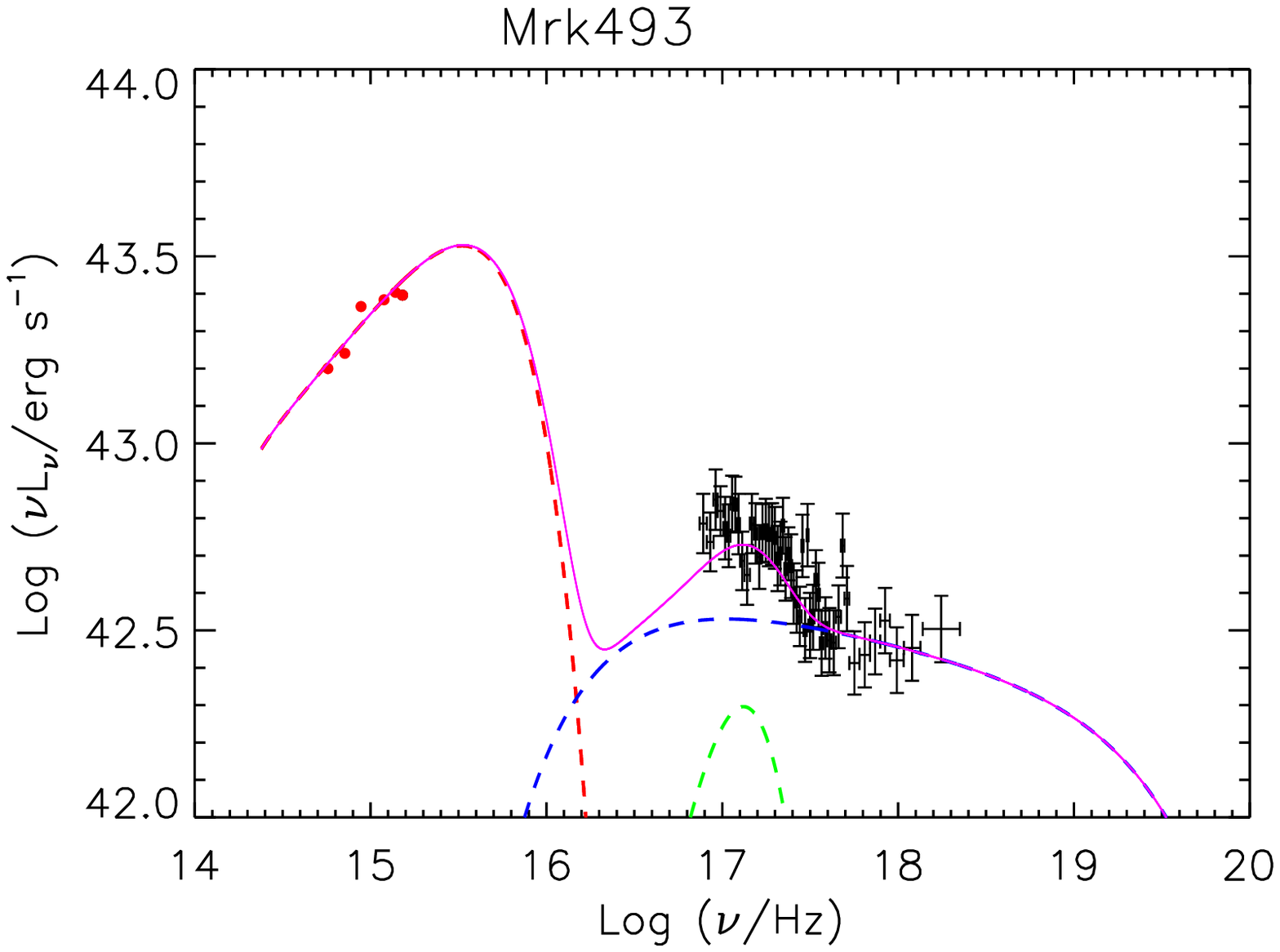}
	\includegraphics[width=5.8cm]{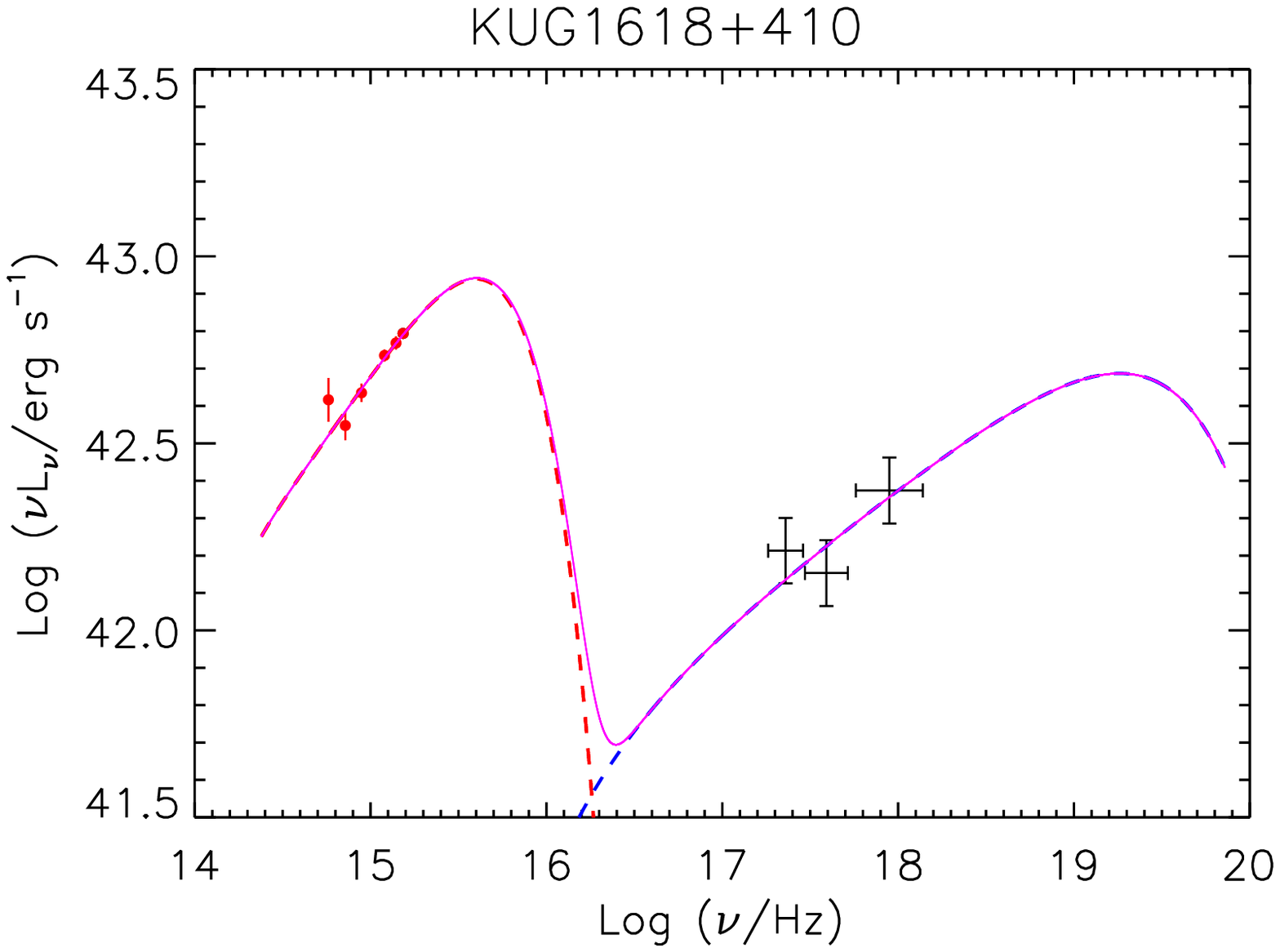}
	\includegraphics[width=5.8cm]{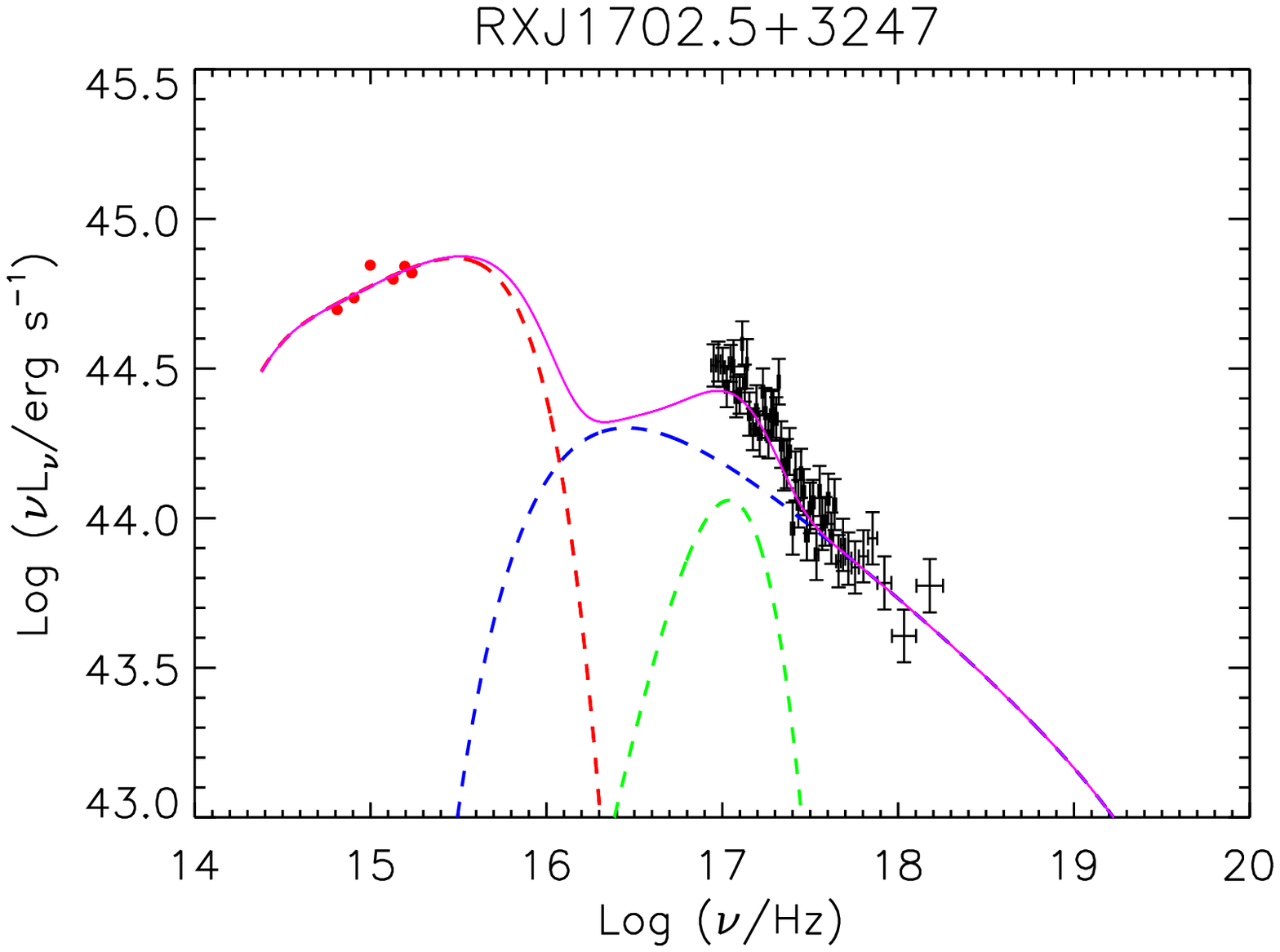}	
	\begin{center}
	Figure~\ref{fig:seds} (continued)
	\end{center}
\end{figure*}

\begin{table*}
\caption{Broadband spectral energy distribution results.}
\begin{threeparttable}
\small
\renewcommand\arraystretch{1.2}
\begin{tabular}{cccccccccccccccccc}
\hline
\hline
\small
No. & Name & $\rm \alpha_{opt-UV}$ & $p$ & $T_{\rm eff}(R_{\rm in})$ & $\rm \alpha_{ox}$ & $\log L_{5100}$ & $\log L_{2-10}$ & $\log L_{\rm bol}$ & $\lambda_{\rm Edd}$ \\
(1)&(2)&(3)&(4)&(5)&(6)&(7)&(8)&(9)&(10)\\
\hline
1& Mrk 1018 & $+0.04$& $0.78_{-0.07}^{+0.07}$& $0.54_{-0.20}^{+0.34}$ & $1.20_{-0.02}^{+0.05}$& $43.57_{-0.06}^{+0.13}$& $43.65_{-0.03}^{+0.03}$& $44.56_{-0.04}^{+0.21}$& $0.011_{-0.006}^{+0.023}$\\
2& MCG +04-22-042& $-0.59$& $0.66_{-0.01}^{+0.02}$& $1.01_{-0.42}^{+0.72}$ & $1.20_{-0.02}^{+0.05}$& $43.37_{-0.05}^{+0.14}$& $43.41_{-0.02}^{+0.02}$& $44.52_{-0.10}^{+0.22}$& $0.050_{-0.028}^{+0.107}$\\
3& Mrk 705& $-0.49$& $0.58_{-0.01}^{+0.01}$& $1.28_{-0.51}^{+0.81}$ & $1.22_{-0.02}^{+0.05}$& $43.22_{-0.06}^{+0.13}$& $43.12_{-0.05}^{+0.05}$& $44.31_{-0.06}^{+0.11}$& $0.096_{-0.050}^{+0.126}$\\
4& RX J1007.1+2203& $-0.38$& $0.59_{-0.01}^{+0.01}$& $1.08_{-0.41}^{+0.69}$& $1.41_{-0.02}^{+0.05}$& $43.47_{-0.06}^{+0.13}$& $42.91_{-0.08}^{+0.08}$& $44.45_{-0.08}^{+0.15}$& $0.068_{-0.036}^{+0.105}$\\
5& CBS 126& $-0.58$& $0.57_{-0.01}^{+0.01}$& $0.79_{-0.30}^{+0.48}$& $1.37_{-0.02}^{+0.05}$& $43.99_{-0.06}^{+0.13}$& $43.47_{-0.06}^{+0.05}$& $44.95_{-0.06}^{+0.12}$& $0.081_{-0.041}^{+0.104}$\\
6& Mrk 141& $-0.55$& $0.63_{-0.05}^{+0.09}$& $0.35_{-0.10}^{+0.17}$& $1.30_{-0.02}^{+0.05}$& $43.27_{-0.05}^{+0.13}$& $42.95_{-0.04}^{+0.04}$& $44.15_{-0.07}^{+0.13}$& $0.005_{-0.003}^{+0.008}$\\
7& Mrk 142&  $-0.39$& $0.59_{-0.01}^{+0.00}$& $1.65_{-0.64}^{+0.94}$& $1.25_{-0.02}^{+0.05}$& $43.39_{-0.05}^{+0.13}$& $43.23_{-0.14}^{+0.06}$& $44.53_{-0.07}^{+0.13}$& $0.184_{-0.091}^{+0.245}$\\
8& Ton 1388& $+0.26$& $0.77_{-0.03}^{+0.04}$& $1.15_{-0.48}^{+0.95}$& $1.51_{-0.02}^{+0.05}$& $45.20_{-0.05}^{+0.13}$& $44.40_{-0.03}^{+0.03}$& $46.43_{-0.24}^{+0.41}$& $0.360_{-0.247}^{+1.332}$\\
9& SBS 1136+594& $-0.19$& $0.62_{-0.01}^{+0.02}$& $0.82_{-0.31}^{+0.53}$& $1.15_{-0.02}^{+0.05}$& $43.61_{-0.05}^{+0.13}$& $43.77_{-0.02}^{+0.02}$& $44.79_{-0.05}^{+0.12}$& $0.065_{-0.034}^{+0.099}$\\
10& PG 1138+222& $+0.02$& $0.67_{-0.01}^{+0.01}$& $1.89_{-0.84}^{+1.47}$& $1.26_{-0.02}^{+0.05}$& $43.91_{-0.06}^{+0.13}$& $43.77_{-0.02}^{+0.02}$& $45.24_{-0.14}^{+0.29}$& $0.352_{-0.213}^{+0.917}$\\
11& KUG 1141+371& $-0.85$& $0.74_{-0.15}^{+0.11}$& $0.24_{-0.06}^{+0.08}$& $1.07_{-0.02}^{+0.05}$& $42.75_{-0.06}^{+0.14}$& $43.11_{-0.05}^{+0.05}$& $44.04_{-0.06}^{+0.13}$& $0.003_{-0.001}^{+0.004}$\\
12& Mrk 1310& $-0.18$& $0.67_{-0.02}^{+0.03}$& $0.88_{-0.36}^{+0.61}$& $1.08_{-0.02}^{+0.05}$& $42.05_{-0.06}^{+0.13}$& $42.56_{-0.05}^{+0.05}$& $43.63_{-0.14}^{+0.22}$& $0.023_{-0.013}^{+0.038}$\\
13& RX J1209.8+3217& $-0.64$& $0.55_{-0.00}^{+0.01}$& $1.11_{-0.41}^{+0.65}$& $1.36_{-0.02}^{+0.05}$& $43.93_{-0.05}^{+0.13}$& $43.13_{-0.10}^{+0.04}$& $44.86_{-0.07}^{+0.12}$& $0.149_{-0.068}^{+0.152}$\\
14& Mrk 50&  $-0.51$& $0.67_{-0.05}^{+0.08}$& $0.42_{-0.13}^{+0.24}$& $1.10_{-0.02}^{+0.05}$& $42.73_{-0.05}^{+0.13}$& $42.97_{-0.02}^{+0.02}$& $43.83_{-0.03}^{+0.09}$& $0.005_{-0.003}^{+0.008}$\\
15& Mrk 771& $-0.09$& $0.67_{-0.01}^{+0.02}$& $1.06_{-0.44}^{+0.79}$& $1.32_{-0.02}^{+0.05}$& $43.84_{-0.06}^{+0.14}$& $43.48_{-0.03}^{+0.03}$& $44.90_{-0.14}^{+0.27}$& $0.075_{-0.045}^{+0.185}$\\
16& PG 1307+085&  $+0.07$& $0.72_{-0.03}^{+0.05}$& $0.76_{-0.28}^{+0.54}$& $1.37_{-0.02}^{+0.05}$& $44.61_{-0.06}^{+0.14}$& $44.29_{-0.03}^{+0.03}$& $45.58_{-0.13}^{+0.31}$& $0.068_{-0.041}^{+0.188}$\\
17& Ton 730& $+0.00$& $0.68_{-0.02}^{+0.03}$& $0.91_{-0.37}^{+0.65}$&  $1.28_{-0.02}^{+0.05}$& $43.69_{-0.06}^{+0.14}$& $43.41_{-0.03}^{+0.03}$& $44.72_{-0.14}^{+0.28}$& $0.043_{-0.026}^{+0.105}$\\
18& RX J1355.2+5612 & $-0.57$& $0.56_{-0.04}^{+0.01}$& $2.14_{-0.71}^{+0.85}$& $1.35_{-0.02}^{+0.05}$& $44.01_{-0.06}^{+0.12}$& $43.53_{-0.13}^{+0.12}$& $44.99_{-0.10}^{+0.11}$& $0.541_{-0.242}^{+0.370}$\\
19& Mrk 1392& $-0.21$& $0.68_{-0.04}^{+0.07}$& $0.53_{-0.18}^{+0.31}$& $1.29_{-0.02}^{+0.05}$& $43.38_{-0.06}^{+0.14}$& $43.16_{-0.03}^{+0.03}$& $44.26_{-0.07}^{+0.19}$& $0.009_{-0.005}^{+0.018}$\\
20& Mrk 290& $-0.62$& $0.58_{-0.01}^{+0.01}$& $0.46_{-0.15}^{+0.24}$&  $1.27_{-0.02}^{+0.05}$& $43.25_{-0.05}^{+0.13}$& $43.11_{-0.03}^{+0.03}$& $44.33_{-0.05}^{+0.10}$& $0.016_{-0.009}^{+0.021}$\\
21& Mrk 493& $-0.12$ & $0.58_{-0.01}^{+0.01}$ & $0.88_{-0.33}^{+0.55}$ & $1.34_{-0.02}^{+0.05}$ & $43.22_{-0.06}^{+0.13}$ & $42.66_{-0.03}^{+0.03}$ & $44.16_{-0.07}^{+0.14}$ & $0.034_{-0.018}^{+0.049}$ \\
22& KUG 1618+410& $-0.39$ & $0.60_{-0.01}^{+0.01}$& $1.00_{-0.39}^{+0.66}$&  $1.18_{-0.02}^{+0.05}$& $42.52_{-0.06}^{+0.14}$& $42.60_{-0.15}^{+0.13}$& $43.67_{-0.11}^{+0.22}$& $0.027_{-0.014}^{+0.049}$\\
23& RX J1702.5+3247 & $-0.73$& $0.54_{-0.01}^{+0.01}$& $0.97_{-0.36}^{+0.54}$&  $1.35_{-0.02}^{+0.05}$& $44.70_{-0.06}^{+0.13}$& $43.93_{-0.04}^{+0.04}$& $45.65_{-0.06}^{+0.11}$& $0.321_{-0.149}^{+0.308}$\\
\hline
\hline
\end{tabular}
{$\bm Notes.$} Column (1): identification number assigned in this paper. Column (2): NED name of the sample objects. Column (3): the optical/UV spectra slope. Column (4): power-law index of the radial temperature profile $p$ in the p-free disk model. Column (5): the effective temperature at the inner disk radius $R_{\rm in}$ in units of $10^5$ K in the p-free disk model. Column (6): the effective optical-to-X-ray spectra index $\alpha_{\rm OX}$. Column (7): logarithmic of the rest-frame $5100$ \angstrom\ monochromatic luminosity in units of erg s$^{-1}$. Column (8): logarithmic of the rest-frame $2-10$ keV luminosity in units of erg s$^{-1}$. Column (9): logarithmic of the rest-frame $0.001-300$ keV bolometric luminosity in units of erg s$^{-1}$. Column (10): Eddington ratio. The errors are given in $1\sigma$.
\end{threeparttable}
\label{tab:sed_analysis}
\end{table*}

\subsection{Bolometric corrections}
\label{sec:bc}

The bolometric luminosities obtained above are used to derive bolometric correction factors at a specific luminosity or band,
\begin{equation}
	\kappa_{\nu}=\frac{L_{\rm bol}}{\nu L_{\nu}}
	\label{eq:knu}
\end{equation}
Here, we focus on two frequently used factors, \kh\ for the $2$ -- $10$ keV X-ray band and \ko\ at the optical $5100$ \angstrom \ wavelength respectively.

\subsubsection{2 -- 10 keV bolometric correction}
\label{sec:kbolx}

\begin{figure}
\includegraphics[width=\columnwidth]{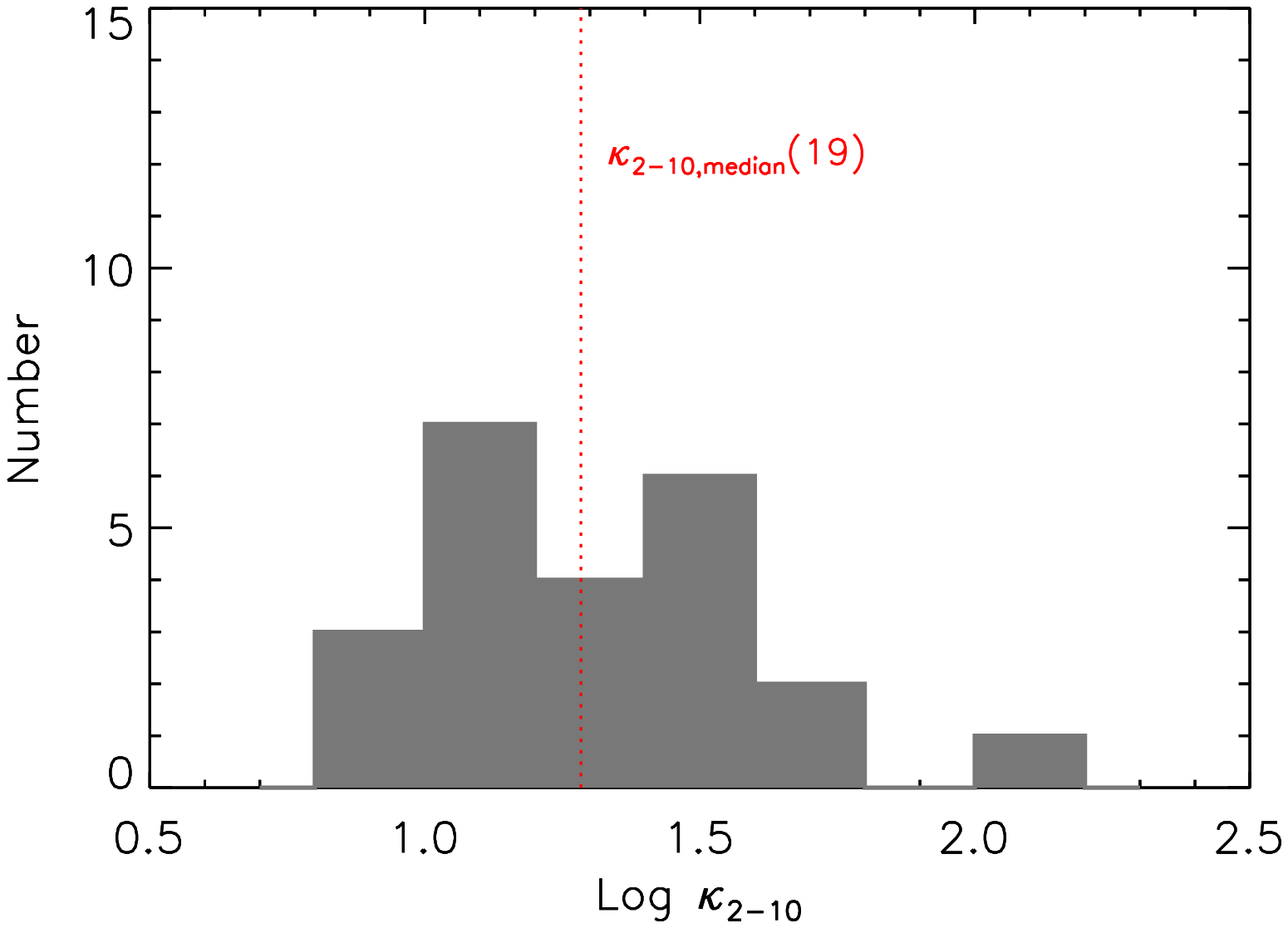}
\caption{The distribution of the hard X-ray bolometric correction factor $\kappa_{2-10}$. The red dotted line represents $\kappa_{2-10}=19$ (the median).}
\label{fig:k210hist}
\end{figure}

\begin{figure}
\includegraphics[width=\columnwidth]{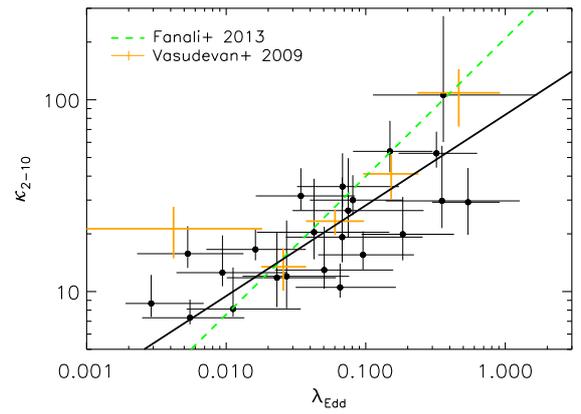}
\caption{Hard X-ray bolometric correction factor $\kappa_{2-10}$ versus Eddington ratio $\lambda_{\rm Edd}$. The black dots represent the sources in our sample, and the errors are given in $1\sigma$. The black solid line is the bisector regression line. The green dashed line is the best-fitting regression line obtained in \citet{2013MNRAS.433..648F}. The yellow crosses are the binned data points in \citet{2009MNRAS.392.1124V}.}
\label{fig:k210edd}
\end{figure}

\begin{figure}
\includegraphics[width=\columnwidth]{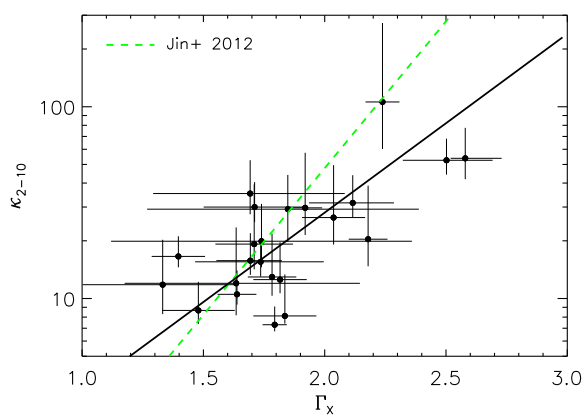}
\caption{Hard X-ray bolometric correction factor $\kappa_{2-10}$ versus hard X-ray photon index $\Gamma_{\rm X}$. The black dots represent the sources in our sample. The error of \gx\ is given in $90\%$ confidence interval, and the error of \kh\ is given in $1\sigma$. The black solid line represents the best-fit regression. The green dashed line is the result in \citet{2012MNRAS.425..907J}.}
\label{fig:k210gammax}
\end{figure}

\begin{figure}
\includegraphics[width=\columnwidth]{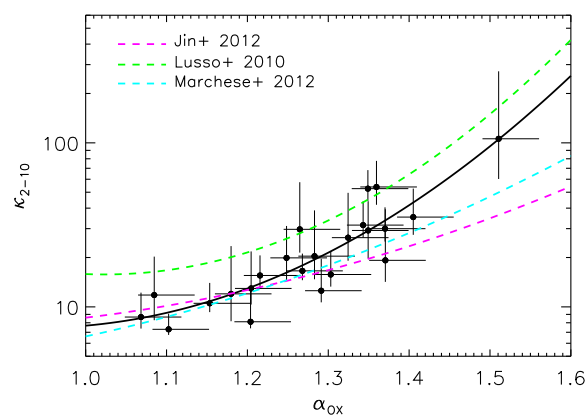}
\caption{Hard X-ray bolometric correction factor $\kappa_{2-10}$ versus effective optical-to-X-ray spectral index $\alpha_{\rm OX}$. The black dots represent the sources in our sample. The errors are given in $1\sigma$. The black solid line represents the best-fitting polynomial. The pink, green and cyan dashed lines represent the results reported in \citet{2010A&A...512A..34L}, \citet{2012MNRAS.425..907J} and \citet{2012A&A...539A..48M} respectively.}
\label{fig:k210aox}
\end{figure}

The $2$ -- $10$ keV bolometric correction reflects the relative contribution of the corona X-ray emission in the total radiation power of AGN. Previous studies suggested dependences of \kh\ on a few parameters including $\rm \lambda_{Edd}$, $\Gamma_{\rm X}$ and $\alpha_{\rm OX}$. \citet{2007MNRAS.381.1235V} found that \kh\ increases with \edd, which was confirmed later in various studies \citep[e.g.,][]{2010A&A...512A..34L, 2012MNRAS.425..907J, 2013MNRAS.433..648F}. The distribution of the derived \kh\ for our sample AGN is shown in Figure \ref{fig:k210hist}, spanning a wide range $\sim10$ -- $100$ with a median of $\sim19$. There is a correlation between \kh\ and \edd\ ($\rho_s=0.74$ and $P=6.4\times10^{-5}$), consistent with those found in previous studies. The \kh\ versus \edd\ is plotted in Figure \ref{fig:k210edd}. A bisector regression analysis using the ordinary least-squares (OLS) algorithm yields a linear relationship Equation \ref{eq:k210edd-bi}, which is overplotted in Figure \ref{fig:k210edd}.
\begin{equation}
	\log\kappa_{2-10}=(0.473\pm0.068)\log\lambda_{\rm Edd}+(1.922\pm0.105)
	\label{eq:k210edd-bi}
\end{equation}

The averaged bolometric correction factors are also obtained for the above groups of the Eddington ratios (low-, \edd$\la0.03$; intermediate-, $0.03\la$\edd$\la0.09$; high-, \edd$\ga0.09$). \kh\ is found to be $8$ -- $15$ for \edd$\la0.03$, $15$ -- $30$ for $0.03\la$\edd$\la0.09$ and $20$ -- $70$ for \edd$\ga0.09$. As in previous work the SSD model was adopted in fitting the optical/UV spectra of AGN, it is of interest to compare our result with previous ones. In Figure \ref{fig:k210edd} the results from \citet{2009MNRAS.392.1124V} and \citet{2013MNRAS.433..648F} are overplotted for comparison. It can be seen that objects with low- and intermediate- \edd\ in our sample have similar \kh\ values to those in \citet{2009MNRAS.392.1124V}, while the high-\edd\ objects have lower \kh\ values in general. The slope of the linear regression is flatter ($\sim3.6\sigma$) than that in \citet{2013MNRAS.433..648F} and predicts lower \kh\ at higher-\edd\ end. This is mainly due to the flatter radial temperature profiles in high-\edd\ objects than those with low- and intermediate-\edd\ (see Section \ref{sec:kbolrevisit} for a detailed discussion).

A relationship between the hard X-ray photon index \gx \ and \kh \ was also found in previous work \citep[e.g.,][]{2012MNRAS.425..907J}. The distribution of \kh \ versus \gx \ for our sample is plotted in Figure \ref{fig:k210gammax}. Only a marginal correlation is found ($\rho_s=0.59$ and $P=3.3\times10^{-3}$). Using a bisector regression analysis gives
\begin{equation}
	\log\kappa_{2-10}=(0.932\pm0.151)\Gamma_{\rm X}-(0.417\pm0.278)
	\label{eq:k210gammax}
\end{equation}

As suggested by \citet{2010A&A...512A..34L}, \aox \ is also an indicator of $\kappa_{2-10}$, as described by a second-order polynomial relation, and this was later confirmed by \citet{2012A&A...539A..48M} and \citet{2012MNRAS.425..907J}. We apply a second-degree polynomial fit to the results in our sample and find
\begin{equation}
	\rm \log\kappa_{2-10}=3.97-6.60\alpha_{OX}+3.52\alpha_{OX}^2
	\label{k210aoxso}
\end{equation}

The $\kappa_{2-10}$ versus $\rm\alpha_{OX}$ distribution is plotted in Figure \ref{fig:k210aox}, along with the relations reported in previous studies. Our fitted curve generally agrees with those derived in \citet{2010A&A...512A..34L} and \citet{2012A&A...539A..48M}, but lower than that in \citet{2012MNRAS.425..907J}. Our result confirms that, compared to $\lambda_{\rm Edd}$ and $\Gamma_{\rm X}$, $\alpha_{\rm OX}$ appears to be a better indicator for $\kappa_{2-10}$. We note that the above correlations (\kh-\edd, \kh-\gx \ and \kh-\aox) can also be inferred from the study of the averaged SED (see Figure \ref{fig:allseds-new}).

\subsubsection{Optical 5100 \angstrom \ bolometric correction}

\begin{figure}
\includegraphics[width=\columnwidth]{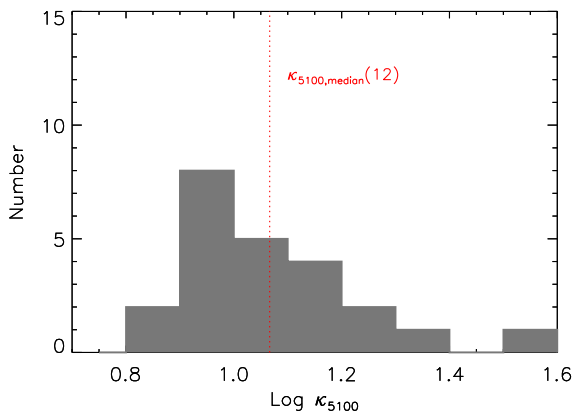}
\caption{The distribution of the optical (5100 \angstrom) bolometric correction factor $\kappa_{5100}$. The red dotted line represents $\kappa_{5100}=12$ (the median).}
\label{fig:k5100hist}
\end{figure}

\begin{figure}
\includegraphics[width=\columnwidth]{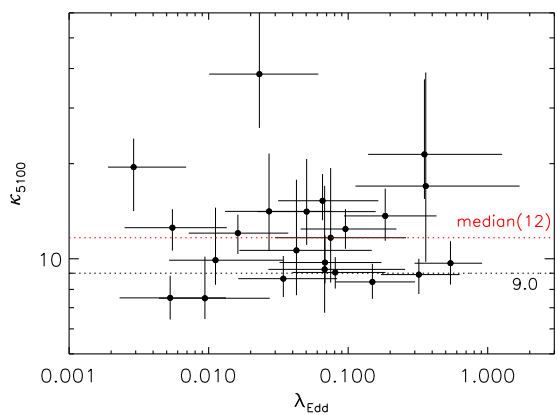}
\caption{Optical (5100 \angstrom) bolometric correction factor $\kappa_{5100}$ versus Eddington ratio $\lambda_{\rm Edd}$. The black dots represent the sources in our sample. The errors are given in $1\sigma$. The red dotted line represents $\kappa_{5100}=12$ (the median). The black dotted line represents $\kappa_{5100}=9.0$ \citep{2000ApJ...533..631K}. }
\label{fig:k5100edd}
\end{figure}

Early studies suggested a constant $5100$ \angstrom\ bolometric correction factor, e.g., $\kappa_{5100}=9.0$ \citep{2000ApJ...533..631K} and $\kappa_{5100}=10.3\pm2.1$ \citep{2006ApJS..166..470R}, with weak or no dependence of \ko \ on \edd. \citet{2007MNRAS.381.1235V} suggested a positive correlation similar to \kh, which was confirmed by \citet{2012MNRAS.425..907J} using a high-\edd \ AGN sample. The distribution of \ko \ in our work is shown in Figure \ref{fig:k5100hist}, within a range of $8$ -- $20$ for most of the AGN. The median value of $12$ is broadly consistent with $10.3\pm2.1$ given by \citet{2006ApJS..166..470R} for quasars. However, no correlation is found between \ko \ and \edd \ for our sample ($\rho_s=-0.031$ and $P=0.89$, see Figure \ref{fig:k5100edd}), which is in contrary to the results in \citet{2007MNRAS.381.1235V} and \citet{2012MNRAS.425..907J}. Similar to the case in \kh, this is also due to the flatter radial temperature profiles in high-\edd\ objects than those with low- and intermediate-\edd\ (see Section \ref{sec:kbolrevisit} for a detailed discussion).

\section{discussion}
\label{sec:discuss}

\subsection{Optical/UV spectral slope}
\label{sec:slope}
The optical/UV spectrum is of great importance to the determination of the broadband SED of AGN. The power-law spectral slopes \aoptuv\ have been measured for many AGN in large samples in numerous studies using photometric and spectroscopic data from various surveys, e.g., SDSS \citep{2001AJ....122..549V, 2007ApJ...668..682D}, FUSE \citep{2004ApJ...615..135S}, HST/COS \citep{2012ApJ...752..162S, 2014ApJ...794...75S}, AKARI \citep{2015ApJ...806..109J}, VLT/X-shooter \citep{2016A&A...585A..87S} and UVOT \citep{2017MNRAS.467.4674L}. Despite large scatter, the optical/UV spectra are found to have shallower slopes in most of the studies. The spectral slopes obtained in our work, with elimination of the effects of the host galaxy contamination and dust extinction, are broadly consistent with previous results. Yet, there may be uncertainties inherent in these measurements which may bias the results, as discussed below.

The contamination of the host galaxy starlight is the main source of uncertainty in measuring AGN optical emission at long wavelengths. \citet{2006ApJ...644..133B} pointed out that even for an aperture of $1''$ for AGN photometry the host galaxy contamination may not be negligible. We try to separate the AGN (nuclear) flux from that of the host galaxy by using the 2-D image decomposition algorithm \textsc{galfit}, as commonly done in previous work \citep[e.g.,][]{2009ApJ...697..160B, 2009MNRAS.399.1553V, 2011ApJ...739...57K}, even though the spatial resolution of the UVOT images is not high (a typical FWHM of $\sim2''$). To check the reliability of our image decomposition analysis, we compare our results with the previous ones for the $7$ objects which were also studied for image decomposition by \citet{2011ApJ...739...57K} using the SDSS images with a better resolution. The magnitudes of the galactic component derived in this work are broadly consistent with those derived in \citet{2011ApJ...739...57K} with differences in the fluxes around $10$ -- $20\%$. We consider the uncertainties to be small and our results to be reliable.

As commonly adopted in AGN studies, we assume that the dust extinction of the optical/UV continuum can well be represented by that in the broad-line region, which can be estimated from the broad-line Balmer decrement. There is good observational basis for this argument \citep[e.g.,][]{2008MNRAS.383..581D, 2017MNRAS.467..226G}. Our sample is selected in such a way that the dust extinction effect is negligible or minor at most. For most ($17/23$) of our sample objects, the Balmer decrements are consistent with the value for which little or no dust extinction is present. For the remaining $6$ sources, there may be a minor effect and the amount of dust reddening is derived by assuming a SMC-like extinction curve in the analysis. The effect of using different extinction curves is also investigated, including the extinction curves of the Milky Way (MW), the average curves of the Large Magellanic Cloud (LMC) and the LMC2 supershell \citep{2003ApJ...594..279G}. For these objects the optical/UV photometric data are corrected for the reddening effect by assuming the above curves. The resulting spectra are fitted with a power-law model as well as the p-free disk model respectively. This results in typical changes in the optical/UV slopes of $\sim5$ per cent only and $\sim2$ per cent in the index $p$. We thus conclude that the use of different extinction curves does not affect our results significantly for those objects in which minor dust reddening might occur. Therefore, in the rest of this work the SMC-like extinction curve is adopted.

\subsection{Accretion disks with non-standard radial temperature profile}
\label{sec:pedd}

It has been demonstrated that the shallower spectral slopes in our sample AGN cannot be attributed mainly to starlight contamination or dust reddening. We consider such spectral shapes intrinsic to the emission from the AGN, which can well be produced by an accretion disk with a radial temperature profile flatter than the canonical $T_{\rm eff}(R)\propto R^{-0.75}$ for the standard SSD model. Theoretically, the spectral slope $\alpha$ of the power-law regime of an accretion disk spectrum and the effective temperature power-law index $p$ are related via $\alpha=3-2/p$ \citep[e.g.,][]{1972A&A....21....1P, 2008bhad.book.....K}. For supermassive black holes the power-law regime of the spectra falls into the optical/UV waveband for a wide range of \mbh\ and \edd, and the slope \aoptuv\ can be reddened depending on the exact value of $p$. \citet{2008RMxAC..32....1G} suggested that $p\approx-0.57$ can explain the typical spectral slopes of $\alpha\approx-0.5$ observed in AGN. In this work, we apply this model, for the first time, to fit the optical/UV spectra for a sample of AGN with simultaneous multi-band photometric data. It is found that, the inferred radial effective temperature profile indices $p$ fall within a range of $0.5$ -- $0.75$ for most of the sample objects in order to explain the observed spectra, flatter than $p=0.75$ for the SSD widely assumed in previous studies. 

Accretion disks with a non-standard radial temperature profile ($p\ne0.75$) may operate if one or more of the assumptions underlying the standard SSD break down. For instance, the local thermal equilibrium condition may not hold, if the dissipated accretion energy can not be radiated away efficiently locally, and consequently the energy/photons are advected into the black hole. Alternatively, the assumption of mass conservation may not hold as, for instance, part of the accreted mass is lost in the accretion process due to various mechanisms (e.g., disk winds/outflows).

\begin{figure}
\includegraphics[width=\columnwidth]{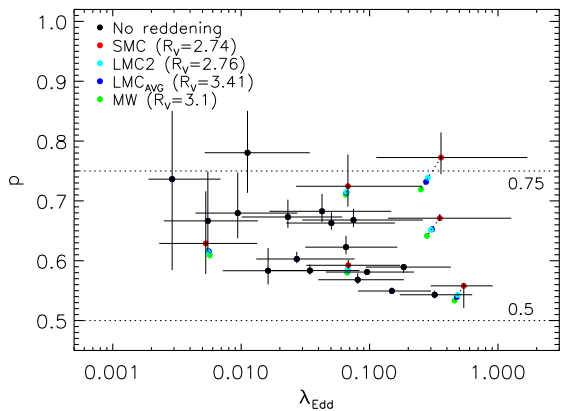}
\caption{Power-law index of the radial temperature profile $p$ versus Eddington ratio \edd. Black dots are the sources with ${\rm H{\alpha}/{\rm H{\beta}}}\le3.06$. Colored dots are the sources with ${\rm H{\alpha}/{\rm H{\beta}}}>3.06$ dereddened using various extinction curves (red: SMC, cyan: LMC2, blue: average LMC, green: MW). The two black dotted lines represent $p=0.75$ (SSD solution) and $p=0.5$ (`slim' disk solution). The errors are given in $1\sigma$.}
\label{fig:p-edd}
\end{figure}

To gain an insight into the cause of the non-standard radial temperature profiles, the relationship is examined between $p$ and the Eddington ratio \edd, which is considered to be the most possible dominant quantity that governs the accretion process of AGN. A marginal correlation is found ($\rho_s=-0.41$, $P=5.0\times 10^{-2}$), as shown in Figure \ref{fig:p-edd}, where the results derived using different dust extinction curves for the sources with intrinsic reddening are marked in different colors. The correlation becomes more prominent when only the sources having no intrinsic dust reddening are used ($\rho_s=-0.71$, $P=1.4\times10^{-3}$).  Such a correlation is qualitatively consistent with the above physical explanations. For instance, when the accretion rate is well below the Eddington limit, the radiative diffusion timescale in the disk is much faster than the accretion timescale. Such an accretion flow is radiatively efficient and has a radial temperature distribution of SSD type ($T_{\rm eff}(R)\propto R^{-0.75}$). However, when the accretion rate approaches the Eddington limit \citep[$\lambda_{\rm Edd}\ga0.3$, e.g.,][]{1988ApJ...332..646A, 2000PASJ...52..499M}, the photon trapping effect in the disk becomes appreciable and the disk becomes a `slim' disk. In this case the radiation energy is trapped and advected inward rather than locally radiated away, resulting in a much flatter radial temperature profile \citep[$T_{\rm eff}(R)\propto R^{-0.5}$, e.g.,][]{1999ApJ...522..839W,1999PASJ...51..725W}. Given the distribution of the Eddington ratio of our sample, the flat temperature profiles in most of the sources can hardly be induced by the photon trapping. Nevertheless, we suggest that this effect may operate in some of the high-\edd\ sources and be partly responsible for the observed $p$-\edd\ correlation.

\begin{figure}
\includegraphics[width=\columnwidth]{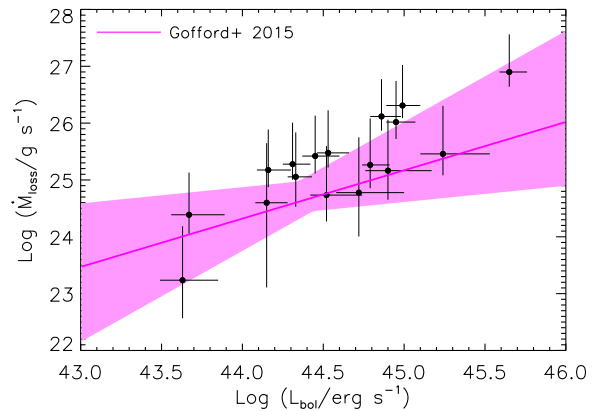}
\caption{Mass loss rate within the disk $\dot{M}_{\rm loss}$ versus bolometric luminosity \lbol. The black crosses are the sample objects (excluding 6 sources in which the SSD solution can be satisfied, see text). The errors are given in $1\sigma$. The pink solid line and shallow region represent the best-fit regression relation and $1\sigma$ error derived in \citet{2015MNRAS.451.4169G}.}
\label{fig:disk-wind}
\end{figure}

Another possible physical explanation is offered by the mass loss process in form of outflows/disk-winds, as a power-law radial temperature profile with an index $p\ne0.75$ implies a non-stationary mass inflow rate $\dot{M}(R)\propto R^{3-4p}$ under the LTE assumption. In fact, observations suggest that outflows may be prevalent or even ubiquitous in bright AGN, and are likely stronger in higher-\edd\ objects \citep[e.g.,][]{2002ApJ...569..641L, 2007ApJ...665..990G, 2014ApJ...786...42Z}. In this picture, the $p$-$\lambda_{\rm Edd}$ relation found in our work can thus be understood, as the $p$ value is determined by the amount of mass loss, i.e., a smaller $p$ value (a steeper mass flow rate distribution and thus a stronger mass loss) tends to be found in higher-\edd\ objects. Interestingly, \citet{2015MNRAS.451.4169G} studied the properties of highly ionized outflows for a sample of $51$ AGN observed with {\it Suzaku} and found that the mass loss rate of the outflow (${\dot{M}}_{\rm loss}$) is well correlated with the bolometric luminosity (\lbol) of the source. Recently, it has been well interpreted under the mechanism of the line-driven disk wind by numerical simulations \citep[e.g.,][]{2017MNRAS.465.2873N, 2018arXiv181101966N}. In this work, by assuming that the deviation of the radial temperature profile is caused by the mass loss process, we can also investigate the ${\dot{M}}_{\rm loss}$ -- \lbol\ relation from the spectral fitting results.

In order to calculate the mass loss rate for the sample AGN, the mass flow rate profile $\dot{M}(R)$ for each object is obtained from the best-fit radial temperature distribution $T_{\rm eff}(R)$ (see Section \ref{sec:pfree}) under the LTE assumption \citep[$T_{\rm eff}(R)=f(\dot{M}(R), M_{\rm BH},R)\propto \dot{M}(R)^{1/4}M_{\rm BH}^{1/4}R^{-3/4}$, e.g.,][]{2002apa..book.....F, 2008bhad.book.....K}. With this accretion rate profile, the mass loss rate is calculated by
\begin{equation}
\dot{M}_{\rm loss}=\dot{M}(R_{\rm out})-\dot{M}(R_{\rm in})
\end{equation}
We note that the mass loss rate is positive when $p<0.75$ and negative when $p>0.75$. For the latter case, this indicates an `inflow' instead of 'outflow'. In this work, we do not take this `inflow' explanation for the sources with $p\ga0.75$, considering the uncertainties inherent in our spectral fitting procedure. The accretion disks in these sources are regarded to be the SSD-type. We also note that although the mass loss rate formally depends both on $p$ and $T_{\rm eff}(R_{\rm in})$, the dependence on $p$ is decisive as it appears in the power-law index of the accretion rate formula thus determines the steepness of the distribution. In other words, no matter how large $T_{\rm eff}(R_{\rm in})$ (the accretion rate at $R_{\rm in}$) is, a $p$ value close to $0.75$ (the SSD solution) stands for a weak/negligible outflow solution ($\dot{M}_{\rm loss}\approx0$). 

The bolometric luminosity \lbol\ is derived by integrating the SED model within the energy range of $0.001$ -- $300$ keV. We note that \lbol\ depends mainly on $T_{\rm eff}(R_{\rm in})$, and the relation can roughly be a fourth power when $L_{\rm bol}\approx L_{\rm disk}$ under high-\edd\ condition. The dependence on $p$, however, is inconspicuous (there is no statistical correlation between \lbol\ and $p$). One should keep in mind that the two free parameters $p$ and $T_{\rm eff}(R_{\rm in})$ is not correlated (the former is determined by the optical/UV spectral slope, and the latter by the observed flux). Figure \ref{fig:disk-wind} shows the obtained relation between ${\dot{M}}_{\rm loss}$ (in units of $\rm g\ s^{-1}$) and the bolometric luminosity \lbol\ for $17$ sources with $p<0.75$ in our sample (in other $6$ sources, the stationary disk solution $p=0.75$ can be satisfied considering the uncertainties). It is clear that the resulting relation is in agreement with the observational result obtained by \citet{2015MNRAS.451.4169G}. Since the mass loss rate and bolometric luminosity derived in our work can be regarded as irrelevant to each other, as we showed above, the relation obtained here is an independent proof that verifies the results in their work. This also suggests that the observed reddened optical/UV spectra in the AGN of our sample are likely a result of the flatter radial temperature profiles of their accretion disks than the standard SSD profile. Such a flat temperature profile may be caused by mass loss in the form of outflows as the mass is accreted inward.

Apart from these two scenarios, the deviation of the radial temperature profile could also result from an accretion disk illuminated by an external radiation source \citep[e.g.,][]{2008ChJAS...8..302C}, or the irradiation of the outer areas by its inner part \citep[e.g.,][]{1993PASJ...45..443S}. By introducing an additional heating term which is more pronounced in the outer areas of the AGN disk, a flattened radial temperature profile can be obtained and lead to a reddened optical/UV spectra \citep[e.g.,][]{2002MNRAS.329..456S, 2004MNRAS.355.1080L}. However, this usually requires the disk to be warped \citep[$H_d\approx R$, $H_d$ is the vertical height of the disk at radius $R$, e.g.,][]{1997MNRAS.292..136P}. In our work, we only consider flat (geometrically thin) accretion disks ($H_d\ll R$). In this case the irradiating disk gives a radial dependence of the temperature identical to the one derived for the SSD \citep[see][Chapter 4.2.2]{2013peag.book.....N}.

\subsection{Implication to the SED modeling of AGN}
\label{sec:overestimate}

\begin{figure}
\includegraphics[width=8cm]{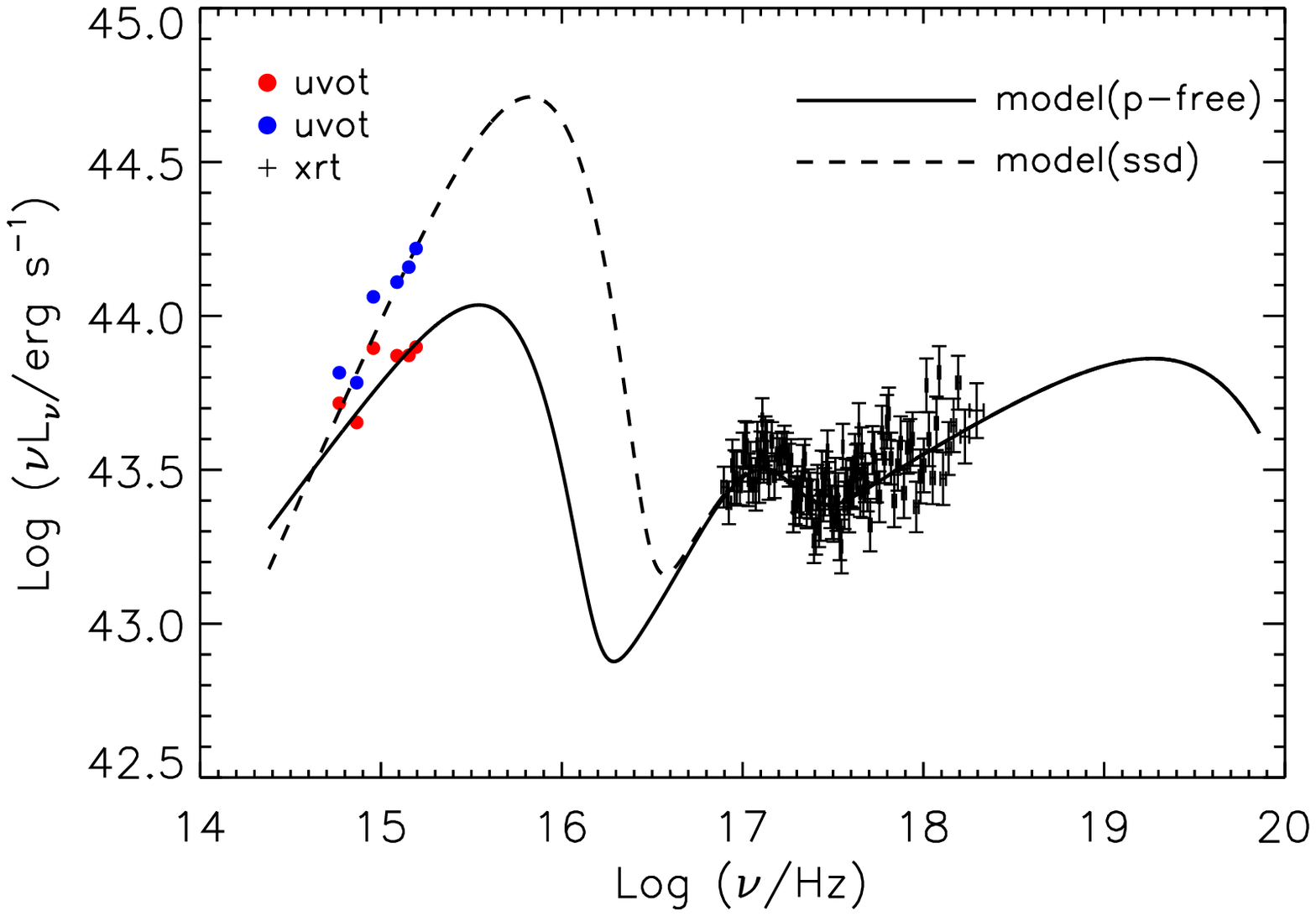}
\includegraphics[width=8cm]{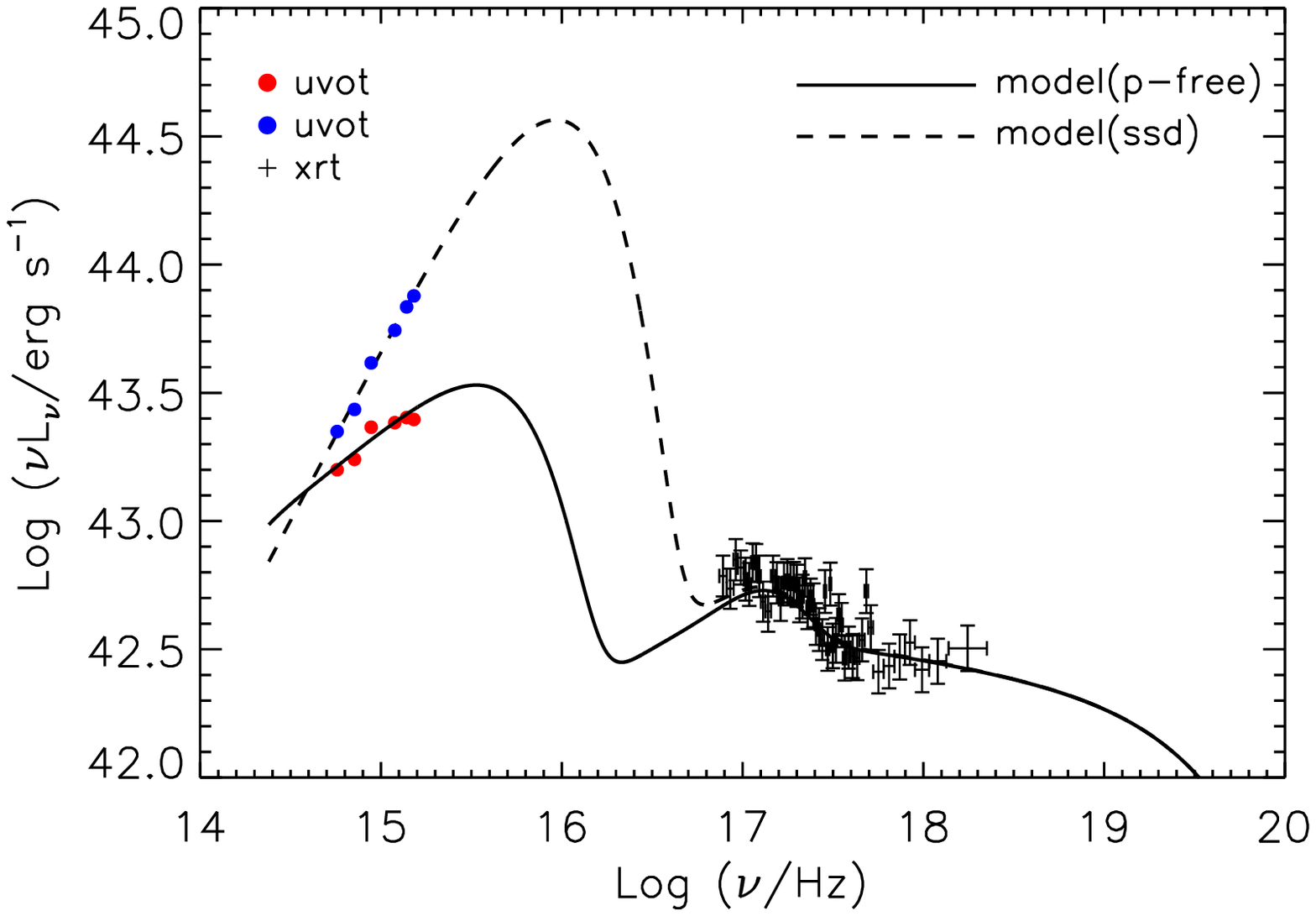}
\caption{Two examples (upper panel: SBS 1136+594, lower panel: Mrk 493) showing the difference in the intrinsic luminosities in optical/UV bands and the broadband SEDs when using two methods in optical/UV spectral fitting, i.e., the p-free disk model employed in this work and the SSD model with the intrinsic dust reddening \ebv\ set as a free parameter. The intrinsic luminosities in the 6 UVOT filters obtained by the above two methods are marked with red dots and blue dots. The two broadband SED models derived by two methods are represented by the solid line and dashed line respectively.}
\label{fig:compare}
\end{figure}

In the modeling of the AGN optical/UV continuum in previous studies, the spectrum from the SSD model is mostly adopted and assumed to be subject to dust reddening with $E(B-V)$ as a free-fitting parameter \citep[e.g.,][]{2009MNRAS.392.1124V, 2009MNRAS.399.1553V, 2016MNRAS.460..212C, 2016MNRAS.458.1839C}. While such a model can reproduce the observed optical/UV spectra reasonably well, it implies an enormous energy budget of the intrinsic source luminosity in the far- and extreme-UV regime, where the multi-temperature blackbody from the accretion disk peaks (the so-called `big blue bump'). On the other hand, the p-free disk model with a flat radial temperature profile leads to a less prominent EUV bump and a lower peak energy (due to lower \teff \ in the inner disk), and thus lower energy output. Given that the bolometric luminosity is dominated by the energy in this regime, the two models may result in somewhat different bolometric luminosities. In another word, if in reality the accretion disk is indeed better described by a non-standard radial temperature profile, the bolometric luminosities calculated assuming the SSD model must be over-estimated systematically to some extent. 

As a demonstration, a quantitive comparison between the broadband SED based on the two models is carried out for two selected objects in our sample with shallower spectral slopes, SBS 1136+594 (\aoptuv$=-0.2$) and Mrk 493 (\aoptuv$=-0.12$). Their optical/UV spectra can be fitted equally well with the SSD model which is reddened using the SMC extinction curve with \ebv\ as a free parameter. The fitted broadband SED model is plotted in Figure \ref{fig:compare}, together with those obtained assuming the p-free disk model above in this work. An excess dust reddening of \ebv\ ($0.08$ and $0.12$, respectively) is required to reproduce the observed optical/UV spectra in both sources, albeit their Balmer decrements ($<3.06$) indicate no reddening. The overall disk luminosities based on the SSD model is higher than that for the p-free disk model by a factor of $3.4$ and $5.9$ respectively. The differences are comparable to those derived in \citet{2012MNRAS.426..656S}, in which the accretion energy carried away by disk winds is considered in calculating the emergent radiation of the accretion disk, effectively similar to the approach adopted in this work. This results in an overestimation of the bolometric luminosity \lbol\ by a factor of $2.1$ and $5.0$, respectively. The difference would be even more pronounced for sources with even more flatter radial temperature profiles (for instance, for RX J1702.5+3247 in which $p\approx0.54$ the difference in \lbol\ is around $8.0$). 

We note that the uncertainties inherent in our method should be considered when assessing the significance of this difference. In this work, we implement the estimation of the uncertainties in two ways. For the optical/UV band, we employ the Monte Carlo method to estimate the uncertainty in $L_{\rm disk}$, which is caused by the systematics of various physical parameters in the p-free disk model ($M_{\rm BH}$, $i$, $a_*$). As for the X-ray (corona) luminosity, the statistical error is calculated using \textsc{xspec}/\texttt{cflux}. By the combination of these two methods, the $1\sigma$ uncertainty in \lbol\ is found to be $0.1$ -- $0.3$ dex for the sample objects (see Table \ref{tab:sed_analysis}). In this sense, the differences in \lbol\ obtained in a moderate fraction of the sample ($8/23$) are somewhat significant as they exceed the uncertainties (the differences are by a factor of $4$ -- $8$). However, we should keep in mind that there are certainly other contributions to the uncertainty in \lbol\ which are not considered in this estimation. For instance, the contribution of the soft excess below the lower end of the {\it Swift}/XRT bandpass is difficult to be robustly constrained by the blackbody model. In fact, there are various physical models in describing this mystical component, such as the thermal comptonisation model \citep[e.g.,][]{1998MNRAS.301..179M, 2012MNRAS.420.1848D} and the blurred ionized reflection model \citep[e.g.,][]{2006MNRAS.365.1067C}. As a pilot test we attempt the thermal comptonisation model \texttt{optxagnf} \citep{2012MNRAS.420.1848D} to some of the sample AGN in which a strong soft X-ray excess exists (e.g., Mrk 142, RX J1355.2+5612, RX J1702.5+3247) and find that the derived bolometric luminosities are higher than those derived in this work by a factor of $3$ -- $4$. This indicates that alternative models of the soft excess may also lead to an overestimation of \lbol. As a first study of the effects caused by using the p-free disk model, we mainly pay our attention to the distinction in the optical/UV band. Further investigations including a more physical model in describing the soft excess will be implemented in future work. 

Our results show that the traditional way in modeling the optical/UV spectra (the SSD model subject to an additional dust reddening) could indeed lead to an overestimation of the disk luminosity when the intrinsic spectra are found to be redder than that of the SSD. As the redder-than-SSD spectra have been widely observed in the literature and the disk luminosity usually dominates the overall energy budget, this also suggests that the bolometric luminosities and hence the Eddington ratios for many AGN may have been overestimated in some of the previous studies (probably by a factor of several).

\subsection{Bolometric correction factors}
\label{sec:kbolrevisit}

The bolometric correction factor $\kappa_{\nu}$ is a convenient quantity used to derive the bolometric luminosity \lbol\ (representing the accretion rate \mdot ) of an AGN based on photometric measurement in a specific waveband. The commonly-used factors are \kh\ for the X-ray band and \ko\ for the optical. There have been studies on \ko\ and \kh\ in the literature, which are found to be dependent on some other parameters in some of the studies. For instance, a strong correlation between \kh\ and \edd\ was reported \citep[e.g.,][]{2007MNRAS.381.1235V, 2009MNRAS.392.1124V, 2009MNRAS.399.1553V, 2010A&A...512A..34L}. The hard X-ray photon index \gx\ \citep[e.g.,][]{2012MNRAS.425..907J} and the optical-to-X-ray effective index \aox\ \citep[e.g.,][]{2010A&A...512A..34L, 2012A&A...539A..48M} were also suggested to correlate with \kh. As for the optical band, a correlation between \ko\ and \edd\ was reported in \citet{2007MNRAS.381.1235V} and later confirmed by \citet{2012MNRAS.425..907J}. These correlations were mostly derived from the broadband SED analysis assuming the SSD model in the optical/UV and EUV bands. As discussed above, for those sources with reddened optical/UV spectra the employment of the p-free disk model implies a much lower EUV bump and hence a smaller \lbol\ compared to the SSD model. It is therefore of interest to re-investigate the bolometric correction factors and their dependence on some of the key parameters.

Compared to \citet{2009MNRAS.399.1553V}, the increase of \kh\ with \edd\ is found to be less steep in our work (see Section \ref{sec:kbolx}). This is because a large fraction of the high-\edd\ sources (\edd$\ga0.1$) have a relatively lowered \kh\ values compared to \citet{2009MNRAS.399.1553V}, while the \kh\ values are similar for the objects with low- and intermediate- \edd\ (see Figure \ref{fig:k210edd}). Since the spectral modeling in the X-ray band is not affected by that in the optical/UV band, the decrease in \ld\ by using the p-free disk model leads to a decrease in \lbol, and hence \kh\ and \edd\ which are both proportional to \lbol. This effect, although present in nearly all AGN of our sample, affects mostly the high-\edd\ sources since they have much flatter radial temperature profiles (smaller $p$ values) than the low- and intermediate-\edd\ objects (see Figure \ref{fig:p-edd}). Compared to the results obtained based on the SSD model, this results in a lower \kh\ for the same \edd\ at the higher-\edd\ end and thus flatten the \kh-\edd\ relation. 

The optical correction factor \ko\ is found to be independent of \edd\ in our work, which differs from previous studies \citep[e.g.,][]{2007MNRAS.381.1235V, 2012MNRAS.425..907J}. The correlations reported in those studies were mainly due to an increase of \ko\ at the higher-\edd\ end (there is little or no correlation among low- and intermediate-\edd\ objects). In our case, however, this trend is weakened since the high-\edd\ objects have significantly lowered \ko\ (while the values of \ko\ in low- and intermediate-\edd\ objects are similar). Unlike \kh, the effect on \ko\ by using different disk models is more complex since both \lopt\ and \lbol\ are inherent to the model and changed (see Figure \ref{fig:compare}). To understand the difference in the high-\edd\ regime between the two models, we calculate the theoretical \ko\ values for the same range of black hole mass \mbh\ ($10^7$ -- $10^8$\msun) and Eddington ratio \edd\ ($0.1$ -- $1.0$) in the framework of the SSD model (the fraction of corona emission in the total luminosity is assumed to be $5\%$, as estimated from the hard X-ray bolometric correction for high-\edd\ sources in \citet{2009MNRAS.392.1124V}). The theoretical values of \ko\ lie between $30$ -- $80$, which are much larger than the values derived based on the p-free disk model in our work ($9$ -- $20$). This is caused by the enormous difference in the radiative energy in the EUV band from the inner disk region ($L\propto T_{\rm eff}(R_{\rm in})^4$), where the two models have very different temperatures. We note that the median value of \ko\ $\approx12$ obtained in our work is generally consistent with those given in some of the previous studies \citep[e.g.,][]{2006ApJS..166..470R}.

\section{summary}
\label{sec:sum}

We present a study on the broadband SED for a well-selected sample of Seyfert $1$ galaxies, by taking advantage of their simultaneous optical, UV and X-ray observations with the Neil Gehrels {\it Swift} Observatory. Their SDSS spectrometric data are also used to derive the optical spectral parameters in order to determine their BH masses and Balmer decrements. The sample is selected in a way that there is little or no intrinsic dust reddening as indicated by the Balmer decrement of the broad emission lines. The starlight contamination of the host galaxies is eliminated to a large extent by performing 2-D nuclear-galaxy image decomposition. 

The main feature of this work that differs from previous studies is that we employ, for the first time, an accretion disk model with a generalized radial temperature profile (rather than the standard SSD profile with the emergent spectra modified by dust reddening) to fit the observed optical/UV spectra of AGN. That is, the power-law index of the radial temperature distribution is treated as a free parameter ($T_{\rm eff}(R)\propto R^{-p}$, i.e., the p-free disk model), which is a special case when $p=0.75$ as commonly adopted in previous studies. The underlying physics of such a model concerns possible breakdown in reality of the assumptions on which the SSD model is based, e.g., the mass conservation of the inward accretion flow or the effective radiative cooling of the disk.

It is found that the majority of our sample AGN show a shallower optical/UV spectral slope ($F_{\nu}\propto\nu^{\alpha}$, $\alpha=-1.0$ -- $+0.3$ with a median of $-0.39$) after correcting for host galaxy starlight and dust reddening wherever applicable. This is broadly consistent with previous results and at odds with the prediction by the SSD solution ($\alpha=+1/3$), implying in general flatter radial temperature profiles than the latter. By fitting the p-free disk model to the optical/UV spectra, the radial temperature power-law indices are derived, $p=0.5$ -- $0.75$ (median $0.63$), smaller than the SSD value $p=0.75$ for most of our sample objects. We suggest that the reddened optical/UV spectra as observed in many AGN are at least in part determined by the intrinsically reddened spectral shape of the disk emission.

The broadband SED of our sample AGN are constructed by integrating the p-free disk model in the optical/UV band and the X-ray radiation model derived from the simultaneous X-ray observations. The averaged SEDs are also presented, whose shapes are found to be dependent strongly on the Eddington ratio. The bolometric luminosities and the bolometric correction factors in the optical \ko\ and X-ray band \kh\ are derived. It is found that, for sources with reddened optical/UV spectra typical of AGN, the SSD plus dust reddening model as commonly adopted in previous work can lead to disk luminosities higher than those derived from the p-free disk model by a factor of several, as well as the bolometric luminosities. The difference is more pronounced in AGN with higher Eddington ratios. Consequently, the values of \kh\ and \ko\ as well as their relations with \edd\ are affected. The hard X-ray bolometric correction factor \kh\ is found to be $8$ -- $15$ for \edd\ $\la0.03$ and $15$ -- $30$ for $0.03\la$ \edd\ $\la0.09$, which is similar to previous results. \kh\ at high \edd\ values (\edd\ $\ga0.1$) are relatively lower ($20$ -- $70$) than previous results. The optical bolometric correction factor \ko\ is found to be independent of the Eddington ratio \edd, with a median value of $\sim12$.

The physical origin of the flattened temperature profile in an accretion disk is complicated and far from fully understood. In this work, we suggest that some kind of mass loss process in form of disk-winds/outflows could be a possible scenario for this issue. A thorough test of this hypothesis will be needed in future work. 

\section*{Acknowledgements}

We thank the anonymous referee for very helpful suggestions. This work is supported by the National Natural Science Foundation of China (grant No.$11473035$, No.$11673026$, No.$11873054$, No.$11803047$ and No.$11773037$) and the Strategic Pioneer Program on Space Science, Chinese Academy of Sciences (grant No.XDA$15052100$). AAB is supported by UKSA. CJ is supported by the Hundred Talents Program of the Chinese Academy of Sciences. We are very grateful to S. Komossa, R. Taam, Xiaobo Dong and Hongyan Zhou for useful comments and suggestions. HQ thanks R. Vasudevan for precious comments in UVOT data procedure, and Zhu Liu for help in X-ray data analysis. This work is mainly based on observations obtained by the Neil Gehrels {\it Swift} Observatory. We acknowledge the entire {\it Swift} team for providing the data that made this work possible. We have also made use of the data products from the SDSS. This research has made use of the NASA/IPAC Extra-galactic Database (NED).





\begin{thebibliography}{}
\bibitem[\protect\citeauthoryear{{Abramowicz}, {Czerny}, {Lasota} \&
  {Szuszkiewicz}}{{Abramowicz} et~al.}{1988}]{1988ApJ...332..646A}
{Abramowicz} M.~A.,  {Czerny} B.,  {Lasota} J.~P.,    {Szuszkiewicz} E.,  1988,
  \apj, 332, 646

\bibitem[\protect\citeauthoryear{{Ai}, {Yuan}, {Zhou}, {Wang} \& {Zhang}}{{Ai}
  et~al.}{2011}]{2011ApJ...727...31A}
{Ai} Y.~L.,  {Yuan} W.,  {Zhou} H.~Y.,  {Wang} T.~G.,    {Zhang} S.~H.,  2011,
  \apj, 727, 31

\bibitem[\protect\citeauthoryear{{Antonucci}}{{Antonucci}}{1993}]{1993ARA&A..31..473A}
{Antonucci} R.,  1993, \araa, 31, 473

\bibitem[\protect\citeauthoryear{{Arnaud}}{{Arnaud}}{1996}]{1996ASPC..101...17A}
{Arnaud} K.~A.,  1996, in {Jacoby} G.~H.,  {Barnes} J.,  eds,  Astronomical
  Society of the Pacific Conference Series Vol. 101, Astronomical Data Analysis
  Software and Systems V. p.~17

\bibitem[\protect\citeauthoryear{{Baron}, {Stern}, {Poznanski} \&
  {Netzer}}{{Baron} et~al.}{2016}]{2016ApJ...832....8B}
{Baron} D.,  {Stern} J.,  {Poznanski} D.,    {Netzer} H.,  2016, \apj, 832, 8

\bibitem[\protect\citeauthoryear{{Bentz}, {Peterson}, {Netzer}, {Pogge} \&
  {Vestergaard}}{{Bentz} et~al.}{2009}]{2009ApJ...697..160B}
{Bentz} M.~C.,  {Peterson} B.~M.,  {Netzer} H.,  {Pogge} R.~W.,
  {Vestergaard} M.,  2009, \apj, 697, 160

\bibitem[\protect\citeauthoryear{{Bentz}, {Peterson}, {Pogge}, {Vestergaard} \&
  {Onken}}{{Bentz} et~al.}{2006}]{2006ApJ...644..133B}
{Bentz} M.~C.,  {Peterson} B.~M.,  {Pogge} R.~W.,  {Vestergaard} M.,    {Onken}
  C.~A.,  2006, \apj, 644, 133

\bibitem[\protect\citeauthoryear{{Breeveld} et~al.,}{{Breeveld}
  et~al.}{2010}]{2010MNRAS.406.1687B}
{Breeveld} A.~A.  et~al., 2010, \mnras, 406, 1687

\bibitem[\protect\citeauthoryear{{Burrows} et~al.,}{{Burrows}
  et~al.}{2005}]{2005SSRv..120..165B}
{Burrows} D.~N.  et~al., 2005, \ssr, 120, 165

\bibitem[\protect\citeauthoryear{{Capellupo}, {Netzer}, {Lira}, {Trakhtenbrot}
  \& {Mej{\'{\i}}a-Restrepo}}{{Capellupo} et~al.}{2016}]{2016MNRAS.460..212C}
{Capellupo} D.~M.,  {Netzer} H.,  {Lira} P.,  {Trakhtenbrot} B.,
  {Mej{\'{\i}}a-Restrepo} J.,  2016, \mnras, 460, 212

\bibitem[\protect\citeauthoryear{{Castell{\'o}-Mor}, {Netzer} \&
  {Kaspi}}{{Castell{\'o}-Mor} et~al.}{2016}]{2016MNRAS.458.1839C}
{Castell{\'o}-Mor} N.,  {Netzer} H.,    {Kaspi} S.,  2016, \mnras, 458, 1839

\bibitem[\protect\citeauthoryear{{Crummy}, {Fabian}, {Gallo} \&
  {Ross}}{{Crummy} et~al.}{2006}]{2006MNRAS.365.1067C}
{Crummy} J.,  {Fabian} A.~C.,  {Gallo} L.,    {Ross} R.~R.,  2006, \mnras, 365,
  1067

\bibitem[\protect\citeauthoryear{{Czerny}}{{Czerny}}{2007}]{2007ASPC..373..586C}
{Czerny} B.,  2007, in {Ho} L.~C.,  {Wang} J.-W.,  eds,  Astronomical Society
  of the Pacific Conference Series Vol. 373, The Central Engine of Active
  Galactic Nuclei. p.~586

\bibitem[\protect\citeauthoryear{{Czerny}, {Goosmann} \& {Janiuk}}{{Czerny}
  et~al.}{2008}]{2008ChJAS...8..302C}
{Czerny} B.,  {Goosmann} R.,    {Janiuk} A.,  2008, Chinese Journal of
  Astronomy and Astrophysics Supplement, 8, 302

\bibitem[\protect\citeauthoryear{{Czerny}, {Hryniewicz}, {Niko{\l}ajuk} \&
  {S{\c a}dowski}}{{Czerny} et~al.}{2011}]{2011MNRAS.415.2942C}
{Czerny} B.,  {Hryniewicz} K.,  {Niko{\l}ajuk} M.,    {S{\c a}dowski} A.,
  2011, \mnras, 415, 2942

\bibitem[\protect\citeauthoryear{{Dadina}}{{Dadina}}{2008}]{2008A&A...485..417D}
{Dadina} M.,  2008, \aap, 485, 417

\bibitem[\protect\citeauthoryear{{Davis} \& {El-Abd}}{{Davis} \&
  {El-Abd}}{2018}]{2018arXiv180905134D}
{Davis} S.~W.,  {El-Abd} S.,  2018, ArXiv e-prints

\bibitem[\protect\citeauthoryear{{Davis}, {Woo} \& {Blaes}}{{Davis}
  et~al.}{2007}]{2007ApJ...668..682D}
{Davis} S.~W.,  {Woo} J.-H.,    {Blaes} O.~M.,  2007, \apj, 668, 682

\bibitem[\protect\citeauthoryear{{Done}, {Davis}, {Jin}, {Blaes} \&
  {Ward}}{{Done} et~al.}{2012}]{2012MNRAS.420.1848D}
{Done} C.,  {Davis} S.~W.,  {Jin} C.,  {Blaes} O.,    {Ward} M.,  2012, \mnras,
  420, 1848

\bibitem[\protect\citeauthoryear{{Done} \& {Jin}}{{Done} \&
  {Jin}}{2016}]{2016MNRAS.460.1716D}
{Done} C.,  {Jin} C.,  2016, \mnras, 460, 1716

\bibitem[\protect\citeauthoryear{{Dong}, {Wang}, {Wang}, {Yuan}, {Zhou}, {Dai}
  \& {Zhang}}{{Dong} et~al.}{2008}]{2008MNRAS.383..581D}
{Dong} X.,  {Wang} T.,  {Wang} J.,  {Yuan} W.,  {Zhou} H.,  {Dai} H.,
  {Zhang} K.,  2008, \mnras, 383, 581

\bibitem[\protect\citeauthoryear{{Dong}, {Zhou}, {Wang}, {Wang}, {Li} \&
  {Zhou}}{{Dong} et~al.}{2005}]{2005ApJ...620..629D}
{Dong} X.-B.,  {Zhou} H.-Y.,  {Wang} T.-G.,  {Wang} J.-X.,  {Li} C.,    {Zhou}
  Y.-Y.,  2005, \apj, 620, 629

\bibitem[\protect\citeauthoryear{{Elvis} et~al.,}{{Elvis}
  et~al.}{1994}]{1994ApJS...95....1E}
{Elvis} M.  et~al., 1994, \apjs, 95, 1

\bibitem[\protect\citeauthoryear{{Fanali}, {Caccianiga}, {Severgnini}, {Della
  Ceca}, {Marchese}, {Carrera}, {Corral} \& {Mateos}}{{Fanali}
  et~al.}{2013}]{2013MNRAS.433..648F}
{Fanali} R.,  {Caccianiga} A.,  {Severgnini} P.,  {Della Ceca} R.,  {Marchese}
  E.,  {Carrera} F.~J.,  {Corral} A.,    {Mateos} S.,  2013, \mnras, 433, 648

\bibitem[\protect\citeauthoryear{{Frank}, {King} \& {Raine}}{{Frank}
  et~al.}{2002}]{2002apa..book.....F}
{Frank} J.,  {King} A.,    {Raine} D.~J.,  2002, {Accretion Power in
  Astrophysics: Third Edition}

\bibitem[\protect\citeauthoryear{{Ganguly}, {Brotherton}, {Cales}, {Scoggins},
  {Shang} \& {Vestergaard}}{{Ganguly} et~al.}{2007}]{2007ApJ...665..990G}
{Ganguly} R.,  {Brotherton} M.~S.,  {Cales} S.,  {Scoggins} B.,  {Shang} Z.,
  {Vestergaard} M.,  2007, \apj, 665, 990

\bibitem[\protect\citeauthoryear{{Gaskell}}{{Gaskell}}{2008}]{2008RMxAC..32....1G}
{Gaskell} C.~M.,  2008, in Revista Mexicana de Astronomia y Astrofisica
  Conference Series. pp 1--11

\bibitem[\protect\citeauthoryear{{Gaskell}}{{Gaskell}}{2017}]{2017MNRAS.467..226G}
{Gaskell} C.~M.,  2017, \mnras, 467, 226

\bibitem[\protect\citeauthoryear{{Gaskell}, {Goosmann}, {Antonucci} \&
  {Whysong}}{{Gaskell} et~al.}{2004}]{2004ApJ...616..147G}
{Gaskell} C.~M.,  {Goosmann} R.~W.,  {Antonucci} R.~R.~J.,    {Whysong} D.~H.,
  2004, \apj, 616, 147

\bibitem[\protect\citeauthoryear{{Gaskell}, {Klimek} \& {Nazarova}}{{Gaskell}
  et~al.}{2007}]{2007arXiv0711.1025G}
{Gaskell} C.~M.,  {Klimek} E.~S.,    {Nazarova} L.~S.,  2007, ArXiv e-prints

\bibitem[\protect\citeauthoryear{{Gebhardt} et~al.,}{{Gebhardt}
  et~al.}{2000}]{2000ApJ...543L...5G}
{Gebhardt} K.  et~al., 2000, \apjl, 543, L5

\bibitem[\protect\citeauthoryear{{Gehrels} et~al.,}{{Gehrels}
  et~al.}{2004}]{2004ApJ...611.1005G}
{Gehrels} N.  et~al., 2004, \apj, 611, 1005

\bibitem[\protect\citeauthoryear{{Gofford}, {Reeves}, {McLaughlin}, {Braito},
  {Turner}, {Tombesi} \& {Cappi}}{{Gofford} et~al.}{2015}]{2015MNRAS.451.4169G}
{Gofford} J.,  {Reeves} J.~N.,  {McLaughlin} D.~E.,  {Braito} V.,  {Turner}
  T.~J.,  {Tombesi} F.,    {Cappi} M.,  2015, \mnras, 451, 4169

\bibitem[\protect\citeauthoryear{{Gordon}, {Clayton}, {Misselt}, {Landolt} \&
  {Wolff}}{{Gordon} et~al.}{2003}]{2003ApJ...594..279G}
{Gordon} K.~D.,  {Clayton} G.~C.,  {Misselt} K.~A.,  {Landolt} A.~U.,
  {Wolff} M.~J.,  2003, \apj, 594, 279

\bibitem[\protect\citeauthoryear{{Greene} \& {Ho}}{{Greene} \&
  {Ho}}{2006}]{2006ApJ...641L..21G}
{Greene} J.~E.,  {Ho} L.~C.,  2006, \apjl, 641, L21

\bibitem[\protect\citeauthoryear{{Grier} et~al.,}{{Grier}
  et~al.}{2013}]{2013ApJ...773...90G}
{Grier} C.~J.  et~al., 2013, \apj, 773, 90

\bibitem[\protect\citeauthoryear{{Grupe}, {Komossa}, {Leighly} \&
  {Page}}{{Grupe} et~al.}{2010}]{2010ApJS..187...64G}
{Grupe} D.,  {Komossa} S.,  {Leighly} K.~M.,    {Page} K.~L.,  2010, \apjs,
  187, 64

\bibitem[\protect\citeauthoryear{{Hirano}, {Kitamoto}, {Yamada}, {Mineshige} \&
  {Fukue}}{{Hirano} et~al.}{1995}]{1995ApJ...446..350H}
{Hirano} A.,  {Kitamoto} S.,  {Yamada} T.~T.,  {Mineshige} S.,    {Fukue} J.,
  1995, \apj, 446, 350

\bibitem[\protect\citeauthoryear{{Hopkins} et~al.,}{{Hopkins}
  et~al.}{2004}]{2004AJ....128.1112H}
{Hopkins} P.~F.  et~al., 2004, \aj, 128, 1112

\bibitem[\protect\citeauthoryear{{Ichimaru}}{{Ichimaru}}{1977}]{1977ApJ...214..840I}
{Ichimaru} S.,  1977, \apj, 214, 840

\bibitem[\protect\citeauthoryear{{Jin}, {Done}, {Ward} \& {Gardner}}{{Jin}
  et~al.}{2017}]{2017MNRAS.471..706J}
{Jin} C.,  {Done} C.,  {Ward} M.,    {Gardner} E.,  2017, \mnras, 471, 706

\bibitem[\protect\citeauthoryear{{Jin}, {Ward} \& {Done}}{{Jin}
  et~al.}{2012}]{2012MNRAS.425..907J}
{Jin} C.,  {Ward} M.,    {Done} C.,  2012, \mnras, 425, 907

\bibitem[\protect\citeauthoryear{{Jun} et~al.,}{{Jun}
  et~al.}{2015}]{2015ApJ...806..109J}
{Jun} H.~D.  et~al., 2015, \apj, 806, 109

\bibitem[\protect\citeauthoryear{{Kaspi}, {Smith}, {Netzer}, {Maoz}, {Jannuzi}
  \& {Giveon}}{{Kaspi} et~al.}{2000}]{2000ApJ...533..631K}
{Kaspi} S.,  {Smith} P.~S.,  {Netzer} H.,  {Maoz} D.,  {Jannuzi} B.~T.,
  {Giveon} U.,  2000, \apj, 533, 631

\bibitem[\protect\citeauthoryear{{Kato}, {Fukue} \& {Mineshige}}{{Kato}
  et~al.}{2008}]{2008bhad.book.....K}
{Kato} S.,  {Fukue} J.,    {Mineshige} S.,  2008, {Black-Hole Accretion Disks
  --- Towards a New Paradigm ---}

\bibitem[\protect\citeauthoryear{{King} \& {Pounds}}{{King} \&
  {Pounds}}{2015}]{2015ARA&A..53..115K}
{King} A.,  {Pounds} K.,  2015, \araa, 53, 115

\bibitem[\protect\citeauthoryear{{Koratkar} \& {Blaes}}{{Koratkar} \&
  {Blaes}}{1999}]{1999PASP..111....1K}
{Koratkar} A.,  {Blaes} O.,  1999, \pasp, 111, 1

\bibitem[\protect\citeauthoryear{{Koss}, {Mushotzky}, {Veilleux}, {Winter},
  {Baumgartner}, {Tueller}, {Gehrels} \& {Valencic}}{{Koss}
  et~al.}{2011}]{2011ApJ...739...57K}
{Koss} M.,  {Mushotzky} R.,  {Veilleux} S.,  {Winter} L.~M.,  {Baumgartner} W.,
   {Tueller} J.,  {Gehrels} N.,    {Valencic} L.,  2011, \apj, 739, 57

\bibitem[\protect\citeauthoryear{{Krawczyk}, {Richards}, {Gallagher},
  {Leighly}, {Hewett}, {Ross} \& {Hall}}{{Krawczyk}
  et~al.}{2015}]{2015AJ....149..203K}
{Krawczyk} C.~M.,  {Richards} G.~T.,  {Gallagher} S.~C.,  {Leighly} K.~M.,
  {Hewett} P.~C.,  {Ross} N.~P.,    {Hall} P.~B.,  2015, \aj, 149, 203

\bibitem[\protect\citeauthoryear{{Krawczyk}, {Richards}, {Mehta}, {Vogeley},
  {Gallagher}, {Leighly}, {Ross} \& {Schneider}}{{Krawczyk}
  et~al.}{2013}]{2013ApJS..206....4K}
{Krawczyk} C.~M.,  {Richards} G.~T.,  {Mehta} S.~S.,  {Vogeley} M.~S.,
  {Gallagher} S.~C.,  {Leighly} K.~M.,  {Ross} N.~P.,    {Schneider} D.~P.,
  2013, \apjs, 206, 4

\bibitem[\protect\citeauthoryear{{Kubota}, {Ebisawa}, {Makishima} \&
  {Nakazawa}}{{Kubota} et~al.}{2005}]{2005ApJ...631.1062K}
{Kubota} A.,  {Ebisawa} K.,  {Makishima} K.,    {Nakazawa} K.,  2005, \apj,
  631, 1062

\bibitem[\protect\citeauthoryear{{Kubota} \& {Makishima}}{{Kubota} \&
  {Makishima}}{2004}]{2004ApJ...601..428K}
{Kubota} A.,  {Makishima} K.,  2004, \apj, 601, 428

\bibitem[\protect\citeauthoryear{{La Mura}, {Popovi{\'c}}, {Ciroi}, {Rafanelli}
  \& {Ili{\'c}}}{{La Mura} et~al.}{2007}]{2007ApJ...671..104L}
{La Mura} G.,  {Popovi{\'c}} L.~{\v C}.,  {Ciroi} S.,  {Rafanelli} P.,
  {Ili{\'c}} D.,  2007, \apj, 671, 104

\bibitem[\protect\citeauthoryear{{Laor} \& {Brandt}}{{Laor} \&
  {Brandt}}{2002}]{2002ApJ...569..641L}
{Laor} A.,  {Brandt} W.~N.,  2002, \apj, 569, 641

\bibitem[\protect\citeauthoryear{{Laor} \& {Davis}}{{Laor} \&
  {Davis}}{2014}]{2014MNRAS.438.3024L}
{Laor} A.,  {Davis} S.~W.,  2014, \mnras, 438, 3024

\bibitem[\protect\citeauthoryear{{Laor} \& {Netzer}}{{Laor} \&
  {Netzer}}{1989}]{1989MNRAS.238..897L}
{Laor} A.,  {Netzer} H.,  1989, \mnras, 238, 897

\bibitem[\protect\citeauthoryear{{Lawther}, {Vestergaard}, {Raimundo} \&
  {Grupe}}{{Lawther} et~al.}{2017}]{2017MNRAS.467.4674L}
{Lawther} D.,  {Vestergaard} M.,  {Raimundo} S.,    {Grupe} D.,  2017, \mnras,
  467, 4674

\bibitem[\protect\citeauthoryear{{Li}}{{Li}}{2007}]{2007ASPC..373..561L}
{Li} A.,  2007, in {Ho} L.~C.,  {Wang} J.-W.,  eds,  Astronomical Society of
  the Pacific Conference Series Vol. 373, The Central Engine of Active Galactic
  Nuclei. p.~561

\bibitem[\protect\citeauthoryear{{Liu} \& {Taam}}{{Liu} \&
  {Taam}}{2009}]{2009ApJ...707..233L}
{Liu} B.~F.,  {Taam} R.~E.,  2009, \apj, 707, 233

\bibitem[\protect\citeauthoryear{{Liu}, {Taam}, {Qiao} \& {Yuan}}{{Liu}
  et~al.}{2015}]{2015ApJ...806..223L}
{Liu} B.~F.,  {Taam} R.~E.,  {Qiao} E.,    {Yuan} W.,  2015, \apj, 806, 223

\bibitem[\protect\citeauthoryear{{Liu}, {Yuan}, {Dong}, {Zhou} \& {Liu}}{{Liu}
  et~al.}{2018}]{2018ApJS..235...40L}
{Liu} H.-Y.,  {Yuan} W.,  {Dong} X.-B.,  {Zhou} H.,    {Liu} W.-J.,  2018,
  \apjs, 235, 40

\bibitem[\protect\citeauthoryear{{Liu} et~al.,}{{Liu}
  et~al.}{2016}]{2016ApJ...822...64L}
{Liu} W.-J.  et~al., 2016, \apj, 822, 64

\bibitem[\protect\citeauthoryear{{Loska}, {Czerny} \& {Szczerba}}{{Loska}
  et~al.}{2004}]{2004MNRAS.355.1080L}
{Loska} Z.,  {Czerny} B.,    {Szczerba} R.,  2004, \mnras, 355, 1080

\bibitem[\protect\citeauthoryear{{Lusso} et~al.,}{{Lusso}
  et~al.}{2010}]{2010A&A...512A..34L}
{Lusso} E.  et~al., 2010, \aap, 512, A34

\bibitem[\protect\citeauthoryear{{Magdziarz}, {Blaes}, {Zdziarski}, {Johnson}
  \& {Smith}}{{Magdziarz} et~al.}{1998}]{1998MNRAS.301..179M}
{Magdziarz} P.,  {Blaes} O.~M.,  {Zdziarski} A.~A.,  {Johnson} W.~N.,
  {Smith} D.~A.,  1998, \mnras, 301, 179

\bibitem[\protect\citeauthoryear{{Malizia}, {Molina}, {Bassani}, {Stephen},
  {Bazzano}, {Ubertini} \& {Bird}}{{Malizia}
  et~al.}{2014}]{2014ApJ...782L..25M}
{Malizia} A.,  {Molina} M.,  {Bassani} L.,  {Stephen} J.~B.,  {Bazzano} A.,
  {Ubertini} P.,    {Bird} A.~J.,  2014, \apjl, 782, L25

\bibitem[\protect\citeauthoryear{{Marchese}, {Della Ceca}, {Caccianiga},
  {Severgnini}, {Corral} \& {Fanali}}{{Marchese}
  et~al.}{2012}]{2012A&A...539A..48M}
{Marchese} E.,  {Della Ceca} R.,  {Caccianiga} A.,  {Severgnini} P.,  {Corral}
  A.,    {Fanali} R.,  2012, \aap, 539, A48

\bibitem[\protect\citeauthoryear{{Marin}}{{Marin}}{2014}]{2014MNRAS.441..551M}
{Marin} F.,  2014, \mnras, 441, 551

\bibitem[\protect\citeauthoryear{{Markwardt}}{{Markwardt}}{2009}]{2009ASPC..411..251M}
{Markwardt} C.~B.,  2009, in {Bohlender} D.~A.,  {Durand} D.,   {Dowler} P.,
  eds,  Astronomical Society of the Pacific Conference Series Vol. 411,
  Astronomical Data Analysis Software and Systems XVIII. p.~251

\bibitem[\protect\citeauthoryear{{Mason} et~al.,}{{Mason}
  et~al.}{2001}]{2001A&A...365L..36M}
{Mason} K.~O.  et~al., 2001, \aap, 365, L36

\bibitem[\protect\citeauthoryear{{Mineshige}, {Hirano}, {Kitamoto}, {Yamada} \&
  {Fukue}}{{Mineshige} et~al.}{1994}]{1994ApJ...426..308M}
{Mineshige} S.,  {Hirano} A.,  {Kitamoto} S.,  {Yamada} T.~T.,    {Fukue} J.,
  1994, \apj, 426, 308

\bibitem[\protect\citeauthoryear{{Mineshige}, {Kawaguchi}, {Takeuchi} \&
  {Hayashida}}{{Mineshige} et~al.}{2000}]{2000PASJ...52..499M}
{Mineshige} S.,  {Kawaguchi} T.,  {Takeuchi} M.,    {Hayashida} K.,  2000,
  \pasj, 52, 499

\bibitem[\protect\citeauthoryear{{Monet}}{{Monet}}{1998}]{1998AAS...19312003M}
{Monet} D.~G.,  1998, in American Astronomical Society Meeting Abstracts.
  p.~1427

\bibitem[\protect\citeauthoryear{{Narayan} \& {Yi}}{{Narayan} \&
  {Yi}}{1994}]{1994ApJ...428L..13N}
{Narayan} R.,  {Yi} I.,  1994, \apjl, 428, L13

\bibitem[\protect\citeauthoryear{{Narayan} \& {Yi}}{{Narayan} \&
  {Yi}}{1995a}]{1995ApJ...444..231N}
{Narayan} R.,  {Yi} I.,  1995a, \apj, 444, 231

\bibitem[\protect\citeauthoryear{{Narayan} \& {Yi}}{{Narayan} \&
  {Yi}}{1995b}]{1995ApJ...452..710N}
{Narayan} R.,  {Yi} I.,  1995b, \apj, 452, 710

\bibitem[\protect\citeauthoryear{{Netzer}}{{Netzer}}{2013}]{2013peag.book.....N}
{Netzer} H.,  2013, {The Physics and Evolution of Active Galactic Nuclei}

\bibitem[\protect\citeauthoryear{{Netzer}}{{Netzer}}{2015}]{2015ARA&A..53..365N}
{Netzer} H.,  2015, \araa, 53, 365

\bibitem[\protect\citeauthoryear{{Nomura} \& {Ohsuga}}{{Nomura} \&
  {Ohsuga}}{2017}]{2017MNRAS.465.2873N}
{Nomura} M.,  {Ohsuga} K.,  2017, \mnras, 465, 2873

\bibitem[\protect\citeauthoryear{{Nomura}, {Ohsuga} \& {Done}}{{Nomura}
  et~al.}{2018}]{2018arXiv181101966N}
{Nomura} M.,  {Ohsuga} K.,    {Done} C.,  2018, arXiv e-prints, p.
  arXiv:1811.01966

\bibitem[\protect\citeauthoryear{{Peng}, {Ho}, {Impey} \& {Rix}}{{Peng}
  et~al.}{2002}]{2002AJ....124..266P}
{Peng} C.~Y.,  {Ho} L.~C.,  {Impey} C.~D.,    {Rix} H.-W.,  2002, \aj, 124, 266

\bibitem[\protect\citeauthoryear{{Poole} et~al.,}{{Poole}
  et~al.}{2008}]{2008MNRAS.383..627P}
{Poole} T.~S.  et~al., 2008, \mnras, 383, 627

\bibitem[\protect\citeauthoryear{{Pringle}}{{Pringle}}{1981}]{1981ARA&A..19..137P}
{Pringle} J.~E.,  1981, \araa, 19, 137

\bibitem[\protect\citeauthoryear{{Pringle}}{{Pringle}}{1997}]{1997MNRAS.292..136P}
{Pringle} J.~E.,  1997, \mnras, 292, 136

\bibitem[\protect\citeauthoryear{{Pringle} \& {Rees}}{{Pringle} \&
  {Rees}}{1972}]{1972A&A....21....1P}
{Pringle} J.~E.,  {Rees} M.~J.,  1972, \aap, 21, 1

\bibitem[\protect\citeauthoryear{{Qiao}, {Liu}, {Panessa} \& {Liu}}{{Qiao}
  et~al.}{2013}]{2013ApJ...777..102Q}
{Qiao} E.,  {Liu} B.~F.,  {Panessa} F.,    {Liu} J.~Y.,  2013, \apj, 777, 102

\bibitem[\protect\citeauthoryear{{Richards} et~al.,}{{Richards}
  et~al.}{2006}]{2006ApJS..166..470R}
{Richards} G.~T.  et~al., 2006, \apjs, 166, 470

\bibitem[\protect\citeauthoryear{{Roming} et~al.,}{{Roming}
  et~al.}{2005}]{2005SSRv..120...95R}
{Roming} P.~W.~A.  et~al., 2005, \ssr, 120, 95

\bibitem[\protect\citeauthoryear{{Salvesen}, {Miller}, {Reis} \&
  {Begelman}}{{Salvesen} et~al.}{2013}]{2013MNRAS.431.3510S}
{Salvesen} G.,  {Miller} J.~M.,  {Reis} R.~C.,    {Begelman} M.~C.,  2013,
  \mnras, 431, 3510

\bibitem[\protect\citeauthoryear{{Sanbuichi}, {Yamada} \& {Fukue}}{{Sanbuichi}
  et~al.}{1993}]{1993PASJ...45..443S}
{Sanbuichi} K.,  {Yamada} T.~T.,    {Fukue} J.,  1993, \pasj, 45, 443

\bibitem[\protect\citeauthoryear{{Schlegel}, {Finkbeiner} \&
  {Davis}}{{Schlegel} et~al.}{1998}]{1998ApJ...500..525S}
{Schlegel} D.~J.,  {Finkbeiner} D.~P.,    {Davis} M.,  1998, \apj, 500, 525

\bibitem[\protect\citeauthoryear{{Schnorr-M{\"u}ller}
  et~al.,}{{Schnorr-M{\"u}ller} et~al.}{2016}]{2016MNRAS.462.3570S}
{Schnorr-M{\"u}ller} A.  et~al., 2016, \mnras, 462, 3570

\bibitem[\protect\citeauthoryear{{Scott}, {Kriss}, {Brotherton}, {Green},
  {Hutchings}, {Shull} \& {Zheng}}{{Scott} et~al.}{2004}]{2004ApJ...615..135S}
{Scott} J.~E.,  {Kriss} G.~A.,  {Brotherton} M.,  {Green} R.~F.,  {Hutchings}
  J.,  {Shull} J.~M.,    {Zheng} W.,  2004, \apj, 615, 135

\bibitem[\protect\citeauthoryear{{Selsing}, {Fynbo}, {Christensen} \&
  {Krogager}}{{Selsing} et~al.}{2016}]{2016A&A...585A..87S}
{Selsing} J.,  {Fynbo} J.~P.~U.,  {Christensen} L.,    {Krogager} J.-K.,  2016,
  \aap, 585, A87

\bibitem[\protect\citeauthoryear{{Shakura} \& {Sunyaev}}{{Shakura} \&
  {Sunyaev}}{1973}]{1973A&A....24..337S}
{Shakura} N.~I.,  {Sunyaev} R.~A.,  1973, \aap, 24, 337

\bibitem[\protect\citeauthoryear{{Shang} et~al.,}{{Shang}
  et~al.}{2005}]{2005ApJ...619...41S}
{Shang} Z.  et~al., 2005, \apj, 619, 41

\bibitem[\protect\citeauthoryear{{Shemmer}, {Brandt}, {Netzer}, {Maiolino} \&
  {Kaspi}}{{Shemmer} et~al.}{2006}]{2006ApJ...646L..29S}
{Shemmer} O.,  {Brandt} W.~N.,  {Netzer} H.,  {Maiolino} R.,    {Kaspi} S.,
  2006, \apjl, 646, L29

\bibitem[\protect\citeauthoryear{{Shields}}{{Shields}}{1978}]{1978Natur.272..706S}
{Shields} G.~A.,  1978, \nat, 272, 706

\bibitem[\protect\citeauthoryear{{Shimura} \& {Takahara}}{{Shimura} \&
  {Takahara}}{1995}]{1995ApJ...445..780S}
{Shimura} T.,  {Takahara} F.,  1995, \apj, 445, 780

\bibitem[\protect\citeauthoryear{{Shull}, {Stevans} \& {Danforth}}{{Shull}
  et~al.}{2012}]{2012ApJ...752..162S}
{Shull} J.~M.,  {Stevans} M.,    {Danforth} C.~W.,  2012, \apj, 752, 162

\bibitem[\protect\citeauthoryear{{Slone} \& {Netzer}}{{Slone} \&
  {Netzer}}{2012}]{2012MNRAS.426..656S}
{Slone} O.,  {Netzer} H.,  2012, \mnras, 426, 656

\bibitem[\protect\citeauthoryear{{Soria} \& {Puchnarewicz}}{{Soria} \&
  {Puchnarewicz}}{2002}]{2002MNRAS.329..456S}
{Soria} R.,  {Puchnarewicz} E.~M.,  2002, \mnras, 329, 456

\bibitem[\protect\citeauthoryear{{Stevans}, {Shull}, {Danforth} \&
  {Tilton}}{{Stevans} et~al.}{2014}]{2014ApJ...794...75S}
{Stevans} M.~L.,  {Shull} J.~M.,  {Danforth} C.~W.,    {Tilton} E.~M.,  2014,
  \apj, 794, 75

\bibitem[\protect\citeauthoryear{{Tananbaum} et~al.,}{{Tananbaum}
  et~al.}{1979}]{1979ApJ...234L...9T}
{Tananbaum} H.  et~al., 1979, \apjl, 234, L9

\bibitem[\protect\citeauthoryear{{Tombesi}, {Cappi}, {Reeves}, {Nemmen},
  {Braito}, {Gaspari} \& {Reynolds}}{{Tombesi}
  et~al.}{2013}]{2013MNRAS.430.1102T}
{Tombesi} F.,  {Cappi} M.,  {Reeves} J.~N.,  {Nemmen} R.~S.,  {Braito} V.,
  {Gaspari} M.,    {Reynolds} C.~S.,  2013, \mnras, 430, 1102

\bibitem[\protect\citeauthoryear{{Vanden Berk} et~al.,}{{Vanden Berk}
  et~al.}{2001}]{2001AJ....122..549V}
{Vanden Berk} D.~E.  et~al., 2001, \aj, 122, 549

\bibitem[\protect\citeauthoryear{{Vasudevan} \& {Fabian}}{{Vasudevan} \&
  {Fabian}}{2007}]{2007MNRAS.381.1235V}
{Vasudevan} R.~V.,  {Fabian} A.~C.,  2007, \mnras, 381, 1235

\bibitem[\protect\citeauthoryear{{Vasudevan} \& {Fabian}}{{Vasudevan} \&
  {Fabian}}{2009}]{2009MNRAS.392.1124V}
{Vasudevan} R.~V.,  {Fabian} A.~C.,  2009, \mnras, 392, 1124

\bibitem[\protect\citeauthoryear{{Vasudevan}, {Mushotzky}, {Winter} \&
  {Fabian}}{{Vasudevan} et~al.}{2009}]{2009MNRAS.399.1553V}
{Vasudevan} R.~V.,  {Mushotzky} R.~F.,  {Winter} L.~M.,    {Fabian} A.~C.,
  2009, \mnras, 399, 1553

\bibitem[\protect\citeauthoryear{{V{\'e}ron-Cetty}, {Joly} \&
  {V{\'e}ron}}{{V{\'e}ron-Cetty} et~al.}{2004}]{2004A&A...417..515V}
{V{\'e}ron-Cetty} M.-P.,  {Joly} M.,    {V{\'e}ron} P.,  2004, \aap, 417, 515

\bibitem[\protect\citeauthoryear{{Vestergaard} \& {Peterson}}{{Vestergaard} \&
  {Peterson}}{2006}]{2006ApJ...641..689V}
{Vestergaard} M.,  {Peterson} B.~M.,  2006, \apj, 641, 689

\bibitem[\protect\citeauthoryear{{Wang} et~al.,}{{Wang}
  et~al.}{2009}]{2009ApJ...707.1334W}
{Wang} J.-G.  et~al., 2009, \apj, 707, 1334

\bibitem[\protect\citeauthoryear{{Wang}, {Szuszkiewicz}, {Lu} \& {Zhou}}{{Wang}
  et~al.}{1999}]{1999ApJ...522..839W}
{Wang} J.-M.,  {Szuszkiewicz} E.,  {Lu} F.-J.,    {Zhou} Y.-Y.,  1999, \apj,
  522, 839

\bibitem[\protect\citeauthoryear{{Wang}, {Zhou}, {Grupe}, {Yuan}, {Dong} \&
  {Lu}}{{Wang} et~al.}{2009}]{2009AJ....137.4002W}
{Wang} T.~G.,  {Zhou} H.~Y.,  {Grupe} D.,  {Yuan} W.,  {Dong} X.~B.,    {Lu}
  H.~L.,  2009, \aj, 137, 4002

\bibitem[\protect\citeauthoryear{{Watarai} \& {Fukue}}{{Watarai} \&
  {Fukue}}{1999}]{1999PASJ...51..725W}
{Watarai} K.-y.,  {Fukue} J.,  1999, \pasj, 51, 725

\bibitem[\protect\citeauthoryear{{Watarai} \& {Mineshige}}{{Watarai} \&
  {Mineshige}}{2001}]{2001PASJ...53..915W}
{Watarai} K.-Y.,  {Mineshige} S.,  2001, \pasj, 53, 915

\bibitem[\protect\citeauthoryear{{Xie}, {Shao}, {Shen}, {Liu} \& {Li}}{{Xie}
  et~al.}{2016}]{2016ApJ...824...38X}
{Xie} X.,  {Shao} Z.,  {Shen} S.,  {Liu} H.,    {Li} L.,  2016, \apj, 824, 38

\bibitem[\protect\citeauthoryear{{Yamada} \& {Fukue}}{{Yamada} \&
  {Fukue}}{1993}]{1993PASJ...45...97Y}
{Yamada} T.,  {Fukue} J.,  1993, \pasj, 45, 97

\bibitem[\protect\citeauthoryear{{You}, {Straub}, {Czerny}, {Sobolewska},
  {R{\'o}{\.z}a{\'n}ska}, {Bursa} \& {Dov{\v c}iak}}{{You}
  et~al.}{2016}]{2016ApJ...821..104Y}
{You} B.,  {Straub} O.,  {Czerny} B.,  {Sobolewska} M.,  {R{\'o}{\.z}a{\'n}ska}
  A.,  {Bursa} M.,    {Dov{\v c}iak} M.,  2016, \apj, 821, 104

\bibitem[\protect\citeauthoryear{{Yuan} \& {Narayan}}{{Yuan} \&
  {Narayan}}{2014}]{2014ARA&A..52..529Y}
{Yuan} F.,  {Narayan} R.,  2014, \araa, 52, 529

\bibitem[\protect\citeauthoryear{{Zhang}, {Wang}, {Wang}, {Xing}, {Zhang},
  {Zhou} \& {Jiang}}{{Zhang} et~al.}{2014}]{2014ApJ...786...42Z}
{Zhang} S.,  {Wang} H.,  {Wang} T.,  {Xing} F.,  {Zhang} K.,  {Zhou} H.,
  {Jiang} P.,  2014, \apj, 786, 42

\bibitem[\protect\citeauthoryear{{Zhou}, {Wang}, {Yuan}, {Lu}, {Dong}, {Wang}
  \& {Lu}}{{Zhou} et~al.}{2006}]{2006ApJS..166..128Z}
{Zhou} H.,  {Wang} T.,  {Yuan} W.,  {Lu} H.,  {Dong} X.,  {Wang} J.,    {Lu}
  Y.,  2006, \apjs, 166, 128


\end{thebibliography}





\bsp	
\label{lastpage}
\end{document}